\documentclass[aps, prd, 10pt, notitlepage, superscriptaddress, nofootinbib,numbers,showpacs]{revtex4-2}

\usepackage[utf8]{inputenc}
\usepackage[T1]{fontenc}
\usepackage{anyfontsize}
\usepackage{amsmath}
\usepackage{amssymb}
\usepackage{amsfonts}
\usepackage{mathrsfs}
\usepackage{enumitem}
\usepackage{amsmath}
\DeclareMathOperator{\sech}{sech}

\usepackage{microtype}
\usepackage[normalem]{ulem}

\usepackage{graphicx}
\usepackage{epsfig}
\usepackage{float}

\usepackage[dvipsnames]{xcolor}
\usepackage{hyperref}
\hypersetup{
    colorlinks=true,
    citecolor=Purple,
    linkcolor=Purple,
    urlcolor=Purple,
    linktocpage=true,
    breaklinks=true
}
\usepackage{soul} 
\usepackage[capitalize]{cleveref}

\begin{document}
\title{Traversable Wormholes Supported by Entropy-Inspired Effective Matter Sectors}

\author{Jonathan A. Rebouças}
\email{jalvesreboucas@ifce.edu.br}
\affiliation{Instituto Federal de Educação, Ciência e Tecnologia do Ceará (IFCE), Iguatu, Brazil}

\author{Francisco Bento Lustosa}
\email{chico.lustosa@uece.br}
\affiliation{Universidade Estadual do Cear\'a (UECE), Faculdade de Educa\c{c}\~ao, Ci\^encias e Letras de Iguatu, Av. D\'ario Rabelo s/n, Iguatu - CE, 63.500-000 - Brazil.}

\author{Celio R. Muniz}
\email{celio.muniz@uece.br}
\affiliation{Universidade Estadual do Cear\'a (UECE), Faculdade de Educa\c{c}\~ao, Ci\^encias e Letras de Iguatu, Av. D\'ario Rabelo s/n, Iguatu - CE, 63.500-000 - Brazil.}


\date{\today}
\begin{abstract}
The thermodynamic interpretation of gravity suggests that modifications of horizon entropy can be encoded in effective geometries and matter distributions. Motivated by the entropy--geometry correspondence, we investigate Morris--Thorne wormholes constructed from density profiles associated with Barrow, Tsallis, Kaniadakis, logarithmic, and exponential black-hole entropies. The wormhole construction retains each entropy-induced density as input to the shape-function equation, while the radial and tangential pressures are reconstructed independently from the wormhole field equations, a barotropic radial equation of state, anisotropic conservation, and throat regularity. Barrow and Tsallis yield algebraic negative-density distributions; Kaniadakis and exponential corrections produce localized profiles; and the logarithmic sector admits both negative-density and positive-density phantom-like regimes. For every regular configuration considered, radial null energy condition violation at the throat follows from the flare-out condition. Its areal-volume integral is finite and negative for every nonzero deformation: it becomes unbounded when the regular Barrow, Tsallis, and Kaniadakis branches approach their undeformed values, but remains finite in the logarithmic and exponential limits. Removing the density deformation also fails to commute with taking the infinite-radius limit in the Barrow, Tsallis, and Kaniadakis sectors, whereas the two operations agree for the logarithmic and exponential sectors in their pole-free domains. The selected entropy-induced densities can therefore be embedded consistently in traversable-wormhole reconstructions, with the pressure sector fixed by the wormhole geometry and the global source behavior controlling both the asymptotic mass and the integrated radial exoticity.
\end{abstract}

\maketitle

\section{Introduction}
\label{sec:introduction}

Traversable wormholes are among the most direct ways in which general relativity exposes the nontrivial relation between geometry, topology, and matter. The idea of connecting distinct asymptotic regions dates back to the Einstein--Rosen bridge, while the modern formulation of traversable wormholes was developed in the Morris--Thorne framework and in related scalar-field solutions such as the Ellis--Bronnikov geometry \cite{EinsteinRosen1935,Ellis1973,Bronnikov1973,MorrisThorne,Visser1989,VisserBook}, as well as in more recent generalizations supported by combined phantom-scalar and electromagnetic sectors \cite{Crispim2026}. A central lesson from these constructions is that a static traversable throat cannot be maintained by ordinary classical matter in general relativity. The flare-out condition at the throat is directly tied to violations of the null energy condition, and this requirement is closely related to topological censorship and to quantum-inequality constraints on negative energy densities \cite{MorrisThorneYurtsever1988,FriedmanTopologicalCensorship,FordRoman1996,HochbergVisser1998,VisserKarDadhich2003}. For this reason, a large part of wormhole physics has focused either on minimizing the amount of exotic matter or on replacing it by effective sources whose origin may be quantum, semiclassical, or geometric \cite{Lobo2005,Sushkov2005,LoboOliveira2009,Lobo2012,Radhakrishnan2024}. Recent developments have also shown that traversable wormhole configurations may be supported by nonstandard but regular matter sectors within general relativity, further motivating the search for effective mechanisms capable of sustaining a throat \cite{KonoplyaZhidenko2022}.

This search for effective sources is not merely a technical device. In several extensions of general relativity, the field equations may be written in a form in which additional curvature, torsion, scalar, or higher-dimensional contributions behave as an effective stress-energy tensor. In this interpretation, the matter sector threading the throat may satisfy more conventional energy requirements while the effective gravitational sector supplies the non-classical contribution required by the wormhole geometry \cite{NojiriOdintsov2011,CapozzielloDeLaurentis2011,Harko2013,BohmerHarkoLobo2012,Kanti2011}. Effective geometric sectors have also been explored in Einstein--Cartan theory, curvature-based extended theories of gravity, and quantum-inspired Casimir wormholes with generalized uncertainty-principle corrections \cite{BattistaDiGreziaManfredoniaMiele2017,DeFalcoBattistaCapozzielloDeLaurentis2021EPJC,BattistaCapozzielloErrehymy2024}. Related reconstructions based on loop-quantum-gravity effective sources further illustrate how quantum-corrected matter profiles can generate traversable wormhole geometries \cite{Cruz2024}. A closely related phenomenological strategy consists in prescribing physically motivated density profiles and then reconstructing the corresponding wormhole geometry. This type of ``inspired'' construction has been widely used, for instance, with noncommutative-geometry motivated sources, Casimir-like negative-energy distributions, and dark-matter halo profiles \cite{Nicolini2006,GarattiniLobo2009,Rahaman2014,Garattini2019,Navarro1996,Rahaman2016}. Recent examples include Yang--Mills Casimir sources, for which traversability, energy conditions, and the Tolman--Oppenheimer--Volkoff equilibrium can be examined explicitly \cite{Santos2024}. Such approaches are useful because they allow one to test whether a given microscopic or effective matter pattern can sustain a traversable throat without first committing to a complete fundamental theory. Related constructions based on quantum-improved dark-matter profiles and matter-first rational approximations of prescribed densities have also been shown to yield controlled wormhole geometries while preserving the intended source sector \cite{Reboucas2026ASG,Reboucas2026Pade}. In addition, wormholes have been investigated as possible black-hole mimickers and in connection with observational diagnostics based on accretion dynamics and the general relativistic Poynting--Robertson effect \cite{DeFalcoBattistaCapozzielloDeLaurentis2020PRD,DeFalcoBattistaCapozzielloDeLaurentis2021PRD}.

At the same time, the thermodynamic interpretation of gravity has become one of the main conceptual bridges between general relativity, quantum theory, and statistical physics. The discovery that black holes carry entropy and temperature established that horizons are thermodynamic systems, while the Euclidean path-integral, Noether-charge, and holographic formulations made clear that gravitational dynamics is deeply connected with boundary information and horizon degrees of freedom \cite{Bekenstein1973,Hawking1975,GibbonsHawking1977,Wald1993,IyerWald1994,tHooft1993,Susskind1995,Bousso2002}. This viewpoint was strengthened by the derivation of Einstein's equations from local horizon thermodynamics, by the interpretation of gravitational field equations as thermodynamic identities, and by entropic-force proposals in which gravity emerges from entropy gradients associated with holographic screens \cite{Jacobson1995,Padmanabhan2010,Verlinde2011,Visser2011}. In cosmological settings, generalized horizon entropies have been used to establish direct correspondences between entropic cosmology and modified gravity, including general modified theories of gravity, $F(T)$ and $F(Q)$ gravity, and horizon-thermodynamic formulations compatible with general equations of state for the matter sector \cite{NojiriOdintsovPaulSenGupta2024,NojiriOdintsov2025PDU,NojiriOdintsovPaulSenGupta2026}. More recent developments based on quantum information and entanglement further suggest that spacetime geometry may be reconstructed, at least in appropriate regimes, from the organization of microscopic degrees of freedom \cite{VanRaamsdonk2010,Faulkner2014,Jacobson2016}. From this broader perspective, gravity may be regarded not only as a classical field, but also as a macroscopic manifestation of an underlying statistical or quantum structure.

Modified black-hole entropies provide a natural arena in which these ideas can be explored. Corrections to the Bekenstein--Hawking area law arise in several contexts, including quantum geometry, logarithmic fluctuations, nonextensive statistical mechanics, relativistic generalized statistics, and non-perturbative quantum corrections \cite{KaulMajumdar2000,DasMajumdarBhaduri2002,LogCorr,Tsallis,TsallisCirto2013,Kaniadakis,Barrow,Saridakis2020,ChatterjeeGhosh2020}. Generalized entropy functionals have also been formulated in broader frameworks in which several commonly used entropy proposals, including Tsallis, R\'enyi, Sharma--Mittal, Barrow, Kaniadakis, and loop-quantum-gravity entropies, arise as particular limits or realizations \cite{NojiriOdintsovFaraoni2022,NojiriOdintsovPaul2022}. Such corrections are usually introduced at the thermodynamic level, but they may also be interpreted as hints that the microscopic structure of the horizon backreacts on the spacetime geometry. A distinct route begins at the level of gravitational dynamics. Applying Jacobson's local horizon-thermodynamic argument to a generalized entropy generically leads to field equations that depend on the horizon area and to an effective gravitational coupling that varies with that area \cite{LuDiGennaroOng2025}. In this approach, the gravity theory is modified first and compatible black-hole solutions are sought afterward. By contrast, the entropy--geometry correspondence associates a chosen entropy functional with a modified black-hole metric and an anisotropic effective matter sector \cite{Anand}. Its thermodynamic, topological, and optical consequences were further investigated in Ref.~\cite{EntropyPaper}. A recent comparison of the two approaches is given in Ref.~\cite{Ong2026GeometryEntropy}. The important point for the present work is that this correspondence provides explicit effective density profiles associated with different entropy deformations.

The purpose of this paper is to investigate whether the density profiles generated by modified black-hole entropies can act as effective source functions for traversable wormholes. Importantly, the construction is entropy-inspired rather than a derivation of wormholes directly from generalized entropy: only the entropy-induced effective density profile is retained, while the radial and tangential pressures are reconstructed independently from the Morris--Thorne field equations and throat conditions. In particular, the constant radial equation-of-state parameter is geometrically selected by the regularity of the redshift derivative at the throat.

The paper is organized as follows. In Sec.~\ref{sec:framework}, we briefly review the entropy--geometry correspondence and introduce the Morris--Thorne wormhole geometry, including the throat, flare-out, asymptotic-flatness, redshift-regularity, and embedding requirements. In Sec.~\ref{sec:general}, we derive the Einstein-tensor components, reconstruct the shape function from a prescribed density, introduce the asymptotic mass and the relevant limit-order distinctions, and then analyze the throat-selected equation-of-state parameter, the pointwise and integrated energy-condition diagnostics, and the Tolman--Oppenheimer--Volkoff balance. In Sec.~\ref{sec:profiles}, we analyze five entropy-inspired sectors: Barrow, Tsallis, Kaniadakis, logarithmic, and exponential. For each case, we discuss the induced geometry, the behavior of the redshift derivative, the embedding structure, the pointwise energy conditions, the integrated radial exoticity, and the equilibrium forces. The asymptotic and exoticity regimes are then compared in Sec.~\ref{sec:mass_comparison}. Finally, in Sec.~\ref{sec:conclusions}, we summarize the main results and outline possible extensions of the present framework.

\section{Entropy-inspired wormhole framework}
\label{sec:framework}

This section introduces the framework used in the paper. We first review the entropy--geometry correspondence and its associated effective matter sector. We then present the Morris--Thorne wormhole geometry, emphasizing the throat, flare-out, horizon-avoidance, and asymptotic-flatness conditions. Finally, we discuss the embedding geometry as a visual diagnostic of the spatial wormhole structure.

\subsection{The entropy--geometry correspondence}

The entropy--geometry correspondence proposed in Ref.~\cite{Anand} provides a constructive way of associating a given black-hole entropy functional with a modified spacetime geometry and, consequently, with an effective matter sector. The starting point is the static and spherically symmetric line element
\begin{equation}
 ds^2=-f(r)dt^2+\frac{dr^2}{f(r)}+r^2d\Omega^2,
\end{equation}
where $f(r)$ is the metric function and $r$ is the areal radius. The event horizon is located at $r=r_+$, defined by
\begin{equation}
 f(r_+)=0.
\end{equation}
The construction assumes that the black-hole entropy can be written as a function of the horizon radius, $S=S(r_+)$. Here $S'(r_+)$, $S''(r_+)$, and higher derivatives denote derivatives with respect to the horizon radius. The parameter $M$ plays the role of the mass parameter, while the auxiliary function $g(r)$ encodes how the mass contribution enters the metric through the ansatz
\begin{equation}
 f(r)=1-Mg(r).
 \label{eq:entropy_metric_ansatz}
\end{equation}
In the Bekenstein--Hawking limit, this prescription must reproduce the usual Schwarzschild behavior. For a generic entropy functional, however, the function $g(r)$ is deformed, and this deformation may be interpreted as the geometric backreaction induced by the modified entropy law.

The key thermodynamic input is the first law of black-hole mechanics,
\begin{equation}
 dM=T_H dS,
\end{equation}
together with the standard expression for the Hawking temperature associated with the metric function,
\begin{equation}
 T_H=\frac{f'(r_+)}{4\pi}.
\end{equation}
Using the horizon condition and the ansatz for $f(r)$, one obtains a relation between the function $g(r)$ and the entropy derivative evaluated at the horizon,
\begin{equation}
 g(r_+)=\frac{4\pi}{S'(r_+)}.
 \label{eq:entropy_horizon_relation}
\end{equation}
The relation in Eq.~\eqref{eq:entropy_horizon_relation} is local to the event horizon and, by itself, does not uniquely determine the radial function away from $r=r_+$. Following the entropy--geometry correspondence of Ref.~\cite{Anand}, we therefore make the explicit global-extension assumption $g(r)\equiv 4\pi/S'(r)$ for generic $r$. This off-horizon prescription is the defining ansatz of the correspondence, rather than a further consequence of the horizon first-law calculation. Combined with the metric ansatz in Eq.~\eqref{eq:entropy_metric_ansatz}, it carries the chosen entropy functional into the static, spherically symmetric exterior geometry and recovers the Schwarzschild metric in the Bekenstein--Hawking limit.
\begin{equation}
 f(r)=1-\frac{4\pi M}{S'(r)}.
\label{eq:entropy_metric}
\end{equation}
Thus, once this global-extension ansatz and an entropy functional are specified, the metric function is fixed within the adopted correspondence. This is the precise sense in which the entropy deformation is used to prescribe the geometric backreaction.

The modified geometry described by Eq.~\eqref{eq:entropy_metric} can be interpreted within Einstein gravity as being sourced by an effective anisotropic fluid. In other words, one may keep the Einstein tensor on the geometric side of the field equations and read the deviations from the Schwarzschild geometry as an effective energy-momentum tensor. For the metric above, this procedure leads to the effective density
\begin{equation}
 \rho(r)=-\frac{M\left[rS''(r)-S'(r)\right]}{2r^2S'(r)^2},
\label{eq:rho_entropic_general}
\end{equation}
the radial pressure
\begin{equation}
 p_r(r)=-\rho(r),
\label{eq:pr_entropic_general}
\end{equation}
and the tangential pressure
\begin{equation}
 p_t(r)=\frac{M\left\{S'(r)\left[rS'''(r)+2S''(r)\right]-2rS''(r)^2\right\}}{4rS'(r)^3}.
\label{eq:pt_entropic_general}
\end{equation}
The relation $p_r=-\rho$ is therefore not imposed as an independent matter equation of state; rather, it follows from the particular black-hole metric ansatz used in the entropy--geometry construction. The tangential pressure, in turn, depends on higher derivatives of the entropy functional and encodes the anisotropic character of the effective source.

This framework gives a clear physical interpretation to modified entropies. If the entropy is changed, the corresponding derivative $S'(r)$ changes the metric function, and the resulting geometry can be represented as being supported by an effective matter distribution. Different entropy functionals therefore generate different density profiles and different anisotropic stress sectors. For instance, power-law, logarithmic, hyperbolic, or exponential corrections to the entropy lead to qualitatively distinct effective sources. In the original black-hole setting, the complete set $\{\rho,p_r,p_t\}$ is fixed once the entropy is chosen.

In the present work, the entropy--geometry correspondence is used as a generator of effective density profiles for a Morris--Thorne reconstruction. The profiles obtained from Eq.~\eqref{eq:rho_entropic_general} are inserted as entropy-inspired source functions, while the radial and tangential pressures are reconstructed within the wormhole spacetime itself. This density-sector projection isolates the geometric role of the entropy-induced radial distribution and makes it possible to test, within the traversable-wormhole framework, whether the corresponding source can satisfy the throat, flare-out, redshift-regularity, and asymptotic-flatness requirements.

\subsection{Wormhole geometry}

We consider the Morris--Thorne line element \cite{MorrisThorne}
\begin{equation}
 ds^2=-e^{2\Phi(r)}dt^2+\frac{dr^2}{1-b(r)/r}
 +r^2\left(d\theta^2+\sin^2\theta\,d\phi^2\right),
 \label{eq:MTmetric2}
\end{equation}
where $r$ is the areal radius, $\Phi(r)$ is the redshift function, and $b(r)$ is the shape function. The function $\Phi(r)$ determines the gravitational redshift between different radial positions and controls the temporal component of the metric. The shape function $b(r)$ determines the spatial geometry of the wormhole slice and, in particular, specifies how the radial coordinate departs from the usual Schwarzschild-like form.

The wormhole throat is defined as the minimum areal radius of the spacetime. It is located at $r=r_0$ and is characterized by
\begin{equation}
 b(r_0)=r_0.
\label{eq:throat}
\end{equation}
The physical domain of the wormhole is therefore restricted to $r\ge r_0$. Outside the throat one must require
\begin{equation}
 b(r)<r,
 \qquad r>r_0,
\end{equation}
so that the radial metric component remains well defined and the spatial geometry has the correct Lorentzian signature. Thus, while the equality $b(r_0)=r_0$ identifies the throat, the inequality $b(r)<r$ guarantees that the region outside the throat is regular in the chosen radial coordinate.

A useful way to understand the throat is through the proper radial distance,
\begin{equation}
 l(r)=\pm\int_{r_0}^{r}\frac{d\tilde r}
 {\sqrt{1-b(\tilde r)/\tilde r}}.
\end{equation}
The two signs describe the two sides of the wormhole, while $l=0$ corresponds to the throat. In terms of this proper coordinate, the throat is a minimum of the areal radius. This means that the geometry must open outward on both sides of $r=r_0$, rather than closing as in an ordinary spherical geometry. This requirement is encoded in the flare-out condition,
\begin{equation}
 \frac{b(r)-rb'(r)}{b(r)^2}>0.
\label{eq:flareout_general2}
\end{equation}
At the throat, this condition reduces to the standard local constraint
\begin{equation}
 b'(r_0)<1.
\label{eq:flareout_throat2}
\end{equation}
This inequality is one of the central geometric requirements for a traversable wormhole. It ensures that the throat is not a marginal surface, but a genuine minimum-radius hypersurface. As will be discussed below, this same condition is directly connected with the violation of the radial null energy condition at the throat.

The redshift function also carries an important geometric constraint. A traversable wormhole must not possess an event horizon at or outside the throat. Since the temporal metric component is given by $g_{tt}=-e^{2\Phi(r)}$, this requires $\Phi(r)$ to remain finite throughout the physical domain. If $\Phi(r)$ diverges negatively, the factor $e^{2\Phi(r)}$ may vanish and a horizon is formed, preventing two-way traversability. Therefore, the regularity of $\Phi(r)$ is not merely a technical condition, but a necessary requirement for the spacetime to describe a traversable wormhole.

In addition to the throat and flare-out conditions, we impose asymptotic flatness. This requires the shape function to become negligible compared with the radial coordinate and the redshift function to approach a constant at spatial infinity. By a suitable normalization of the time coordinate, this constant may be set to zero, so that
\begin{equation}
 \frac{b(r)}{r}\rightarrow 0,
\qquad
\Phi(r)\rightarrow 0,
\qquad r\rightarrow\infty.
\label{eq:asymptotic_conditions}
\end{equation}
These conditions ensure that far from the throat the geometry approaches the Minkowski spacetime. In the following sections, each density profile will therefore be tested against the same set of geometric requirements: the existence of a throat, the flare-out condition, the absence of horizons through a regular redshift function, and the vanishing of the shape-function-to-radius ratio.

\subsection{Embedding geometry}
\label{sec:embedding_geometry}

To visualize the spatial geometry of the wormhole, we consider an equatorial and constant-time slice, defined by $t=\mathrm{const.}$ and $\theta=\pi/2$. The induced two-dimensional line element is
\begin{equation}
 dl^2=\frac{dr^2}{1-b(r)/r}+r^2d\varphi^2.
 \label{eq:slice_metric2}
\end{equation}
This geometry can be embedded in a three-dimensional Euclidean space with line element
\begin{equation}
 dl^2=dz^2+dr^2+r^2d\varphi^2.
 \label{eq:euclidean_embedding_metric2}
\end{equation}
Matching the radial parts of the two metrics gives
\begin{equation}
 \frac{dz}{dr}=\pm\sqrt{\frac{b(r)}{r-b(r)}}.
 \label{eq:embedding2}
\end{equation}
The two signs correspond to the two asymptotic regions connected by the throat.

At the throat, $r=b(r)$, and therefore $dz/dr$ diverges. This divergence does not indicate a physical singularity. It only means that the embedded surface has a vertical tangent at the minimum areal radius, which is precisely the expected geometric behavior of a wormhole throat.

The opening of the embedded surface is controlled by
\begin{equation}
 \frac{d^2r}{dz^2}
 =
 \frac{b(r)-rb'(r)}{2b(r)^2}.
 \label{eq:embedding_flareout2}
\end{equation}
At the throat, this becomes
\begin{equation}
 \left.\frac{d^2r}{dz^2}\right|_{r_0}
 =
 \frac{1-b'(r_0)}{2r_0}.
 \label{eq:embedding_flareout_throat2}
\end{equation}
Thus, the condition $b'(r_0)<1$ guarantees that the embedded surface opens outward at the throat. In the entropy-inspired sectors analyzed below, the embedding diagrams will be used not only as visual illustrations, but also as geometric representations of how each effective density profile shapes the spatial wormhole slice.

\section{General field equations and diagnostic quantities}
\label{sec:general}

This section presents the field equations and diagnostic quantities used in the analysis. We derive the effective matter variables from the Morris--Thorne geometry and show how the density profile fixes the shape function. We then use the asymptotic behavior of this function to define the mass of the two wormhole ends and to explain why removing an entropy deformation need not commute with taking the infinite-radius limit. Finally, we discuss how redshift regularity constrains the radial equation-of-state parameter and introduce the pointwise energy conditions, an integrated measure of radial exoticity, and the anisotropic TOV balance used to characterize each sector.

\subsection{Einstein tensor and effective matter variables}

For the Morris--Thorne metric \eqref{eq:MTmetric2}, we write the anisotropic effective source in mixed coordinate components as
\begin{equation}
 T^{\mu}_{\ \nu}=\mathrm{diag}\left(-\rho,p_r,p_t,p_t\right),
\end{equation}
where $\rho(r)$ is the energy density, $p_r(r)$ is the radial pressure, and $p_t(r)$ is the tangential pressure. In units $G=c=1$, Einstein's equations are
\begin{equation}
 G^{\mu}_{\ \nu}=8\pi T^{\mu}_{\ \nu}.
\end{equation}
The non-vanishing independent components of the Einstein tensor are
\begin{align}
 -G^{t}_{\ t} = & \frac{b'(r)}{r^2} = 8\pi\rho(r),\label{eq:rho_MT2}\\
 G^{r}_{\ r} = &
 -\frac{b(r)}{r^3}
 +\frac{2}{r}\left(1-\frac{b(r)}{r}\right)\Phi'(r) = 8\pi p_r(r), \label{eq:pr_MT2}\\
 G^{\theta}_{\ \theta} = &
\, G^{\phi}_{\ \phi}
 =
 \left(1-\frac{b(r)}{r}\right)
 \left[
 \Phi''+(\Phi')^2
 -\frac{b'r-b}{2r(r-b)}\Phi'
 -\frac{b'r-b}{2r^2(r-b)}
 +\frac{\Phi'}{r}
 \right] = 8\pi p_t(r). \label{eq:pt_MT2}
\end{align}

These expressions show explicitly how the wormhole geometry fixes the effective matter variables. Conversely, once a density profile is prescribed, the shape function can be reconstructed from Eq.~\eqref{eq:rho_MT2}. This is the strategy adopted below for the entropy-inspired sectors.

\subsection{Shape function and geometric constraints}

Once the density profile is specified, the shape function follows from
\begin{equation}
 b(r)=r_0+8\pi\int_{r_0}^{r}\rho(\tilde r)\,\tilde r^2 d\tilde r.
 \label{eq:b_general2}
\end{equation}

Using Eq.~\eqref{eq:rho_MT2}, Eq.~\eqref{eq:flareout_throat2} may also be written as
\begin{equation}
 8\pi r_0^2\rho(r_0)<1.
 \label{eq:flareout_density}
\end{equation}
This form is particularly useful in the present construction because the density profile is the quantity imported from the entropy--geometry correspondence.

The physical domain is taken to be $r\ge r_0$. For $r>r_0$, one must also require
\begin{equation}
 b(r)<r,
\end{equation}
so that the radial metric component remains regular and the coordinate $r$ keeps its interpretation as an areal radius outside the throat.

\subsection{Asymptotic mass and deformation limits}
\label{sec:asymptotic_mass_general}

The shape function contains information beyond the local throat geometry. In particular, the condition $b(r)/r\to0$ ensures the required asymptotic falloff, but it does not imply that the spacetime is massless. When $b(r)$ approaches a finite constant, this constant fixes the leading asymptotic correction to the spatial metric and therefore determines the Arnowitt--Deser--Misner (ADM) mass. Taking the upper limit of Eq.~\eqref{eq:b_general2}, we define
{
\begin{align}
 b_\infty
 &\equiv \lim_{r\to\infty}b(r)
 =r_0+8\pi\int_{r_0}^{\infty}\rho(r)r^2\,dr,
 \label{eq:binf_general}\\
 M_{\mathrm{ADM}}^{+}=M_{\mathrm{ADM}}^{-}
 &=\frac{b_\infty}{2}
 =\frac{r_0}{2}+4\pi\int_{r_0}^{\infty}\rho(r)r^2\,dr.
 \label{eq:madm_general}
\end{align}
}
The superscripts label the two asymptotically flat ends obtained by joining identical copies of the branch $r\geq r_0$. The two masses are therefore equal in the reflection-symmetric construction considered here. Equation~\eqref{eq:madm_general} agrees with the standard Morris--Thorne asymptotic prescription \cite{CataldoCidLabrana2026} and separates the boundary term $r_0/2$, fixed by the throat condition, from the improper integral of the prescribed density. The parameter $M$ inherited from the black-hole profile controls the amplitude of this density, but it is not generally the ADM mass of the reconstructed wormhole.

The role of reflection symmetry in this mass assignment can be illustrated by the spinning Ellis family. In that case, the mass at each end is identified after the corresponding metric has been written in an asymptotically Minkowskian frame; the two masses agree for symmetric solutions but may differ when reflection symmetry is relaxed \cite{ChewKleihausKunz2016}. The present static construction uses identical copies with the same time normalization and asymptotic shape coefficient, so the equality in Eq.~\eqref{eq:madm_general} likewise follows from reflection symmetry. This comparison also distinguishes the limiting configurations: spinning Ellis wormholes approach an extremal Kerr geometry as the phantom charge vanishes, whereas the entropy-deformation limits considered below remain within the static wormhole family.

Writing the mass as an improper integral also shows why a deformation limit requires some care. When an entropy parameter approaches its Bekenstein--Hawking value, the corresponding effective density may vanish at every fixed radius. We shall call this local operation the undeformed density limit. The radial domain, however, is unbounded, and pointwise suppression does not guarantee that the integrated source also vanishes. A profile may become progressively weaker at finite radius while retaining a nonzero contribution over increasingly large distances. In that case, removing the deformation before taking $r\to\infty$ gives a different result from taking the asymptotic limit first.

The two orders must therefore be evaluated from each explicit density and shape function. This comparison concerns the pointwise source, the improper mass integral, and the asymptotic coefficient; it does not by itself define a complete geometric limit of the spacetime family, which requires additional identifications \cite{Geroch1969,BengtssonHolstJakobsson2014}. With the global role of $b(r)$ established, we now return to the local matter closure and to the regularity of the redshift function at the throat.

\subsection{Equation of state and redshift regularity}

To close the system, we impose a barotropic radial equation of state,
\begin{equation}
 p_r(r)=w\rho(r).
 \label{eq:eos2}
\end{equation}
The prescribed density first fixes the shape function through Eq.~\eqref{eq:b_general2}. Equation~\eqref{eq:eos2} then closes the radial sector, but its constant parameter is not inherited from the black-hole pressure in Eq.~\eqref{eq:pr_entropic_general}; it is selected by the throat-regularity condition in Eq.~\eqref{eq:w_th_general}. Once $b(r)$, $w$, and $\Phi(r)$ are fixed, the angular field equation~\eqref{eq:pt_MT2} determines $p_t(r)$, while Eq.~\eqref{eq:pt_from_TOV2} provides the equivalent conservation check. This sequential closure reconstructs the wormhole stress tensor without introducing a second equation of state for the tangential sector.

There is also a geometric reason not to impose $w=-1$ from the outset. Combining Eqs.~\eqref{eq:rho_MT2}, \eqref{eq:pr_MT2}, and \eqref{eq:eos2}, one obtains
\begin{equation}
 \Phi'(r)=
 \frac{b(r)+w r b'(r)}
 {2r\left[r-b(r)\right]}.
 \label{eq:redshift_general2}
\end{equation}
At the throat, the denominator vanishes because $b(r_0)=r_0$. In order for this apparent divergence to be removable, the numerator must vanish at the same point:
\begin{equation}
 b(r_0)+w r_0 b'(r_0)=0.
 \label{eq:redshift_regular_condition1}
\end{equation}
Using the throat condition, this gives the throat-selected value
\begin{equation}
 w=w_{\rm th}\equiv -\frac{1}{b'(r_0)}
 =
 -\frac{1}{8\pi r_0^2\rho(r_0)}.
 \label{eq:w_th_general}
\end{equation}
Thus, for a constant equation-of-state parameter, $w$ is not a free parameter once the density profile and the throat radius are specified. Equivalently, if one insists on fixing $w$ in advance, the parameters of the density profile must be adjusted so that the regularity condition is satisfied.

This result also clarifies why the vacuum-like choice $w=-1$ is generally not appropriate for a regular traversable wormhole in the present construction. If $w=-1$ were imposed together with the throat regularity condition, one would obtain $b'(r_0)=1$, which saturates the flare-out condition instead of satisfying it strictly. Therefore, the relation $p_r=-\rho$ belongs naturally to the black-hole effective sector of the entropy--geometry correspondence, but it is not the generic regular choice for the wormhole branch considered here.

When Eq.~\eqref{eq:w_th_general} is imposed, Eq.~\eqref{eq:redshift_general2} takes the indeterminate form $0/0$ at the throat. This is not, by itself, a pathology. It means that the value of $\Phi'(r)$ at $r=r_0$ must be defined by the corresponding limit:
\begin{equation}
 \Phi'(r_0)=
 \lim_{r\to r_0}
 \frac{b(r)+w_{\rm th}r b'(r)}
 {2r\left[r-b(r)\right]}.
 \label{eq:phi_prime_limit}
\end{equation}
If $b(r)$ is sufficiently smooth around the throat, with
\begin{equation}
 B_1=b'(r_0),
 \qquad
 B_2=b''(r_0),
\end{equation}
this limit becomes
\begin{equation}
 \Phi'(r_0)=
 \frac{
 B_1+w_{\rm th}\left(B_1+r_0B_2\right)
 }
 {2r_0\left(1-B_1\right)}.
 \label{eq:phi_prime_limit_explicit}
\end{equation}
The apparent $0/0$ behavior is therefore removable provided that $b'(r_0)\neq0$, $b'(r_0)\neq1$, and $b''(r_0)$ remains finite. The condition $b'(r_0)\neq0$ excludes the limit in which the entropy-inspired source vanishes at the throat, while $b'(r_0)\neq1$ excludes the marginal flare-out case.

Away from the throat, the redshift function is obtained by quadrature,
\begin{equation}
 \Phi(r)=\Phi(r_\star)+
 \int_{r_\star}^{r}
 \frac{b(\tilde r)+w_{\rm th}\tilde r b'(\tilde r)}
 {2\tilde r\left[\tilde r-b(\tilde r)\right]}
 d\tilde r,
 \qquad r_\star>r_0,
 \label{eq:redshift_integral2}
\end{equation}
where the additive constant is fixed by the asymptotic normalization. The condition that $\Phi(r)$ remain finite throughout the physical domain must be checked for every entropy-inspired sector.

\subsection{Energy conditions}

The energy conditions are written as
\begin{align}
 \text{NEC}:\quad &\rho+p_r\ge 0, \, \rho+p_t\ge 0,\nonumber\\
 \text{WEC}: \quad &\rho\ge 0, \, \rho+p_r\ge 0, \, \rho+p_t\ge 0,\nonumber\\
 \text{DEC}: \quad & \rho\ge |p_r|, \, \rho\ge |p_t|,\nonumber\\
 \text{SEC}: \quad & \rho+p_r\ge 0, \, \rho+p_t\ge 0, \, \rho+p_r+2p_t\ge 0.
 \label{eq:energy_conditions2}
\end{align}
For the barotropic equation of state,
\begin{equation}
 \rho+p_r=(1+w)\rho,
 \qquad
 \rho-p_r=(1-w)\rho.
 \label{eq:NEC_w2}
\end{equation}
However, for a regular traversable configuration, the value of $w$ must be the throat-selected one. Therefore,
\begin{equation}
 \rho+p_r=
 \left[
 1-\frac{1}{b'(r_0)}
 \right]\rho(r).
 \label{eq:NEC_wth_general}
\end{equation}
At the throat this becomes
\begin{equation}
 \left.\left(\rho+p_r\right)\right|_{r_0}
 =
 \frac{b'(r_0)-1}{8\pi r_0^2}.
 \label{eq:NEC_throat_geometry}
\end{equation}
Hence, the flare-out condition $b'(r_0)<1$ automatically implies violation of the radial NEC at the throat. This is an important point: after imposing redshift regularity, the radial NEC violation is no longer controlled by an arbitrary choice of $w$, but follows directly from the local geometry of the wormhole throat.

In addition to the radial and tangential NECs, we shall monitor the strong energy condition combination
\begin{equation}
 \rho+p_r+2p_t.
 \label{eq:SEC_combination_general}
\end{equation}
This quantity is useful because it probes the combined effect of radial tension, tangential stresses, and redshift gradients. Two sectors may display similar radial NEC violation while having quite different tangential responses. The SEC combination therefore helps to distinguish how the exoticity required at the throat is redistributed by each entropy-inspired source.

\subsection{Integrated radial exoticity}
\label{sec:integrated_exoticity}

The pointwise conditions identify where the effective source is exotic, but they do not measure its accumulated radial contribution. Following the volume-integral approach to wormhole matter \cite{VisserKarDadhich2003,NandiZhangKumar2004}, we define an integrated diagnostic for either asymptotic end:
{
\begin{equation}
 \mathcal{I}_{\mathrm{NEC}}^{+}
 =
 \mathcal{I}_{\mathrm{NEC}}^{-}
 \equiv
 4\pi\int_{r_0}^{\infty}
 \left[\rho(r)+p_r(r)\right]r^2\,dr .
 \label{eq:integrated_nec_definition}
\end{equation}
}
The equality follows from the reflection-symmetric construction, in which the same branch is used on both sides of the throat. Thus Eq.~\eqref{eq:integrated_nec_definition} is the contribution from one end, while the total over the two ends is twice this value. The measure and normalization are part of the definition: here the integral is used as a common diagnostic in the areal coordinate of Eq.~\eqref{eq:MTmetric2}, rather than as a universal matter charge or as the null-geodesic averaged null energy condition (ANEC) integral.

Because the equation-of-state coefficient is constant along each regular solution, Eqs.~\eqref{eq:NEC_w2}, \eqref{eq:rho_MT2}, and \eqref{eq:binf_general} give the exact identity
{
\begin{equation}
 \mathcal{I}_{\mathrm{NEC}}^{+}
 =
 \mathcal{I}_{\mathrm{NEC}}^{-}
 =
 \frac{1+w_{\rm th}}{2}\left(b_\infty-r_0\right),
 \qquad
 \mathcal{I}_{\mathrm{NEC}}^{\mathrm{tot}}
 =2\mathcal{I}_{\mathrm{NEC}}^{+}.
 \label{eq:integrated_nec_identity}
\end{equation}
}
This relation provides an analytic cross-check of the direct integration in every sector and connects the accumulated radial NEC violation with the same asymptotic shape coefficient that enters the mass analysis. For every admissible nonzero deformation considered below, the integrand is regular at the throat and decays sufficiently fast for the integral to converge. The explicit values and their undeformed limits will be compared only after all five sectors have been developed.

\subsection{Equilibrium and TOV balance}

The conservation equation for the anisotropic effective source,
\begin{equation}
 \nabla_\mu T^{\mu}_{\ r}=0,
\end{equation}
leads to the generalized Tolman--Oppenheimer--Volkoff equation
\begin{equation}
 p_r'(r)+\left[\rho(r)+p_r(r)\right]\Phi'(r)
 +\frac{2}{r}\left[p_r(r)-p_t(r)\right]=0.
 \label{eq:TOV2}
\end{equation}
Solving for the tangential pressure gives
\begin{equation}
 p_t(r)=p_r(r)+\frac{r}{2}
 \left[
 p_r'(r)+\left(\rho+p_r\right)\Phi'(r)
 \right].
 \label{eq:pt_from_TOV2}
\end{equation}
With the barotropic relation and the throat-selected value of the equation-of-state parameter, this becomes
\begin{equation}
 p_t(r)=w_{\rm th}\rho(r)
 +\frac{r}{2}
 \left[
 w_{\rm th}\rho'(r)
 +\left(1+w_{\rm th}\right)\rho(r)\Phi'(r)
 \right].
 \label{eq:pt_wth_general}
\end{equation}
This expression is particularly useful in the present work because, once the density and the redshift derivative are known, the tangential pressure can be reconstructed without directly using the more cumbersome angular component of Einstein's equations.

The TOV equation can also be written as a balance among three contributions,
\begin{equation}
 F_h+F_g+F_a=0,
 \label{eq:forces_general_balance}
\end{equation}
where
\begin{equation}
 F_h=-\frac{dp_r}{dr},
 \qquad
 F_g=-\left(\rho+p_r\right)\Phi',
 \qquad
 F_a=\frac{2}{r}\left(p_t-p_r\right).
 \label{eq:forces_general}
\end{equation}
After imposing throat regularity, these become
\begin{equation}
 F_h=-w_{\rm th}\rho'(r),
 \label{eq:Fh_wth}
\end{equation}
\begin{equation}
 F_g=-\left(1+w_{\rm th}\right)\rho(r)\Phi'(r),
 \label{eq:Fg_wth}
\end{equation}
and
\begin{equation}
 F_a=w_{\rm th}\rho'(r)
 +\left(1+w_{\rm th}\right)\rho(r)\Phi'(r).
 \label{eq:Fa_wth}
\end{equation}
The equilibrium condition is therefore satisfied by construction, but its decomposition into hydrostatic, gravitational, and anisotropic terms remains physically informative. It allows us to identify whether a given entropy-inspired density profile sustains the throat mainly through pressure gradients, through the redshift response, or through anisotropic stresses.

\section{Entropy-inspired density sectors}
\label{sec:profiles}

We now apply the general construction to five entropy-inspired density profiles: Barrow, Tsallis, Kaniadakis, logarithmic, and exponential. These sectors were selected from the modified-entropy families analyzed in Refs.~\cite{Anand,EntropyPaper}, since their associated effective densities provide branches compatible with the formation of traversable wormhole geometries under the Morris--Thorne reconstruction adopted here. The selection is therefore not arbitrary: it is guided by the entropy--geometry correspondence and by the requirement that the resulting density profile, together with the throat-regularity condition, be able to sustain the flare-out geometry.

For each sector, we first analyze the geometry induced by the corresponding density profile, including the shape function, asymptotic behavior, flare-out condition, redshift derivative, and embedding structure. We then study the pointwise energy conditions, evaluate the integrated radial NEC diagnostic, and examine the anisotropic Tolman--Oppenheimer--Volkoff balance. This organization relates the local violation required at the throat to its accumulated radial contribution before comparing the five sectors globally.

\subsection{Barrow-inspired sector}

\subsubsection{Geometry induced by the Barrow-inspired profile}

\begin{figure}[!htp]
    \centering
    \includegraphics[width=0.6\linewidth]{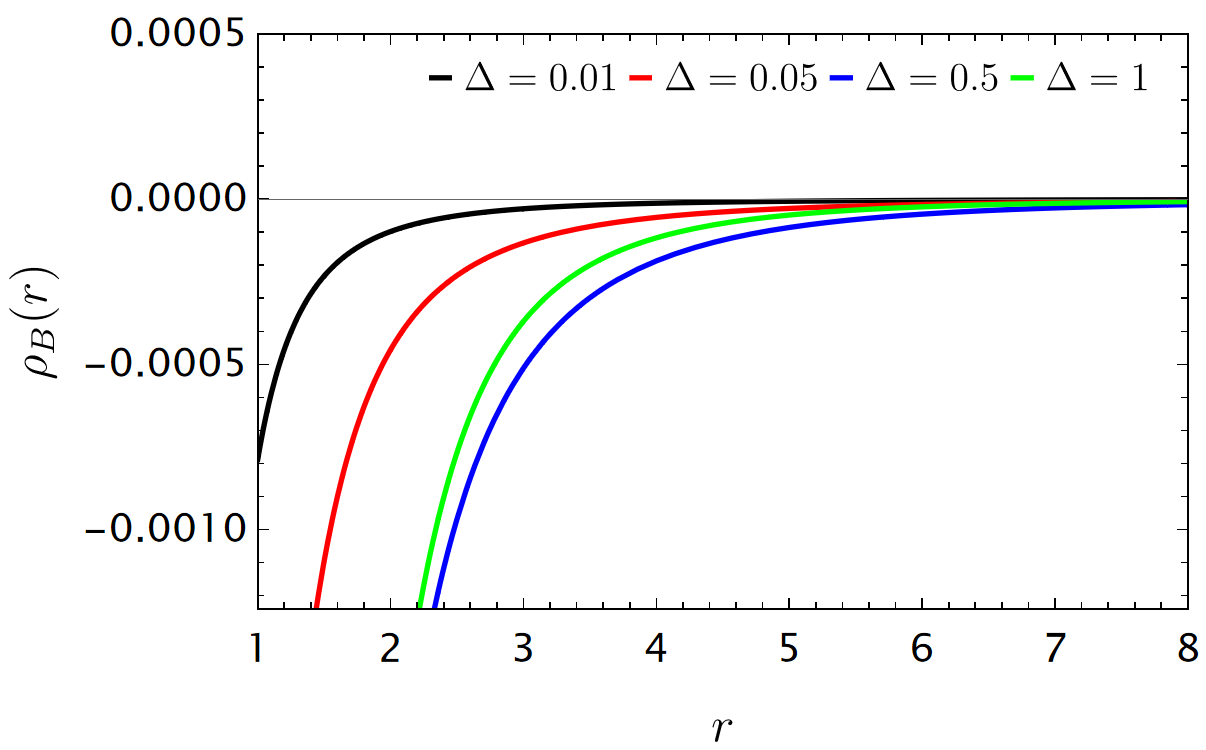}
    \caption{The Barrow-inspired density profile, $\rho_B(r)$, for different values of the deformation parameter $\Delta$ as a function of the radial coordinate $r$, with $M=1$ and $r_0=1$ fixed. The curves correspond to $\Delta=0.01$, $0.1$, $0.5$, and $1.0$.}
    \label{fig:rhobarrow}
\end{figure}

Barrow entropy introduces a fractal deformation of the horizon area law through the parameter $\Delta$ \cite{Barrow}. The corresponding effective density extracted from Ref.~\cite{Anand} is
\begin{equation}
 \rho_B(r)=-\frac{\pi^{-\Delta/2}\Delta M}{2\pi(\Delta+2)}r^{-\Delta-3},
 \qquad 0<\Delta\le 1.
 \label{eq:rho_B2}
\end{equation}
For $M>0$ and $\Delta>0$, the density is strictly negative. Figure~\ref{fig:rhobarrow} includes small values of $\Delta$ to show how the source is suppressed at fixed radius as the deformation is removed.

The shape function follows from Eq.~\eqref{eq:b_general2} and can be integrated analytically:
\begin{equation}
 b_B(r)=r_0+\frac{4M\pi^{-\Delta/2}}{\Delta+2}\left(r^{-\Delta}-r_0^{-\Delta}\right).
 \label{eq:b_B2}
\end{equation}
As we will see, in contrast to the Tsallis-inspired profile, whose falloff is governed by the non-extensive parameter $\delta$, the present branch is controlled by the fractal deformation parameter $\Delta$, leading to a distinct algebraic decay of both the density and the shape function.

The Barrow solution provides the first application of the limit distinction introduced in Sec.~\ref{sec:asymptotic_mass_general}. If $\Delta\to0$ is taken at fixed $r$, Eq.~\eqref{eq:rho_B2} vanishes and Eq.~\eqref{eq:b_B2} approaches $r_0$. If the large-radius limit is taken first, the same shape function gives $b_\infty\to r_0-2M$ and $M_{\mathrm{ADM}}^{\pm}\to r_0/2-M$. The limits therefore do not commute. The algebraic source becomes weak at every fixed radius, but its contribution remains spread over the unbounded integration domain.

\begin{figure}[!htp]
    \centering
    \includegraphics[width=0.49\linewidth]{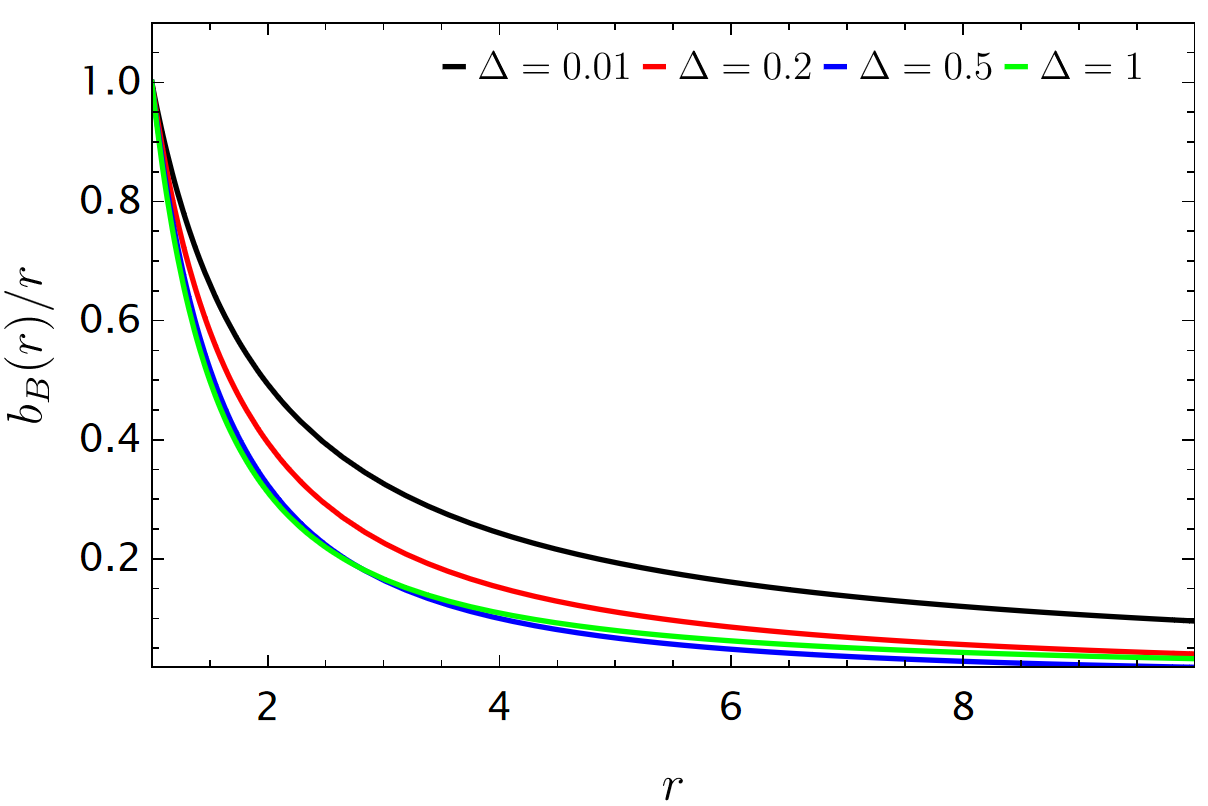}
    \includegraphics[width=0.49\linewidth]{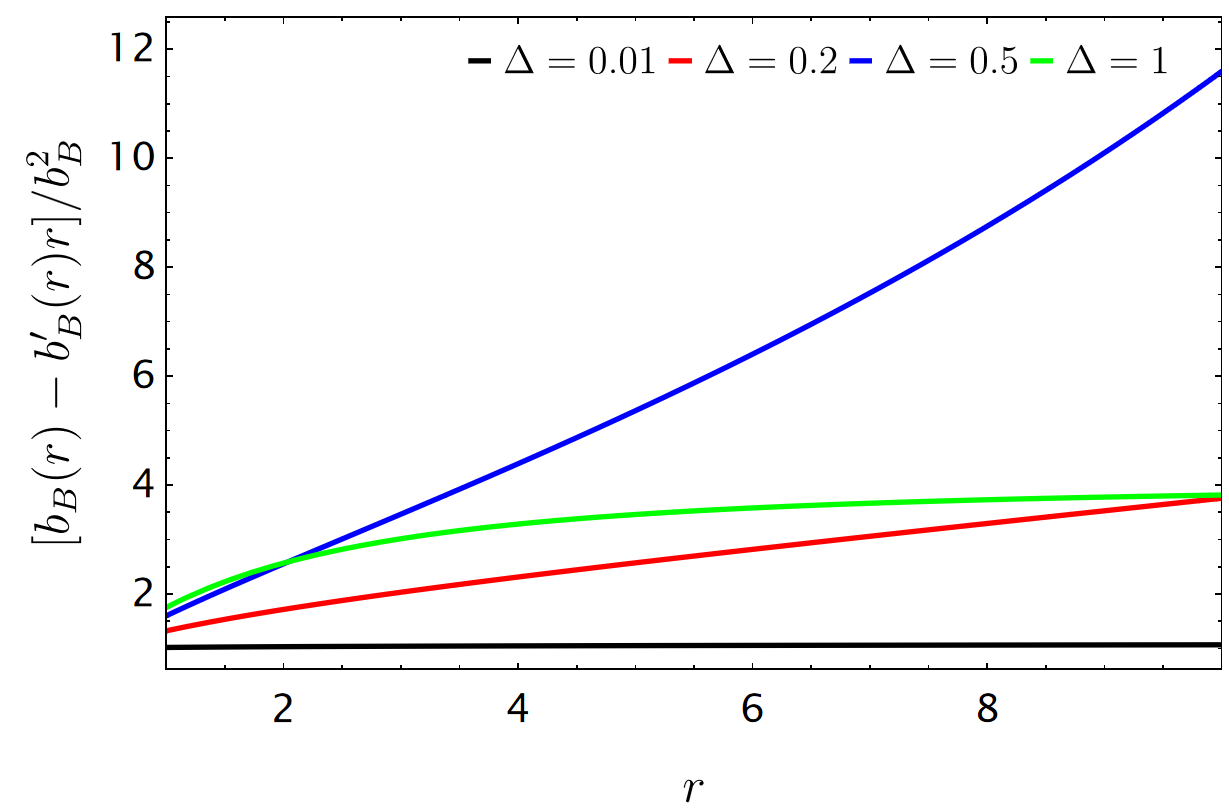}
    \caption{The geometric behavior of the Barrow-inspired wormhole for different values of the deformation parameter $\Delta$, with $M=1$ and $r_0=1$ fixed. In the left panel, we show the ratio $b_B(r)/r$ as a function of the radial coordinate $r$. In the right panel, we present the flare-out function $[b_B(r)-b_B'(r)r]/b_B^2$ as a function of $r$. The curves correspond to $\Delta=0.01$, $0.2$, $0.5$, and $1.0$.}
    \label{fig:geobarrow}
\end{figure}

The corresponding geometric behavior is displayed in Fig.~\ref{fig:geobarrow}. The left panel shows that the ratio $b_B(r)/r$ decreases and tends to zero for every value of $\Delta$ considered, indicating asymptotic flatness of this branch. The right panel verifies that $b_B(r)-r b'_B(r)$ remains positive in the physical region outside the throat.

Using Eq.~\eqref{eq:redshift_general2} together with Eq.~\eqref{eq:b_B2}, the derivative of the redshift function can be written explicitly as
\begin{equation}
 \Phi_B'(r)=
 \frac{
 r_0+\dfrac{4M\pi^{-\Delta/2}}{\Delta+2}
 \left[(1-\Delta w)\,r^{-\Delta}-r_0^{-\Delta}\right]
 }
 {
 2r\left[
 r-r_0-\dfrac{4M\pi^{-\Delta/2}}{\Delta+2}
 \left(r^{-\Delta}-r_0^{-\Delta}\right)
 \right]
 }.
 \label{eq:phi_B2}
\end{equation}
Although Eq.~\eqref{eq:phi_B2} is the most useful expression for the subsequent analysis, it is instructive to note that the redshift function can be explicitly integrated for particular values of the Barrow parameter. In general, for an arbitrary unspecified value of $\Delta$, the integral does not lead to a compact elementary expression. For rational or fixed values of $\Delta$, however, the integration can be carried out case by case, although the resulting expressions may become unnecessarily lengthy.

As a simple representative example, consider the maximal Barrow deformation, $\Delta=1$. Defining the dimensionless radial coordinate
\begin{equation}
 u=\frac{r}{r_0},
\end{equation}
together with
\begin{equation}
 \beta_B=\frac{4M}{3\sqrt{\pi}\,r_0^2},
\end{equation}
the redshift function can be written as
\begin{equation}
 \Phi_B(u)=
 \frac{1-\beta_B}{2\beta_B}
 \ln\left(\frac{u}{u+\beta_B}\right),
 \qquad \Delta=1,
 \label{eq:Phi_B_Delta1}
\end{equation}
where the integration constant has been fixed by the asymptotic normalization $\Phi_B(\infty)=0$.
This explicit case illustrates that the apparent singular behavior of $\Phi'_B(r)$ at the throat is removable once the throat-regularity condition is imposed.

It is worth emphasizing that, for the geometric and matter-sector calculations, the relevant quantities are not the absolute values of $\Phi_B(r)$, but rather its derivatives. Indeed, the radial pressure depends on $\Phi_B'(r)$, while the tangential pressure depends on both $\Phi_B'(r)$ and $\Phi_B''(r)$. In this sense, once Eq.~\eqref{eq:phi_B2} is known, the full geometric and physical analysis can be carried out without requiring a closed elementary expression for $\Phi_B(r)$ itself. The regularity of this derivative at the throat will play a central role in the next subsection, where the equation-of-state parameter is no longer treated as arbitrary, but is fixed by the throat-regularity condition.

\begin{figure}[!htp]
    \centering
    \includegraphics[width=0.58\linewidth]{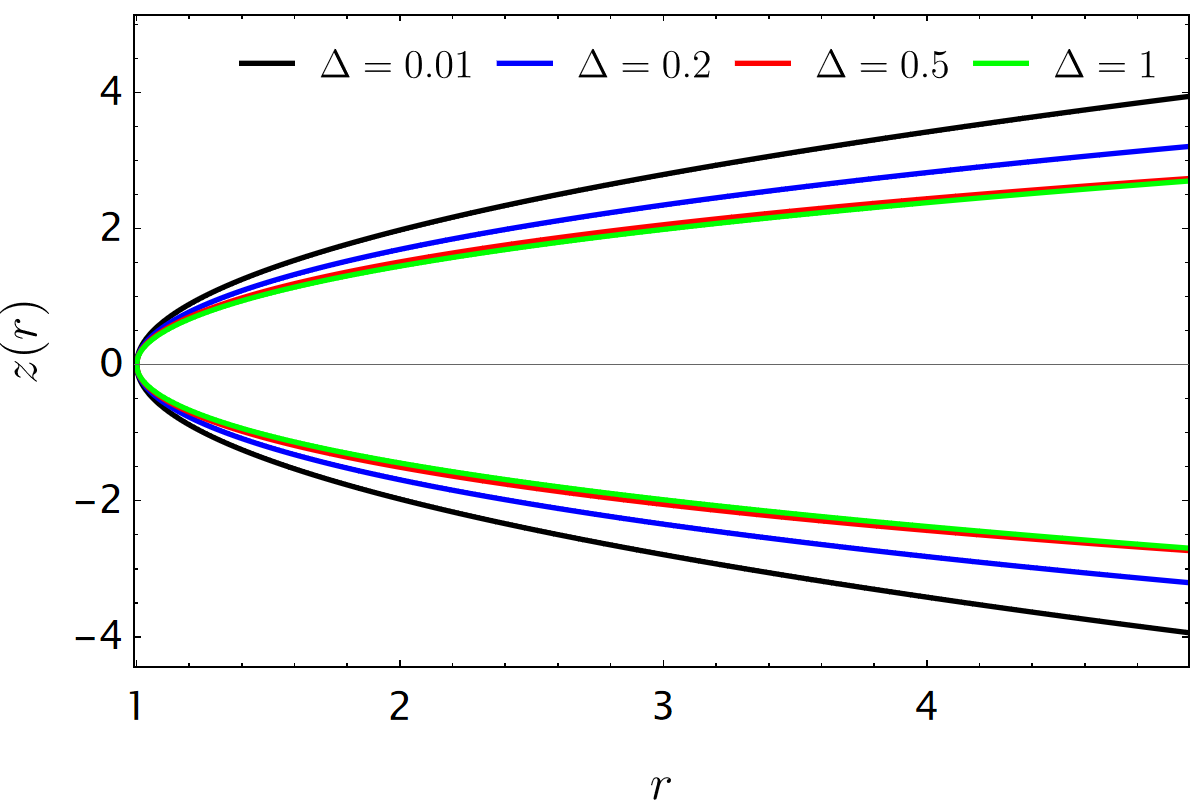}
    \includegraphics[width=0.4\linewidth]{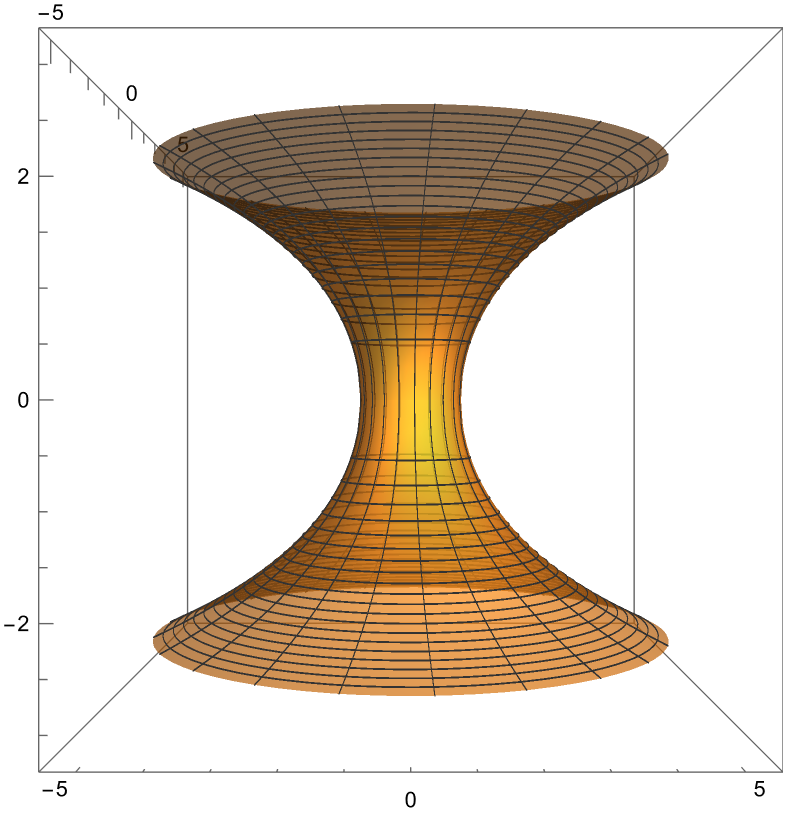}
    \caption{Embedding structure of the Barrow-inspired wormhole geometry for different values of the deformation parameter $\Delta$, with $M=1$ and $r_0=1$ fixed. In the left panel, we show the embedding function $z(r)$ as a function of the radial coordinate $r$ for $\Delta=0.01$, $0.2$, $0.5$, and $1.0$. In the right panel, we present the corresponding three-dimensional embedded surface for the representative case $\Delta=0.5$, with $M=1$ and $r_0=1$ fixed.}
    \label{fig:zperfilbarrow}
\end{figure}

A complementary visualization of the spatial geometry is given by the embedding diagram in Fig.~\ref{fig:zperfilbarrow}. The left panel shows the embedding profile $z(r)$ for different values of $\Delta$, while the right panel presents the corresponding three-dimensional surface for a representative case. The divergence of $dz/dr$ at the throat is not a physical singularity, but rather indicates that the embedded surface has a vertical tangent at the minimum radius. This is the expected geometric signature of a wormhole throat. The different curves in the left panel show how the Barrow deformation changes the opening of the embedded surface: larger values of $\Delta$ modify the spatial slice more strongly near the throat, while smaller values approach the weakly deformed regime.

\subsubsection{Energy conditions and equilibrium in the Barrow-inspired sector}

The radial pressure is fixed by the chosen equation of state,
\begin{equation}
 p_{r,B}(r)=w_B\rho_B(r).
\end{equation}
However, as discussed in the general construction, the parameter $w_B$ cannot be chosen independently if the redshift function is required to be regular at the throat. For the Barrow-inspired shape function, one has
\begin{equation}
 w_B\equiv w_{\rm th}^{(B)}
 =
 \frac{\pi^{\Delta/2}r_0^{1+\Delta}(\Delta+2)}
 {4M\Delta}.
 \label{eq:wth_B}
\end{equation}
Thus, for each value of the Barrow parameter $\Delta$, the regular wormhole branch selects a different value of the radial equation-of-state parameter. In particular, for $M>0$, $r_0>0$, and $0<\Delta\leq 1$, one finds $w_B>0$. Since $\rho_B<0$, this corresponds to a negative radial pressure, namely a radial tension. As $\Delta\to0$ along regular configurations, Eq.~\eqref{eq:wth_B} diverges, but this is a divergence of the constant coefficient rather than of the throat pressure, which approaches the finite value $-1/(8\pi r_0^2)$. The exact endpoint does not belong to the fixed-$r_0$ family with finite constant $w_B$.

Using Eq.~\eqref{eq:pt_from_TOV2} together with Eqs.~\eqref{eq:rho_B2} and \eqref{eq:phi_B2}, with $w=w_B$, the tangential pressure can be written explicitly as
\begin{equation}
 p_{t,B}(r)=
 \frac{\pi^{-\Delta/2}\Delta M}{4\pi(\Delta+2)}\,r^{-\Delta-3}
 \left[
 w_B(\Delta+1)-(1+w_B)r\Phi_B'(r)
 \right].
\end{equation}
Thus, the Barrow-inspired source remains anisotropic, with the difference between $p_{r,B}$ and $p_{t,B}$ controlled jointly by the fractal deformation parameter $\Delta$ and by the redshift response. The quantity $\Phi_B'(r)$ is understood with the regularity prescription imposed at the throat, so that its value at $r=r_0$ is obtained by the finite limiting procedure discussed above.

Since $\rho_B<0$ for $M>0$ and $\Delta>0$, the weak energy condition is violated throughout this branch. The radial NEC is
\begin{equation}
 \rho_B+p_{r,B}
 =-
 \frac{4M\pi^{-\Delta/2}\Delta+(\Delta+2)r_0^{\Delta+1}}
 {8\pi(\Delta+2)}\,r^{-\Delta-3}.
\end{equation}
Because $w_B>0$ in the negative-density Barrow branch, the radial NEC is violated for all $r$ in the physical domain. At the throat, this violation is not a consequence of an arbitrary choice of $w_B$, but follows directly from the flare-out condition:
\begin{equation}
 \left.\left(\rho_B+p_{r,B}\right)\right|_{r_0}
 =-
 \frac{4M\pi^{-\Delta/2}\Delta+(\Delta+2)r_0^{\Delta+1}}
 {8\pi(\Delta+2)r_0^{\Delta+3}}<0.
\end{equation}
This shows that, once the redshift regularity condition is imposed, the violation of the radial NEC becomes geometrically tied to the existence of the throat.

\begin{figure}[!htp]
    \centering
    \includegraphics[width=0.49\linewidth]{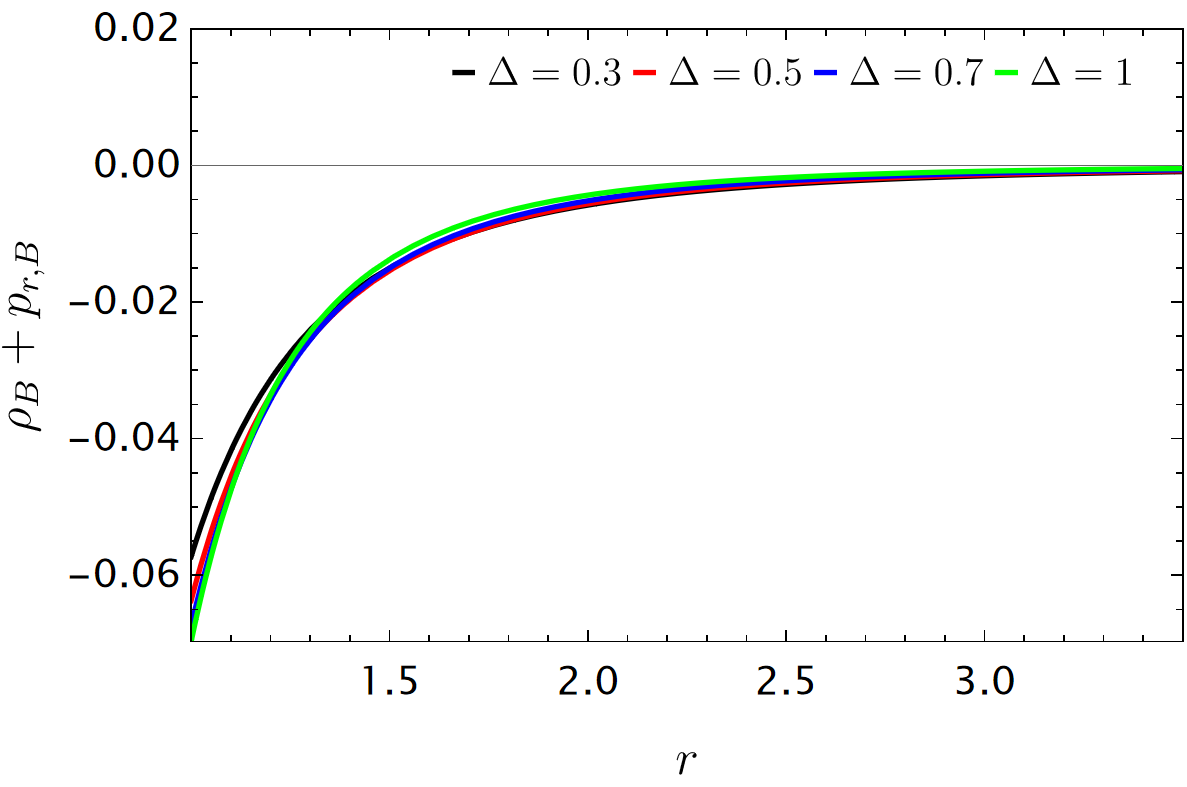}
    \includegraphics[width=0.49\linewidth]{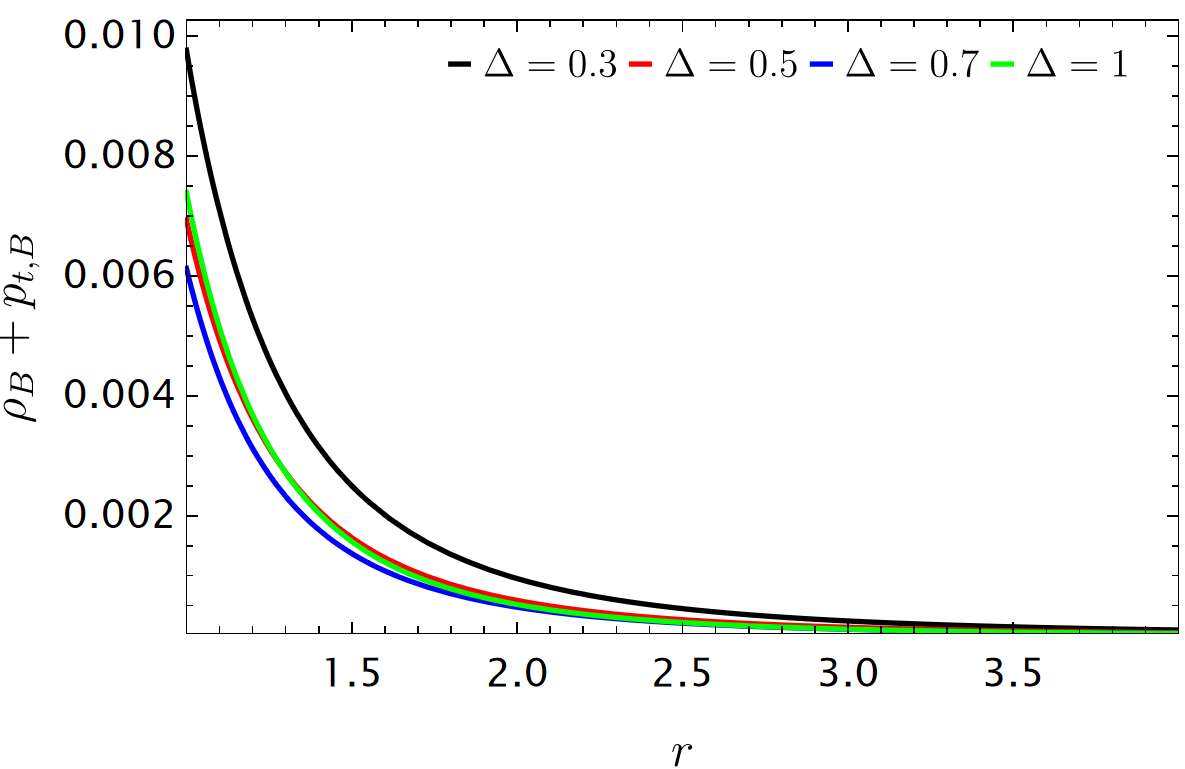}
    \includegraphics[width=0.55\linewidth]{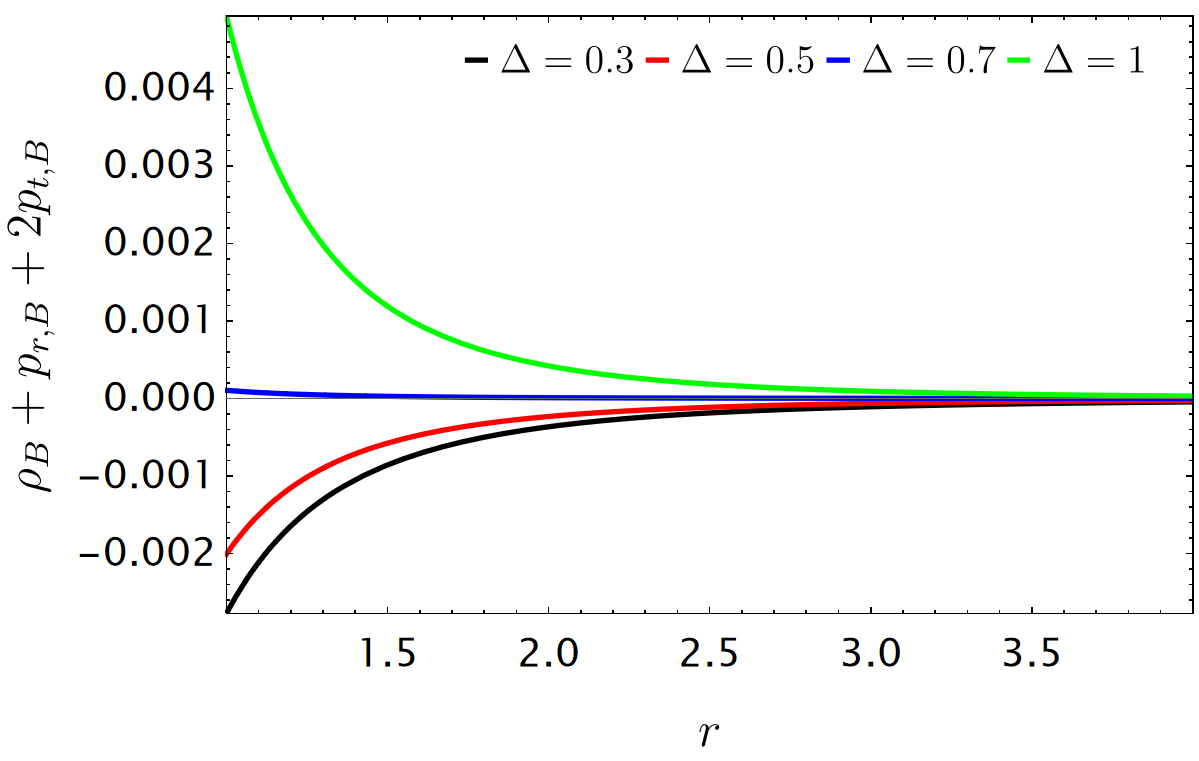}
    \caption{Behavior of the energy conditions for the Barrow-inspired wormhole, with $M=1$ and $r_0=1$ fixed. The top-left panel shows the behavior of the radial NEC, the top-right panel shows the behavior of the tangential NEC, and the bottom panel shows the behavior of the SEC. We set $\Delta = 0.3$, $0.5$, $0.7$, and $1.0$. In all panels, the parameter $w_B$ is fixed for each value of $\Delta$ by the throat-regularity condition \eqref{eq:wth_B}.}
    \label{fig:necbarrow}
\end{figure}

Figure~\ref{fig:necbarrow} displays the behavior of the radial NEC, tangential NEC, and SEC combination for the Barrow-inspired branch. In all panels, the parameter $w_B$ is not fixed independently, but is determined for each curve by Eq.~\eqref{eq:wth_B}. The top-left panel shows that the radial NEC is violated throughout the plotted domain, as required for a traversable throat. The top-right panel shows that the tangential NEC can remain positive, indicating that the exoticity is mainly radial rather than isotropically distributed. The SEC panel reveals a more sensitive dependence on $\Delta$: smaller values of the Barrow deformation tend to favor SEC violation, whereas larger values can restore the SEC combination in the near-throat region. This illustrates that the SEC provides information beyond the radial NEC, since it probes the combined contribution of the radial tension, tangential pressure, and redshift gradient.

The accumulated radial violation follows by substituting Eqs.~\eqref{eq:wth_B} and \eqref{eq:b_B2} into the general identity~\eqref{eq:integrated_nec_identity}. For either asymptotic end, one obtains
{
\begin{equation}
 \mathcal{I}_{\mathrm{NEC},B}^{\pm}
 =-\frac{2M\pi^{-\Delta/2}r_0^{-\Delta}}{\Delta+2}
  -\frac{r_0}{2\Delta}.
 \label{eq:integrated_nec_B}
\end{equation}
}
Both contributions are negative for $M>0$, $r_0>0$, and $\Delta>0$, and the weighted integrand decays as $r^{-1-\Delta}$, so the integral converges for every regular Barrow configuration. The first term is the density contribution associated with the asymptotic shape function and approaches $-M$ as $\Delta\to0$. The second term originates from the throat-selected radial pressure and becomes unbounded as $-r_0/(2\Delta)$. Therefore, although the density vanishes at every fixed radius, the integrated radial exoticity diverges along the regular undeformed limit because the algebraic tail approaches the marginal $r^{-1}$ behavior.

Analyzing the equilibrium condition, we have
\begin{equation}
 F_{h,B}+F_{g,B}+F_{a,B}=0,
\end{equation}
where the three force contributions are
\begin{equation}
 F_{h,B}(r)=
 -\frac{w_B\pi^{-\Delta/2}\Delta M(\Delta+3)}
 {2\pi(\Delta+2)}\,r^{-\Delta-4},
\end{equation}
\begin{equation}
 F_{g,B}(r)=
 \frac{(1+w_B)\pi^{-\Delta/2}\Delta M}
 {2\pi(\Delta+2)}\,r^{-\Delta-3}\Phi_B'(r),
\end{equation}
and
\begin{equation}
 F_{a,B}(r)=
 \frac{\pi^{-\Delta/2}\Delta M}{2\pi(\Delta+2)}\,r^{-\Delta-4}
 \left[
 w_B(\Delta+3)-(1+w_B)r\Phi_B'(r)
 \right].
\end{equation}
These expressions show explicitly that the hydrostatic and anisotropic contributions inherit the algebraic decay of the Barrow-inspired density, while the gravitational contribution is modulated by the regularized redshift derivative. Since $p_{t,B}$ is obtained from the TOV equation, the balance equation is satisfied by construction; nevertheless, the decomposition into $F_{h,B}$, $F_{g,B}$, and $F_{a,B}$ is useful for identifying how the equilibrium is distributed among hydrostatic gradients, redshift effects, and pressure anisotropy.

\begin{figure}[!htp]
    \centering
    \includegraphics[width=0.6\linewidth]{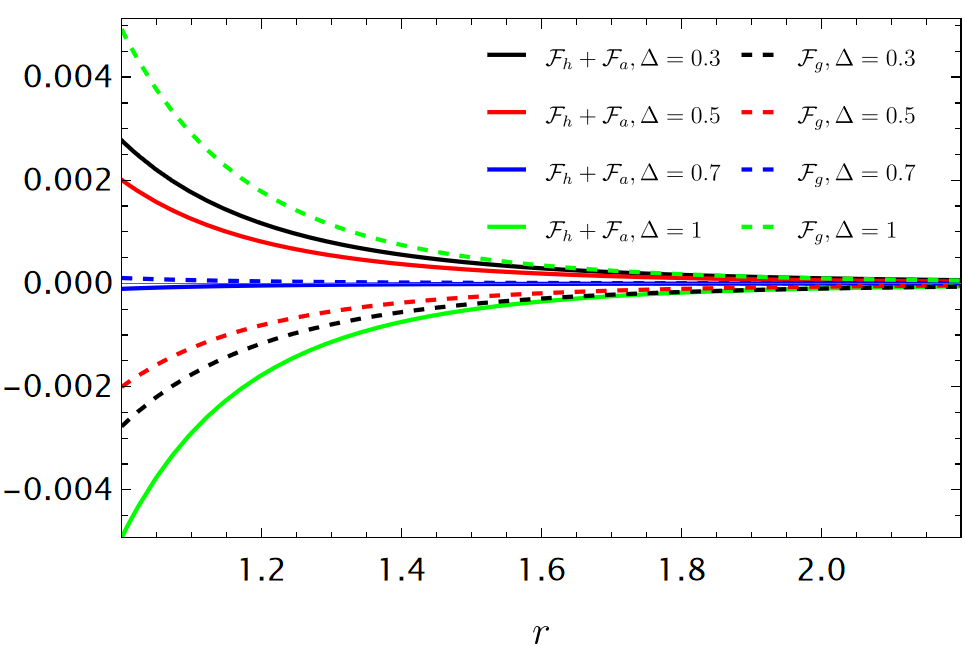}
    \caption{Equilibrium forces for the Barrow-inspired wormhole as functions of the radial coordinate $r$, with $M=1$ and $r_0=1$ fixed. Solid curves represent the combined anisotropic and hydrostatic contribution $F_{a,B}+F_{h,B}$, while dashed curves represent the gravitational contribution $F_{g,B}$. We set $\Delta=0.3$, $0.5$, $0.7$, and $1.0$, with $w_B$ fixed for each curve by the throat-regularity condition.}
    \label{fig:tovbarrow}
\end{figure}

In Figure~\ref{fig:tovbarrow} we show the TOV force balance for the same Barrow-inspired configurations. The solid curves represent the combined contribution $F_{h,B}+F_{a,B}$, while the dashed curves represent the gravitational contribution $F_{g,B}$. For each value of $\Delta$, these two contributions have opposite signs and comparable magnitudes, confirming the equilibrium relation. The sign of each contribution is not the same for all plotted values of $\Delta$, which is why the line style, rather than the vertical position of the curve, is used to identify the force sector. The dependence on $\Delta$, however, is not strictly monotonic near the throat. As $\Delta$ increases from smaller to intermediate values, the force amplitude tends to decrease, suggesting that the reduction of the throat-selected parameter $w_B$ partially weakens the response of the pressure sector. Nevertheless, this trend is reversed as $\Delta$ approaches its maximal value. In particular, the curve with $\Delta=1$ displays a stronger near-throat force separation, indicating that the increased localization of the Barrow-inspired density becomes dominant over the decrease of $w_B$.

This turning behavior is physically relevant because it shows that the equilibrium is not governed only by the magnitude of the density or only by the equation-of-state parameter. Rather, it results from the combined effect of the Barrow deformation, the throat-regularity condition, and the regularized redshift derivative. The enhancement observed for $\Delta=1$ should therefore be interpreted as a nonlinear near-throat response of the anisotropic equilibrium, not as a numerical artifact. As $r$ increases, all force contributions rapidly approach zero, consistently with the algebraic decay of the Barrow-inspired source and the asymptotic weakening of the effective matter sector.

\subsection{Tsallis-inspired sector}

\subsubsection{Geometry induced by the Tsallis-inspired profile}

The Tsallis entropy modifies the standard additive structure of horizon thermodynamics through the non-extensive parameter $\delta$ \cite{Tsallis}. In the entropy--geometry correspondence of Ref.~\cite{Anand}, the associated effective density is
\begin{equation}
 \rho_T(r)=-\frac{M\pi^{-\delta}(\delta-1)}{2\delta}\,r^{-2\delta-1},
 \label{eq:rho_T2}
\end{equation}
which is negative in the branch $\delta>1$ considered in the present work. In contrast to the Barrow-inspired profile, whose decay is governed by the fractal parameter $\Delta$, the Tsallis-inspired density is controlled by the non-extensive exponent $\delta$. Thus, the parameter $\delta$ affects both the amplitude of the source and the rate at which it decays away from the throat.

\begin{figure}[!htp]
    \centering
    \includegraphics[width=0.6\linewidth]{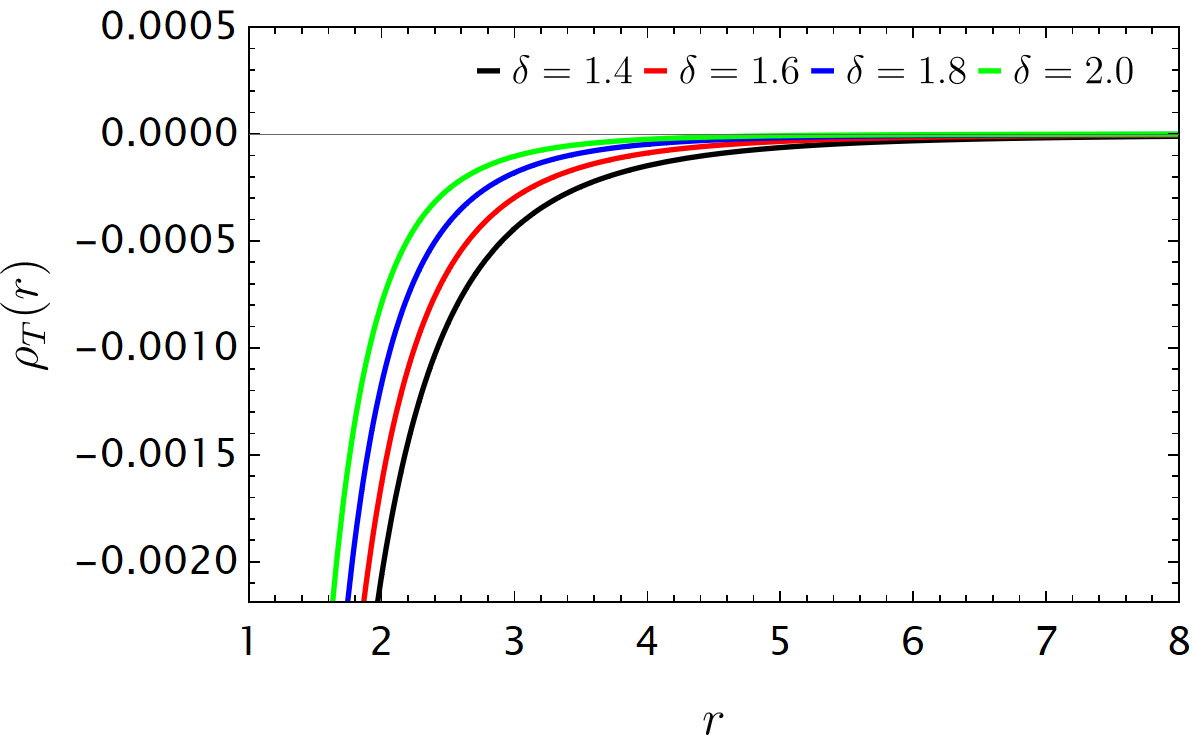}
    \caption{The Tsallis-inspired density profile, $\rho_T(r)$, for different values of the deformation parameter $\delta$ as a function of the radial coordinate $r$, with $M=1$ and $r_0=1$ fixed. The curves correspond to $\delta=1.4$, $1.6$, $1.8$, and $2.0$.}
    \label{fig:rhotsallis}
\end{figure}

The behavior of $\rho_T(r)$ is shown in Fig.~\ref{fig:rhotsallis}. The values of $\delta$ used in this figure are chosen within the branch that supports the wormhole geometry and remain sufficiently above $\delta=1$, where the effective density is suppressed. As $\delta$ increases, the source becomes more concentrated near the throat and decays more rapidly with $r$, reflecting the stronger power-law suppression induced by the non-extensive deformation.

The corresponding shape function follows from Eq.~\eqref{eq:b_general2} and can be integrated in closed form as
\begin{equation}
 b_T(r)=r_0+\frac{2M\pi^{1-\delta}}{\delta}\left(r^{2-2\delta}-r_0^{2-2\delta}\right).
 \label{eq:b_T2}
\end{equation}
This expression shows that, similarly to the Barrow branch, the throat geometry is determined by a negative density profile with algebraic decay. However, the radial dependence is now dictated by $2-2\delta$ rather than by $-\Delta$, which makes the Tsallis sector more sensitive to changes in the entropy parameter.

The Tsallis density and shape function show the same limit-order effect through a different power law. At fixed $r$, taking $\delta\to1$ makes Eq.~\eqref{eq:rho_T2} vanish and Eq.~\eqref{eq:b_T2} approach $r_0$. Taking $r\to\infty$ first instead gives $b_\infty\to r_0-2M$ and $M_{\mathrm{ADM}}^{\pm}\to r_0/2-M$. Hence, the two limits again fail to commute, now because the factor that suppresses the source is accompanied by a radial power that becomes progressively long ranged as $\delta\to1$.

\begin{figure}[!htp]
    \centering
    \includegraphics[width=0.49\linewidth]{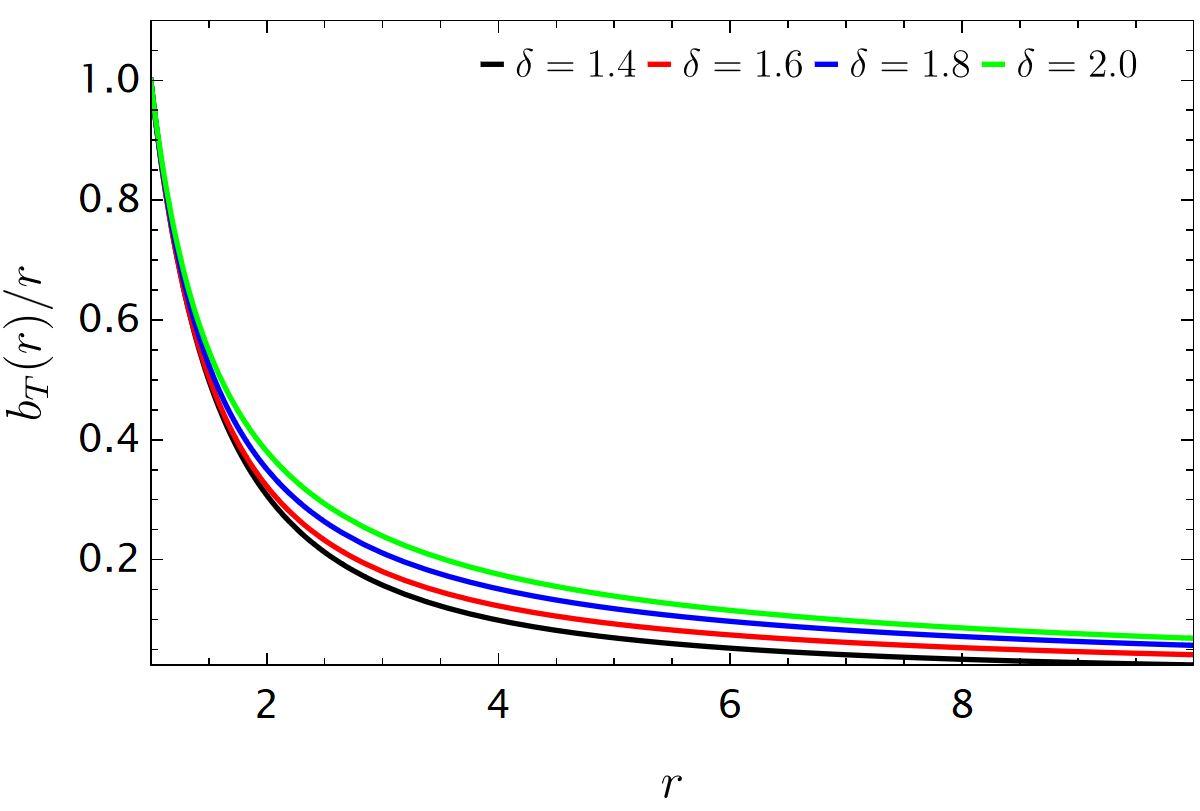}
     \includegraphics[width=0.49\linewidth]{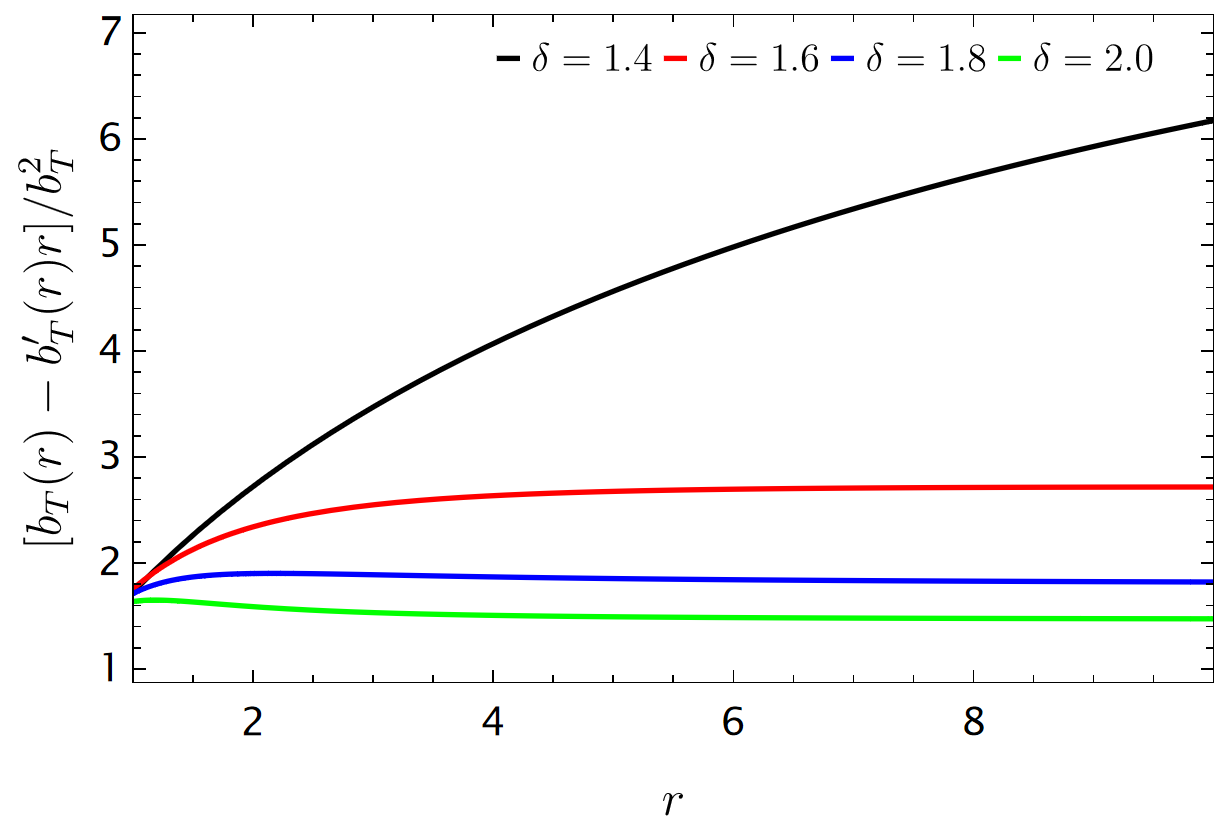}
    \caption{The geometric behavior of the Tsallis-inspired wormhole for different values of the deformation parameter $\delta$, with $M=1$ and $r_0=1$ fixed. In the left panel, we show the ratio $b_T(r)/r$ as a function of the radial coordinate $r$. In the right panel, we present the flare-out function $[b_T(r)-b_T'(r)r]/b_T^2$ as a function of $r$. The curves correspond to $\delta=1.4$, $1.6$, $1.8$, and $2.0$.}
    \label{fig:geotsallis}
\end{figure}

The geometric behavior associated with Eq.~\eqref{eq:b_T2} is displayed in Fig.~\ref{fig:geotsallis}. The left panel shows that the ratio $b_T(r)/r$ decreases and tends to zero for all values of $\delta$ considered. Therefore, the asymptotic flatness condition associated with the shape function is satisfied throughout this branch. The right panel verifies the flare-out behavior. For the chosen values of $\delta$, the quantity $b_T(r)-r b'_T(r)$ remains positive in the physical region, indicating that the spatial geometry opens outward from the throat.

Substituting Eq.~\eqref{eq:b_T2} into the general redshift equation \eqref{eq:redshift_general2}, one obtains
\begin{equation}
 \Phi_T'(r)=
 \frac{
 r_0+\dfrac{2M\pi^{1-\delta}}{\delta}
 \left[
 \bigl(1-2w(\delta-1)\bigr)\,r^{2-2\delta}
 -r_0^{2-2\delta}
 \right]
 }
 {
 2r\left[
 r-r_0-\dfrac{2M\pi^{1-\delta}}{\delta}
 \left(r^{2-2\delta}-r_0^{2-2\delta}\right)
 \right]
 }.
 \label{eq:phi_T2}
\end{equation}
Compared with the Barrow-inspired case, the Tsallis branch preserves the same qualitative mechanism: a negative entropy-inspired density induces a shape function compatible with a wormhole throat. The distinction lies in how the entropy parameter controls the radial hierarchy of the source. While $\Delta$ modifies the Barrow profile through a relatively mild algebraic exponent, the Tsallis parameter $\delta$ changes the decay through the power $2-2\delta$, making the localization of the geometry more sensitive to the non-extensive deformation. As in the Barrow sector, the regularity of Eq.~\eqref{eq:phi_T2} at the throat will be addressed in the next subsection, where the equation-of-state parameter is fixed by the throat-regularity condition.

Although Eq.~\eqref{eq:phi_T2} will be used as the working expression in the subsequent analysis, the corresponding redshift function can also be obtained explicitly for fixed values of the non-extensive parameter. As in the Barrow sector, a compact elementary primitive is not available for an arbitrary unspecified value of $\delta$. Nevertheless, for rational values of $\delta$ the integral can be reduced to a rational form and evaluated case by case. This is particularly relevant here because the values used in the numerical analysis, $\delta=1.4,1.6,1.8,2.0$, are fixed rational values.

As a representative example, let us consider $\delta=2$. Introducing the dimensionless coordinate $u=r/r_0$ and the parameter $\beta_T=M/(\pi r_0^3)$, which follows from the Tsallis shape function for $\delta=2$, the redshift function can be written, after imposing the throat-regularity condition and the asymptotic normalization $\Phi_T(\infty)=0$, as
\begin{equation}
\begin{split}
 \Phi_T(u)=\;&
 \frac{1-\beta_T}{2\beta_T}\ln u
 -\frac{1-\beta_T}{4\beta_T}
 \ln\left(u^2+\beta_T u+\beta_T\right)  \\
 &+\frac{1-\beta_T}{2\sqrt{\beta_T(4-\beta_T)}}
 \left[
 \tan^{-1}\left(\frac{2u+\beta_T}
 {\sqrt{\beta_T(4-\beta_T)}}\right)
 -\frac{\pi}{2}
 \right],
 \qquad \delta=2 .
\end{split}
\label{eq:Phi_T_delta2}
\end{equation}
This expression is written for $0<\beta_T<4$, which includes the numerical choice $M=r_0=1$, for which $\beta_T=1/\pi$. For other parameter ranges, the same integration can be expressed in terms of logarithmic or hyperbolic functions after the corresponding factorization of the quadratic denominator.

This explicit case shows an important difference with respect to the Barrow example. While the Barrow expression for $\Delta=1$ reduces to a single logarithm, the Tsallis case already combines logarithmic and arctangent terms for $\delta=2$. This reflects the different algebraic structure of the Tsallis shape function, whose radial dependence involves the power $2-2\delta$. Therefore, even though the Barrow and Tsallis sectors are both governed by algebraic density profiles, their reconstructed redshift functions need not have the same analytic complexity.

\begin{figure}[!htp]
    \centering
    \includegraphics[width=0.58\linewidth]{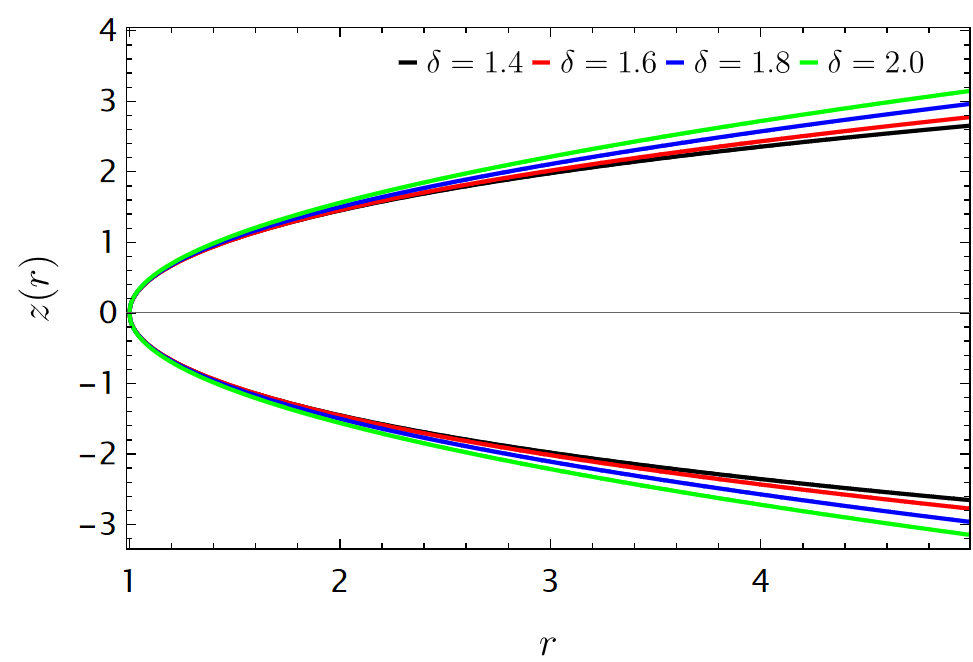}
    \includegraphics[width=0.4\linewidth]{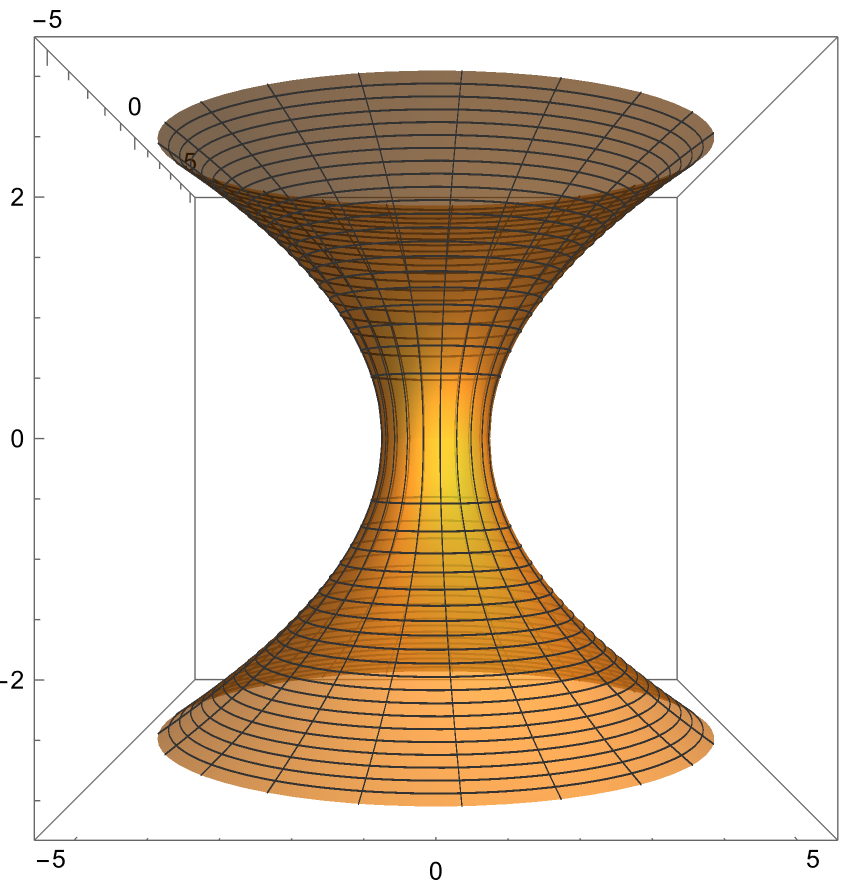}
    \caption{Embedding structure of the Tsallis-inspired wormhole geometry for different values of the deformation parameter $\delta$, with $M=1$ and $r_0=1$ fixed. In the left panel, we show the embedding function $z(r)$ as a function of the radial coordinate $r$ for $\delta=1.4$, $1.6$, $1.8$, and $2.0$. In the right panel, we present the corresponding three-dimensional embedded surface for the representative case $\delta=2.0$, with $M=1$ and $r_0=1$ fixed.}
    \label{fig:embeddtsallis}
\end{figure}

A complementary view of the Tsallis-inspired spatial geometry is provided by the embedding diagram in Fig.~\ref{fig:embeddtsallis}. The left panel shows the embedding function $z(r)$ for the same values of $\delta$, while the right panel presents a representative three-dimensional embedded surface. The vertical tangent of the embedded surface at the throat is the expected geometric signature of the minimum-radius surface and does not indicate a physical singularity. The dependence on $\delta$ shows how the non-extensive deformation changes the opening of the spatial slice: larger values of $\delta$ tend to produce a more localized modification near the throat, consistently with the sharper decay of the density profile.

\subsubsection{Energy conditions and equilibrium in the Tsallis-inspired sector}

The radial pressure is fixed by the chosen equation of state,
\begin{equation}
 p_{r,T}(r)=w_T\rho_T(r).
\end{equation}
As in the Barrow-inspired sector, the equation-of-state parameter cannot be chosen independently if the redshift function is required to be regular at the throat. For the Tsallis-inspired shape function, one obtains
\begin{equation}
 b_T'(r)=
 -\frac{4M\pi^{1-\delta}(\delta-1)}{\delta}\,
 r^{1-2\delta}.
\end{equation}
Therefore, the throat-regularity condition $w=-1/b'(r_0)$ gives
\begin{equation}
 w_T\equiv w_{\rm th}^{(T)}
 =
 \frac{\pi^{\delta-1}r_0^{2\delta-1}\delta}
 {4M(\delta-1)}.
 \label{eq:wth_T}
\end{equation}
For the branch considered here, $\delta>1$, one has $w_T>0$ for $M>0$ and $r_0>0$. Since $\rho_T<0$, this again corresponds to a negative radial pressure, or radial tension. As $\delta\to1$, Eq.~\eqref{eq:wth_T} diverges, while the throat pressure approaches the same finite value found in the Barrow branch along regular configurations. The values used below remain away from this endpoint of the constant-coefficient family.

Using Eq.~\eqref{eq:pt_from_TOV2} together with Eqs.~\eqref{eq:rho_T2} and \eqref{eq:phi_T2}, with $w=w_T$, the tangential pressure can be written explicitly as
\begin{equation}
 p_{t,T}(r)=
 \frac{M\pi^{-\delta}(\delta-1)}{4\delta}\,r^{-2\delta-1}
 \left[
 w_T(2\delta-1)-(1+w_T)r\Phi_T'(r)
 \right].
\end{equation}
Thus, the Tsallis-inspired matter sector remains anisotropic. The difference between $p_{r,T}$ and $p_{t,T}$ is now governed by the non-extensive parameter $\delta$, by the throat-selected parameter $w_T$, and by the redshift response. Compared with the Barrow case, the anisotropy is more directly affected by the power-law exponent of the density, since $\delta$ controls both the amplitude of the source and its radial decay.

Since $\rho_T<0$ for $\delta>1$, the weak energy condition is violated throughout this branch. The radial NEC is
\begin{equation}
 \rho_T+p_{r,T}
 =-
 \frac{4M\pi^{1-\delta}(\delta-1)+\delta r_0^{2\delta-1}}
 {8\pi\delta}\,r^{-2\delta-1}.
\end{equation}
Because $w_T>0$ in the negative-density Tsallis branch, the radial NEC is violated for all $r$ in the physical domain. At the throat, this violation is again tied to the flare-out condition,
\begin{equation}
 \left.\left(\rho_T+p_{r,T}\right)\right|_{r_0}
 =-
 \frac{4M\pi^{1-\delta}(\delta-1)+\delta r_0^{2\delta-1}}
 {8\pi\delta r_0^{2\delta+1}}<0.
\end{equation}
This is the sector-specific realization of the general result derived in Eq.~\eqref{eq:NEC_throat_geometry}: after imposing throat regularity, the radial NEC violation follows from the flare-out geometry rather than from an arbitrary choice of the equation-of-state parameter.

\begin{figure}[!htp]
    \centering
    \includegraphics[width=0.49\linewidth]{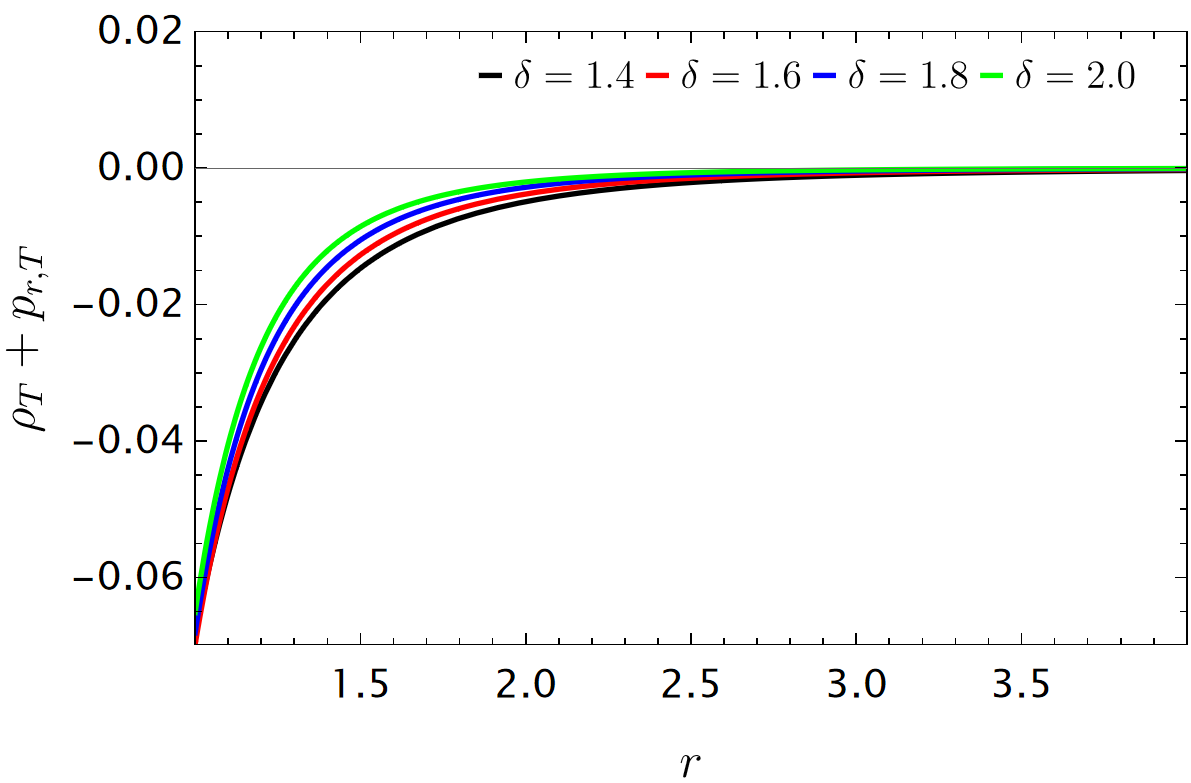}
    \includegraphics[width=0.49\linewidth]{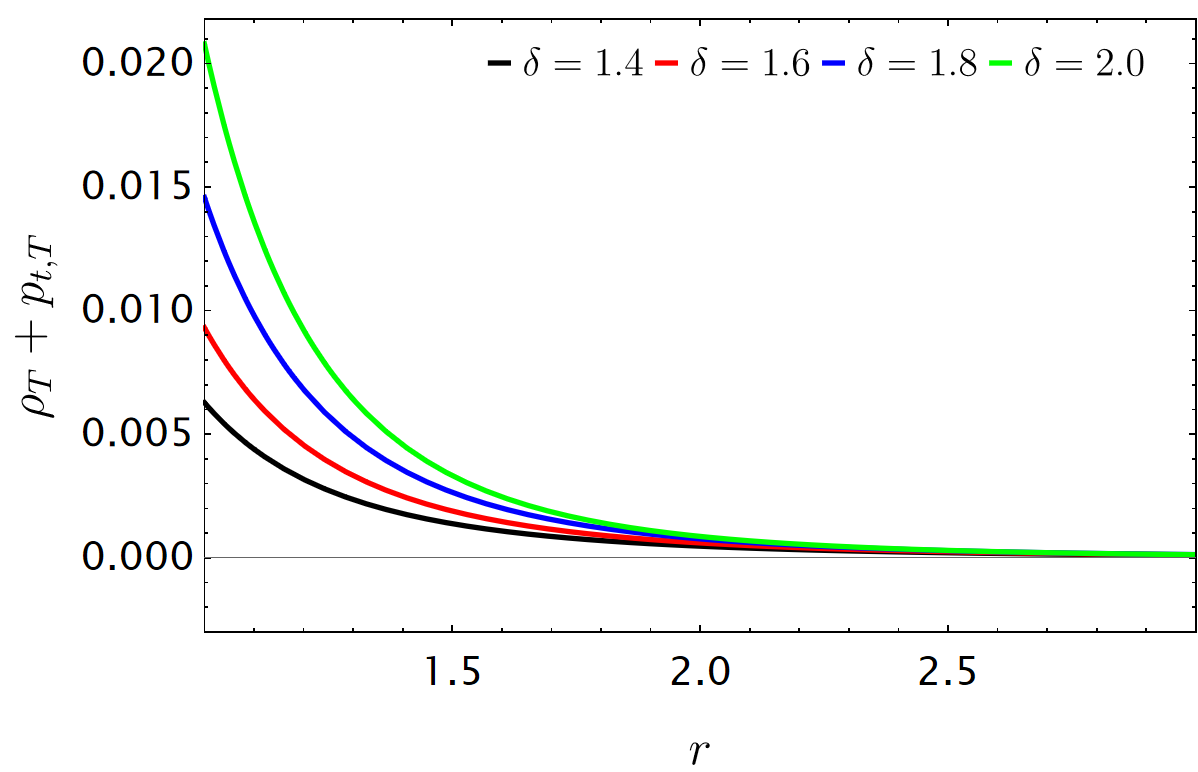}
    \includegraphics[width=0.55\linewidth]{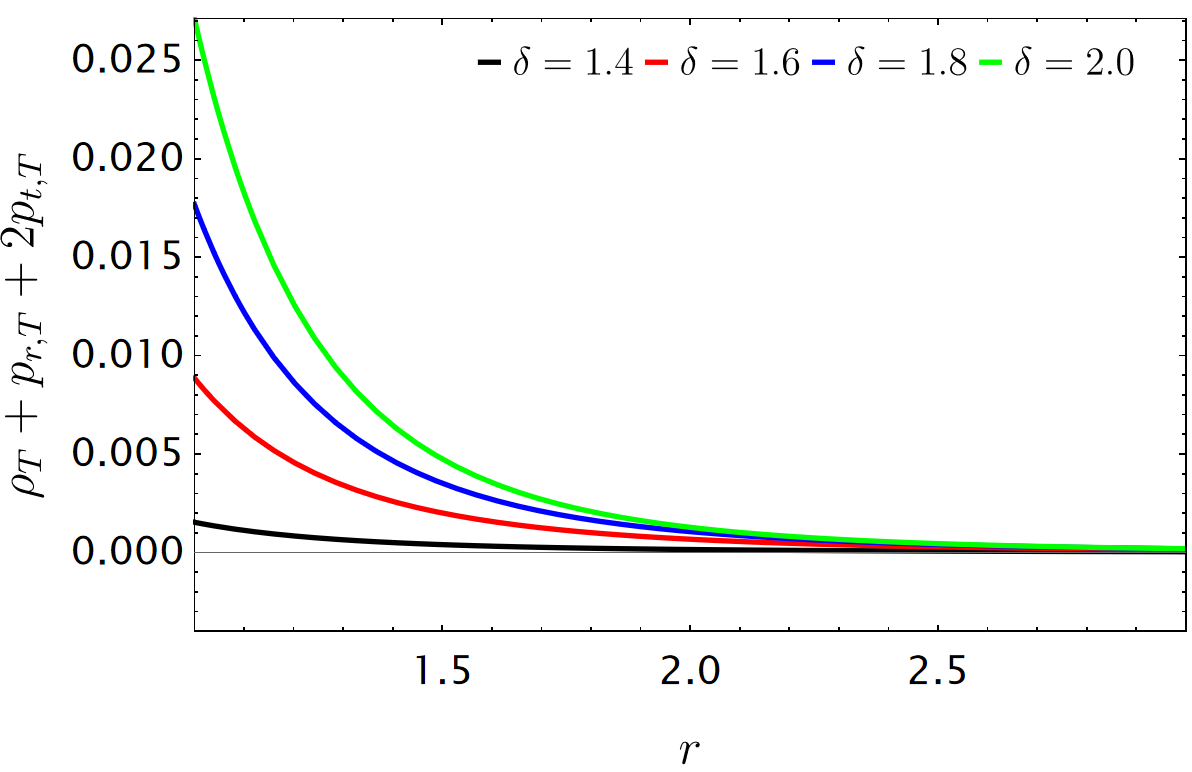}
    \caption{Behavior of the energy conditions for the Tsallis-inspired wormhole, with $M=1$ and $r_0=1$ fixed. The top-left panel shows the behavior of the radial NEC, the top-right panel shows the behavior of the tangential NEC, and the bottom panel shows the behavior of the SEC. We set $\delta = 1.4$, $1.6$, $1.8$, and $2.0$. In all panels, the parameter $w_T$ is fixed for each value of $\delta$ by the throat-regularity condition \eqref{eq:wth_T}.}
    \label{fig:nectsallis}
\end{figure}

Figure~\ref{fig:nectsallis} displays the behavior of the radial NEC, tangential NEC, and SEC combination for the Tsallis-inspired sector. In all panels, the parameter $w_T$ is not fixed independently, but is determined for each curve by Eq.~\eqref{eq:wth_T}. The top-left panel shows that the radial NEC is violated throughout the plotted domain, as required by the throat geometry. The top-right panel shows that the tangential NEC remains positive for the values of $\delta$ considered, indicating that the exoticity is concentrated mainly in the radial direction. This behavior is similar to the Barrow branch, but the SEC response is qualitatively different. In the Tsallis case, the combination $\rho_T+p_{r,T}+2p_{t,T}$ remains positive in the plotted range and becomes more pronounced near the throat as $\delta$ increases. This indicates that the tangential sector contributes more strongly to the total pressure combination, compensating part of the radial exoticity in the SEC diagnostic. Thus, while the full SEC is still not satisfied because the radial NEC is violated, the SEC combination reveals that the Tsallis branch has a stronger tangential support than the Barrow-inspired one.

Using Eqs.~\eqref{eq:wth_T} and \eqref{eq:b_T2} in Eq.~\eqref{eq:integrated_nec_identity}, the one-ended integrated quantity becomes
{
\begin{equation}
 \mathcal{I}_{\mathrm{NEC},T}^{\pm}
 =-\frac{M\pi^{1-\delta}r_0^{2-2\delta}}{\delta}
  -\frac{r_0}{4(\delta-1)}.
 \label{eq:integrated_nec_T}
\end{equation}
}
This expression is finite and negative for every $\delta>1$. The radial measure changes the power-law NEC profile into an integrand proportional to $r^{1-2\delta}$, which is integrable precisely on this branch. As $\delta\to1^{+}$, the density contribution approaches $-M$, whereas the pressure contribution diverges as $-r_0/[4(\delta-1)]$. The exponent simultaneously approaches the marginal value $-1$, showing that the unbounded integrated exoticity is the global counterpart of the increasingly long-ranged Tsallis profile, not a divergence of any fixed nonzero-deformation solution.

Analyzing the equilibrium condition, we have
\begin{equation}
 F_{h,T}+F_{g,T}+F_{a,T}=0,
\end{equation}
where the three force contributions are
\begin{equation}
 F_{h,T}(r)=
 -\frac{M\pi^{-\delta}(\delta-1)w_T(2\delta+1)}
 {2\delta}\,r^{-2\delta-2},
\end{equation}
\begin{equation}
 F_{g,T}(r)=
 \frac{M\pi^{-\delta}(\delta-1)(1+w_T)}
 {2\delta}\,r^{-2\delta-1}\Phi_T'(r),
\end{equation}
and
\begin{equation}
 F_{a,T}(r)=
 \frac{M\pi^{-\delta}(\delta-1)}{2\delta}\,r^{-2\delta-2}
 \left[
 w_T(2\delta+1)-(1+w_T)r\Phi_T'(r)
 \right].
\end{equation}
These expressions show that the hydrostatic and anisotropic sectors inherit the Tsallis power-law decay, while the gravitational contribution is controlled by the same redshift derivative that regularizes the throat. In comparison with the Barrow branch, the parameter $\delta$ affects the TOV balance more directly through the radial exponent $2\delta+2$ appearing in the hydrostatic and anisotropic terms. Consequently, increasing $\delta$ not only changes the strength of the source near the throat, but also makes the force contributions decay more rapidly away from it.

\begin{figure}[!htp]
    \centering
    \includegraphics[width=0.6\linewidth]{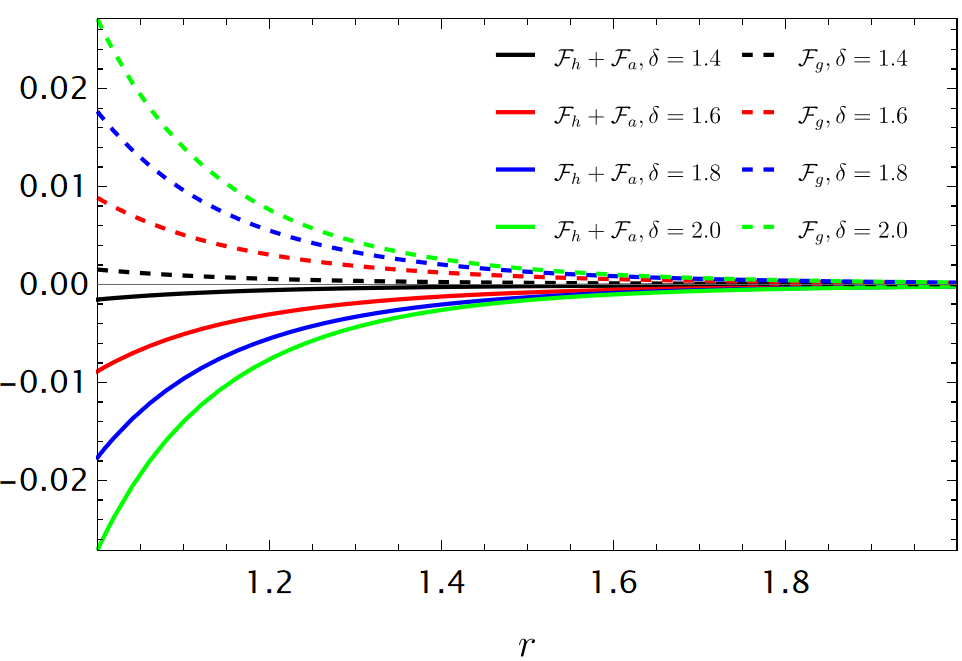}
    \caption{Equilibrium forces for the Tsallis-inspired wormhole as functions of the radial coordinate $r$, with $M=1$ and $r_0=1$ fixed. Solid curves represent the combined anisotropic and hydrostatic contribution $F_{a,T}+F_{h,T}$, while dashed curves represent the gravitational contribution $F_{g,T}$. We set $\delta=1.4$, $1.6$, $1.8$, and $2.0$, with $w_T$ fixed for each curve by the throat-regularity condition \eqref{eq:wth_T}.}
    \label{fig:TOVtsallis}
\end{figure}

Figure~\ref{fig:TOVtsallis} shows the TOV force balance for the Tsallis-inspired configurations. The solid curves represent the combined contribution $F_{h,T}+F_{a,T}$, while the dashed curves represent the gravitational contribution $F_{g,T}$. For each value of $\delta$, the two contributions have opposite signs and comparable magnitudes, reflecting the equilibrium condition. Unlike the Barrow-inspired case, where the force amplitude showed a non-monotonic behavior as the entropy parameter approached its maximal value, the Tsallis branch displays a more systematic enhancement of the near-throat force separation as $\delta$ increases. This is consistent with the stronger localization of the density profile and with the positive behavior of the SEC combination observed in Fig.~\ref{fig:nectsallis}. The tangential sector therefore plays a more prominent role in distributing the equilibrium, making the Tsallis-inspired branch more sharply supported near the throat while still allowing the forces to decay rapidly in the asymptotic region.

\subsection{Kaniadakis-inspired sector}

\subsubsection{Geometry induced by the Kaniadakis-inspired profile}

The Kaniadakis entropy introduces a relativistically motivated deformation of the standard statistical framework through the parameter $\kappa$ \cite{Kaniadakis}. In the entropy--geometry correspondence of Ref.~\cite{Anand}, the associated effective density is
\begin{equation}
 \rho_\kappa(r)=-\frac{\kappa M}{2r}\tanh(\pi\kappa r^2)\,\operatorname{sech}(\pi\kappa r^2),
 \qquad \kappa>0.
 \label{eq:rho_K2}
\end{equation}
For $M>0$ and $\kappa>0$, this branch is negative for all $r>0$. This already places the Kaniadakis-inspired sector within the class of effective sources capable of supporting the throat. However, its radial structure is qualitatively different from the Barrow- and Tsallis-inspired cases. While those sectors are governed by power-law profiles, the Kaniadakis density is controlled by hyperbolic functions, vanishing in the small-$r$ limit and being strongly suppressed at large distances.

\begin{figure}[!htp]
    \centering
    \includegraphics[width=0.49\linewidth]{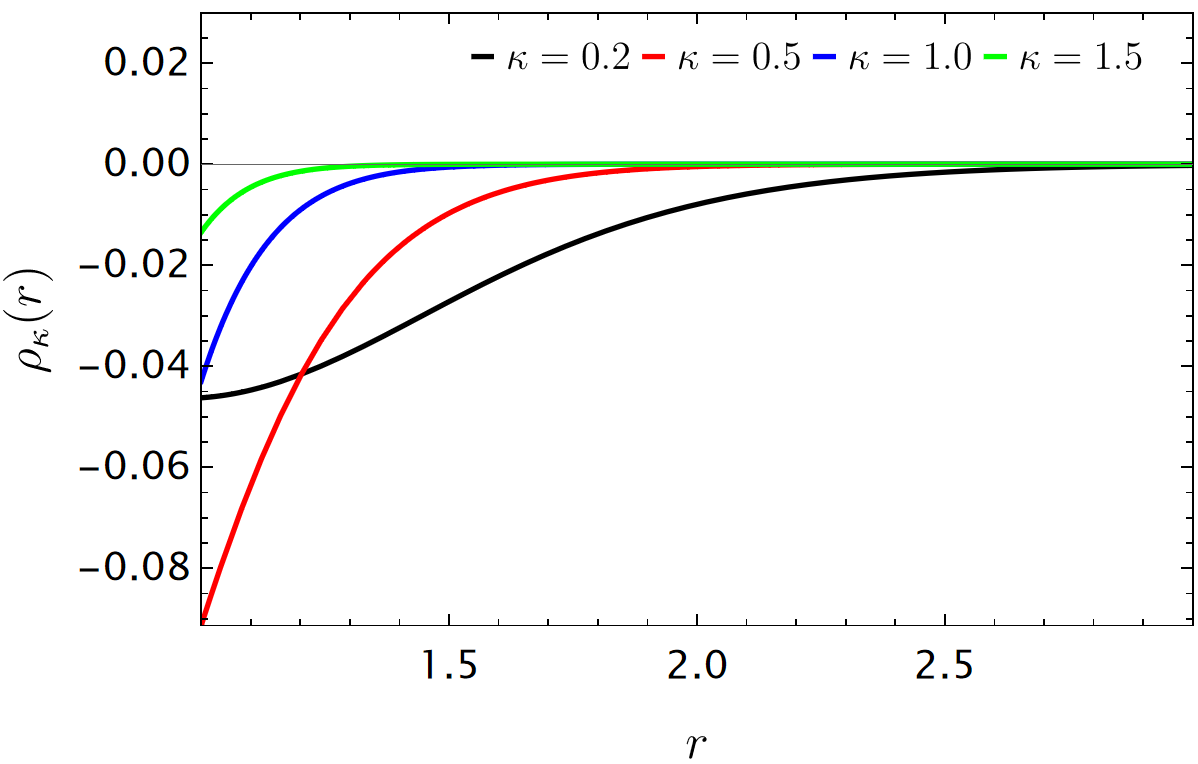}
    \includegraphics[width=0.49\linewidth]{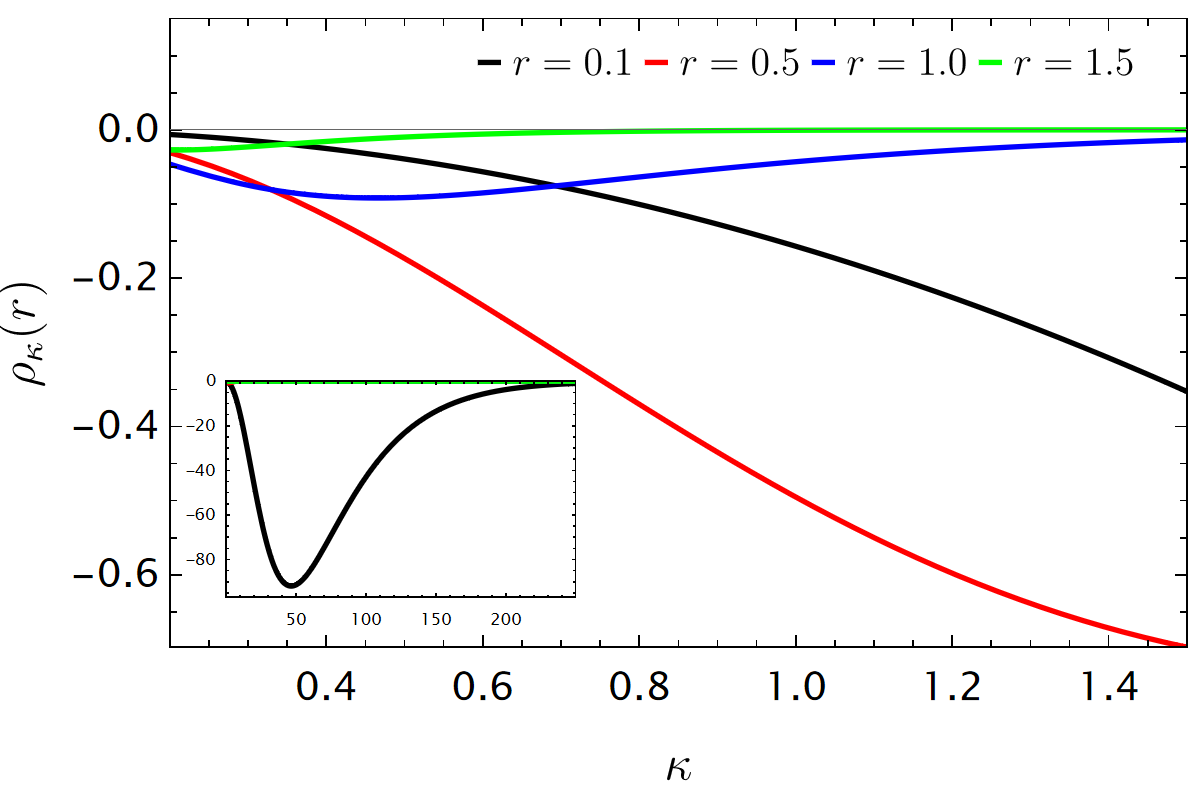}
    \caption{The Kaniadakis-inspired density profile, $\rho_\kappa(r)$, with $M=1$ and $r_0=1$ fixed. The left panel shows the radial behavior of $\rho_\kappa(r)$ for different values of the deformation parameter $\kappa$. The right panel shows the dependence of the density profile on $\kappa$ for fixed values of the radial coordinate $r$. The inset highlights the profile over a wider range of $\kappa$.}
    \label{fig:rhokaniadakis}
\end{figure}

The behavior of $\rho_\kappa(r)$ is displayed in Fig.~\ref{fig:rhokaniadakis}. The left panel shows the radial profile for different values of $\kappa$, emphasizing that the source is concentrated around a finite region rather than controlled by a purely algebraic tail. The right panel shows the dependence of the density on $\kappa$ for fixed values of $r$. This plot highlights an important feature of the Kaniadakis branch: for a fixed radius, the density is suppressed both when $\kappa$ is very small and when $\kappa$ becomes large. Thus, the deformation does not simply amplify the source monotonically. Instead, there is an intermediate range of $\kappa$ where the effective density becomes most relevant. This behavior has no direct analogue in the Barrow and Tsallis sectors, where the entropy parameters mainly control the strength and decay of power-law profiles.

The corresponding shape function follows from Eq.~\eqref{eq:b_general2} and can be integrated analytically as
\begin{equation}
 b_\kappa(r)=r_0+2M\left[\operatorname{sech}(\pi\kappa r^2)-\operatorname{sech}(\pi\kappa r_0^2)\right].
 \label{eq:b_K2}
\end{equation}
Thus, unlike the Barrow and Tsallis sectors, where the shape function is governed by powers of $r$, the Kaniadakis-inspired geometry is controlled by a rapidly suppressed hyperbolic profile. This makes the source effectively localized around the throat region and causes the geometry to approach its asymptotic regime more rapidly.

The small-$\kappa$ limit shows that rapid decay at fixed parameter does not guarantee a uniform deformation limit. At fixed $r$, Eqs.~\eqref{eq:rho_K2} and \eqref{eq:b_K2} approach zero density and $r_0$, respectively. At the same time, the characteristic support moves outward on the scale $\kappa^{-1/2}$. Taking the large-radius limit first therefore retains $b_\infty\to r_0-2M$ and $M_{\mathrm{ADM}}^{\pm}\to r_0/2-M$. The two limits do not commute because a finite integrated contribution follows the source toward increasingly large radii.

\begin{figure}[!htp]
    \centering
    \includegraphics[width=0.49\linewidth]{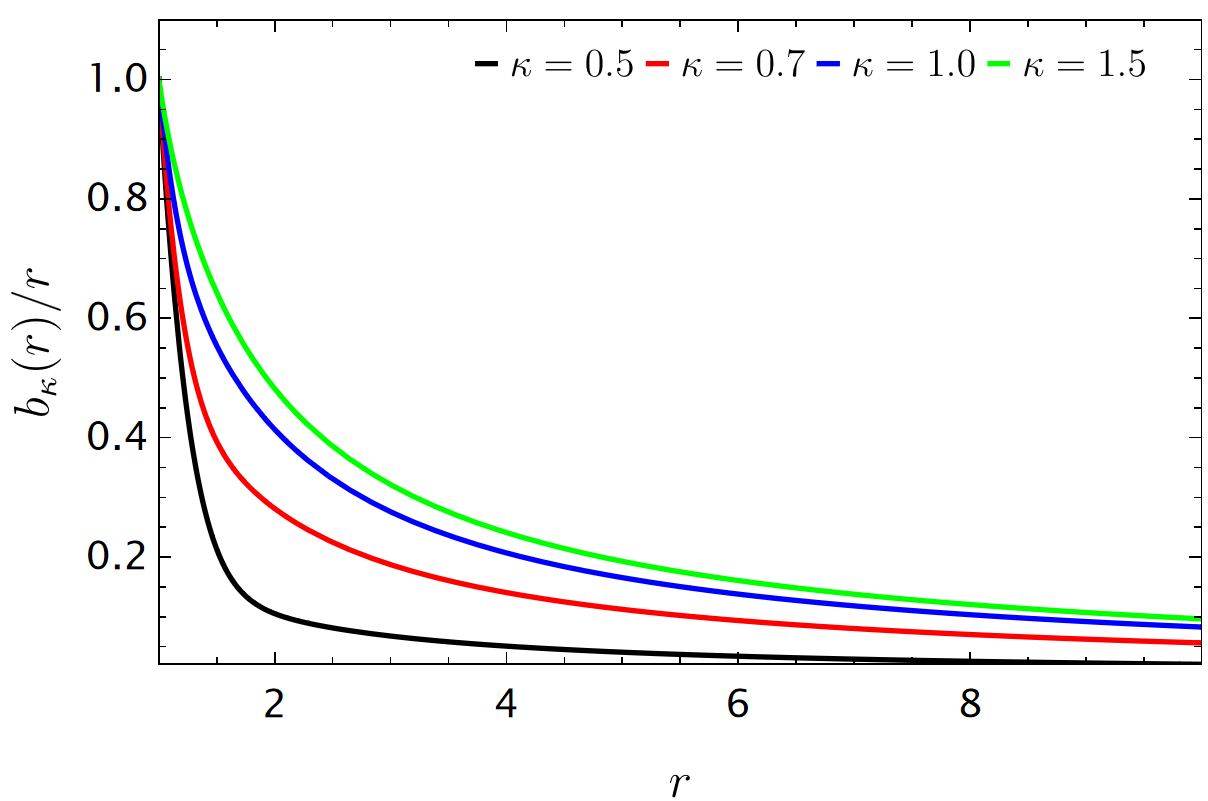}
    \includegraphics[width=0.49\linewidth]{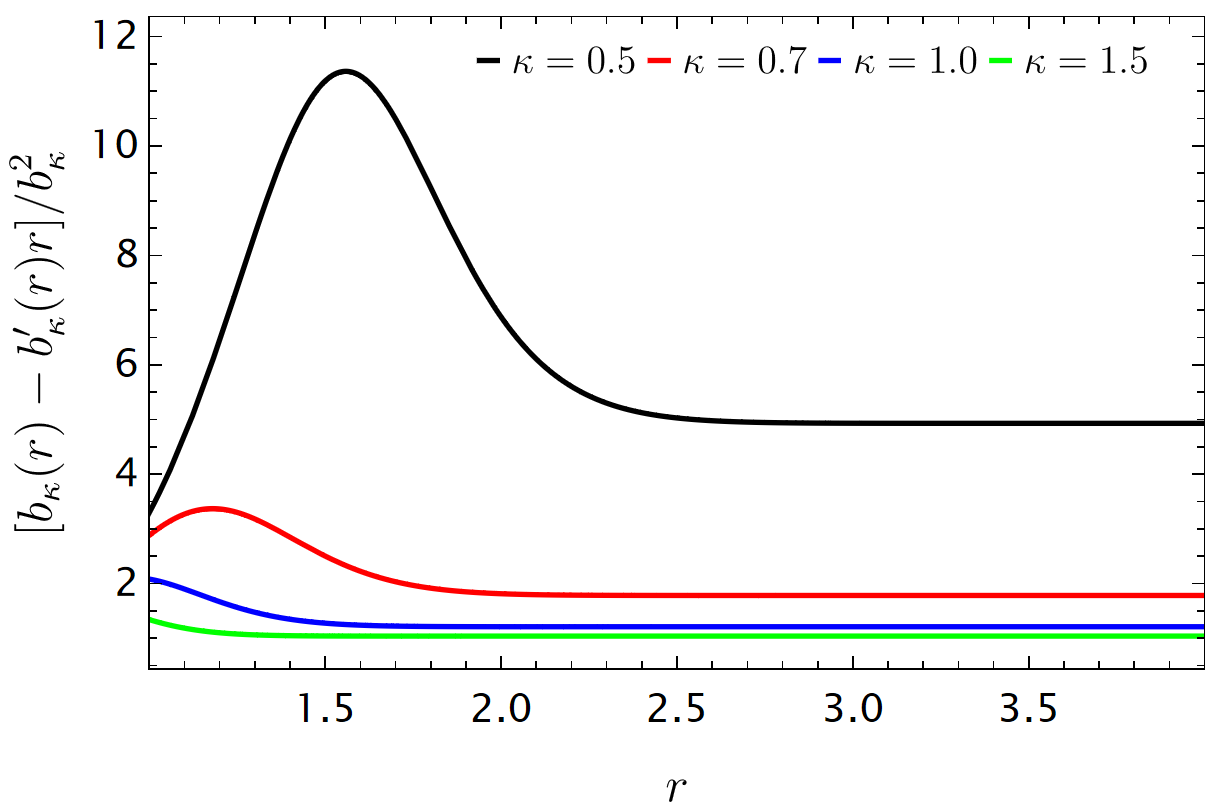}
    \caption{The geometric behavior of the Kaniadakis-inspired wormhole for different values of the parameter $\kappa$, with $M=1$ and $r_0=1$ fixed. In the left panel, we show the ratio $b_\kappa(r)/r$ as a function of the radial coordinate $r$. In the right panel, we present the flare-out function $[b_\kappa(r)-b_\kappa'(r)r]/b_\kappa^2$ as a function of $r$. The curves correspond to $\kappa=0.5$, $0.7$, $1.0$, and $1.5$.}
    \label{fig:geokaniadakis}
\end{figure}

The corresponding geometric behavior is shown in Fig.~\ref{fig:geokaniadakis}. The left panel shows that the ratio $b_\kappa(r)/r$ decreases and tends to zero for all values of $\kappa$ considered, confirming that the shape-function part of the asymptotic flatness condition is satisfied. The right panel verifies the flare-out behavior. For the plotted values of $\kappa$, the quantity $b_\kappa(r)-r b'_\kappa(r)$ remains positive in the physical region, indicating that the spatial geometry opens outward from the throat. Compared with the Tsallis case, the variation with the entropy parameter is less associated with a change in a radial power and more with a redistribution of the source around a finite-width region near the throat.

Substituting Eq.~\eqref{eq:b_K2} into the general redshift equation \eqref{eq:redshift_general2}, one obtains
\begin{equation}
 \Phi_\kappa'(r)=
 \frac{
 r_0+2M\left[\operatorname{sech}(\pi\kappa r^2)-\operatorname{sech}(\pi\kappa r_0^2)\right]
 -4\pi\kappa w M r^2\tanh(\pi\kappa r^2)\operatorname{sech}(\pi\kappa r^2)
 }
 {
 2r\left[
 r-r_0-2M\left(\operatorname{sech}(\pi\kappa r^2)-\operatorname{sech}(\pi\kappa r_0^2)\right)
 \right]
 }.
 \label{eq:phi_K2}
\end{equation}
Compared with the Barrow- and Tsallis-inspired geometries, the Kaniadakis sector is less sensitive to asymptotic algebraic tails and more strongly controlled by the finite-width region selected by the hyperbolic functions. At the same time, unlike the logarithmic and exponential branches, no pole-removal condition is required in the physical domain. The regularity of Eq.~\eqref{eq:phi_K2} at the throat will be treated in the next subsection, where the equation-of-state parameter is fixed by the same throat-regularity prescription used in the previous sectors. In addition, in contrast to the Barrow and Tsallis sectors, the Kaniadakis redshift function does not admit a closed elementary form, even for fixed positive values of $\kappa$. This is due to the simultaneous presence of $\tanh(\pi\kappa r^2)$ and $\operatorname{sech}(\pi\kappa r^2)$ in Eq.~\eqref{eq:phi_K2}, which prevents the reduction of the integral to a simple algebraic or rational form. Nevertheless, $\Phi_\kappa(r)$ can be consistently obtained by numerical integration for any admissible value of $\kappa>0$, with the integration constant fixed by the asymptotic normalization. This does not affect the geometric or matter-sector analysis, since the relevant quantities are determined by $\Phi_\kappa'(r)$ and by its finite limiting behavior at the throat.

\begin{figure}[!htp]
    \centering
    \includegraphics[width=0.58\linewidth]{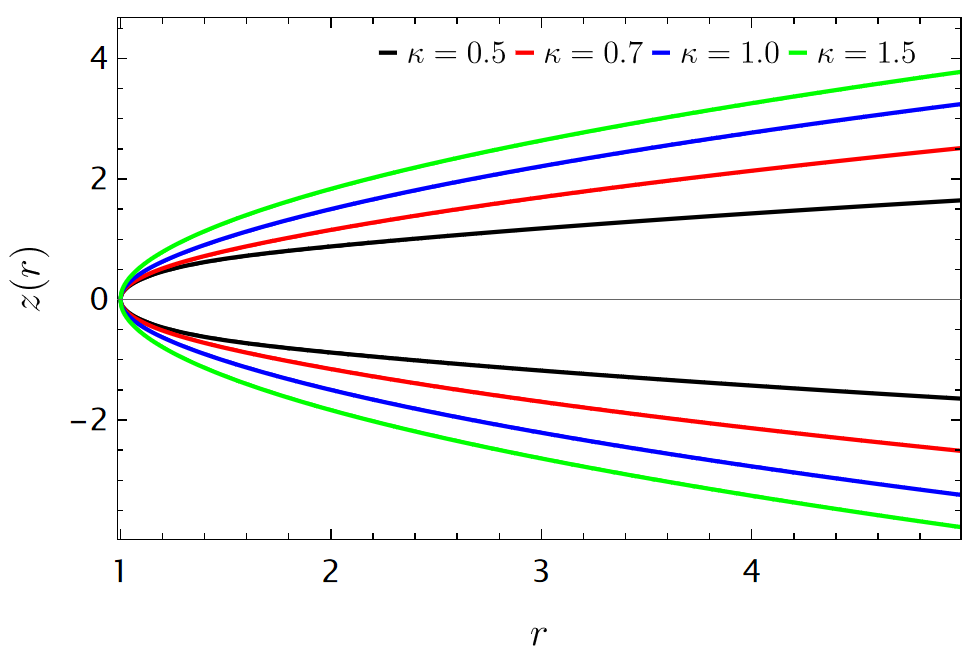}
    \includegraphics[width=0.4\linewidth]{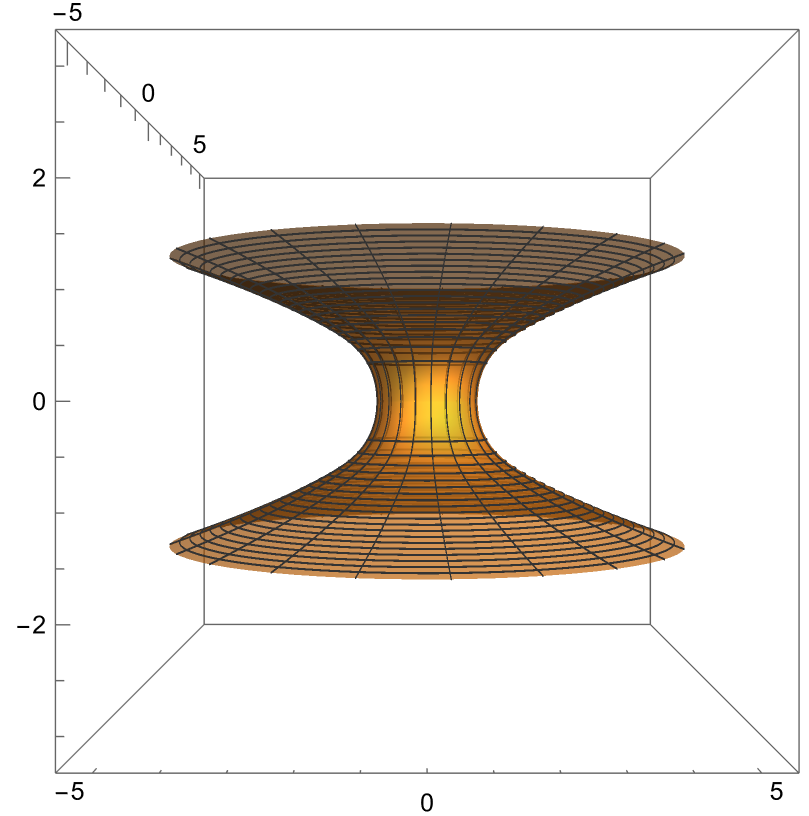}
    \caption{Embedding structure of the Kaniadakis-inspired wormhole geometry for different values of the parameter $\kappa$, with $M=1$ and $r_0=1$ fixed. In the left panel, we show the embedding function $z(r)$ as a function of the radial coordinate $r$ for $\kappa=0.5$, $0.7$, $1.0$, and $1.5$. In the right panel, we present the corresponding three-dimensional embedded surface for the representative case $\kappa=0.5$, with $M=1$ and $r_0=1$ fixed.}
    \label{fig:embeddedkaniadakis}
\end{figure}

The embedding diagram in Fig.~\ref{fig:embeddedkaniadakis} provides a complementary visualization of this localized geometry. The left panel shows the embedding function $z(r)$ for different values of $\kappa$, while the right panel presents a representative three-dimensional embedded surface. As in the previous sectors, the vertical tangent at the throat is a geometric signature of the minimum-radius surface and does not represent a physical singularity. The Kaniadakis deformation changes the opening of the embedded surface by modifying the width and intensity of the localized support near the throat, rather than by changing a power-law decay. This makes the embedding profile particularly useful for distinguishing the Kaniadakis branch from the algebraic Barrow and Tsallis branches.

\subsubsection{Energy conditions and equilibrium in the Kaniadakis-inspired sector}

The radial pressure is fixed by the chosen equation of state,
\begin{equation}
 p_{r,\kappa}(r)=w_\kappa\rho_\kappa(r).
\end{equation}
As in the previous sectors, the equation-of-state parameter is not arbitrary once the regularity of the redshift derivative at the throat is imposed. For the Kaniadakis-inspired shape function, one has
\begin{equation}
 b_\kappa'(r)=
 -4\pi\kappa M r\,
 \tanh(\pi\kappa r^2)\operatorname{sech}(\pi\kappa r^2).
\end{equation}
Therefore, the throat-regularity condition $w=-1/b'(r_0)$ gives
\begin{equation}
 w_\kappa\equiv w_{\rm th}^{(\kappa)}
 =
 \frac{1}
 {4\pi\kappa M r_0
 \tanh(\pi\kappa r_0^2)\operatorname{sech}(\pi\kappa r_0^2)}.
 \label{eq:wth_K}
\end{equation}
For $M>0$, $r_0>0$, and $\kappa>0$, one finds $w_\kappa>0$. Since $\rho_\kappa<0$, this corresponds again to a negative radial pressure, namely a radial tension. However, differently from the Barrow and Tsallis branches, the dependence of $w_\kappa$ on the entropy parameter is not governed by a simple algebraic factor. The same hyperbolic functions that localize the density also determine the amount of radial tension required at the throat. As $\kappa\to0$ along the regular branch, $w_\kappa$ diverges while the throat pressure remains finite; the singular behavior belongs to the constant coefficient, not to the pressure itself.

Using Eq.~\eqref{eq:pt_from_TOV2} together with Eqs.~\eqref{eq:rho_K2} and \eqref{eq:phi_K2}, with $w=w_\kappa$, the tangential pressure can be written explicitly as
\begin{align}
 p_{t,\kappa}(r)=\;&
 -\frac{w_\kappa\kappa M}{4r}
 \tanh(\pi\kappa r^2)\operatorname{sech}(\pi\kappa r^2)
 \nonumber\\[0.3em]
 &-\frac{\pi w_\kappa\kappa^2Mr}{2}
 \left[
 \operatorname{sech}^3(\pi\kappa r^2)
 -\operatorname{sech}(\pi\kappa r^2)\tanh^2(\pi\kappa r^2)
 \right]
 \nonumber\\[0.3em]
 &-\frac{(1+w_\kappa)\kappa M}{4}
 \tanh(\pi\kappa r^2)\operatorname{sech}(\pi\kappa r^2)\,
 \Phi_\kappa'(r).
\end{align}
In this form, the anisotropy is no longer controlled by a pure radial power, as in the Barrow and Tsallis sectors, but by localized combinations of hyperbolic functions. Thus, the distinction between radial and tangential stresses is naturally concentrated around the near-throat region and becomes progressively suppressed away from it. This is one of the main qualitative differences of the Kaniadakis-inspired branch: the pressure anisotropy is shaped by the width and intensity of a localized source rather than by an algebraic decay exponent.

Because $\rho_\kappa<0$ for $r>0$, the weak energy condition is violated throughout the physical branch. The radial NEC is
\begin{equation}
 \rho_\kappa+p_{r,\kappa}
 =-
 \frac{\tanh(\pi\kappa r^2)\sech(\pi\kappa r^2)}
 {8\pi r r_0\tanh(\pi\kappa r_0^2)\sech(\pi\kappa r_0^2)}
 \left[1+4\pi\kappa M r_0\tanh(\pi\kappa r_0^2)\sech(\pi\kappa r_0^2)\right].
\end{equation}
Since $w_\kappa>0$, the radial NEC is violated for all $r$ in the physical domain. At the throat, this violation is again geometrically tied to the flare-out condition,
\begin{equation}
 \left.\left(\rho_\kappa+p_{r,\kappa}\right)\right|_{r_0}
 =-
 \frac{1+4\pi\kappa M r_0\tanh(\pi\kappa r_0^2)\sech(\pi\kappa r_0^2)}
 {8\pi r_0^2}<0.
\end{equation}
Thus, the Kaniadakis branch again realizes the general throat identity \eqref{eq:NEC_throat_geometry}: the required radial exoticity is fixed by the local flare-out geometry once the redshift-regularity condition has selected $w_\kappa$.

\begin{figure}[!htp]
    \centering
    \includegraphics[width=0.49\linewidth]{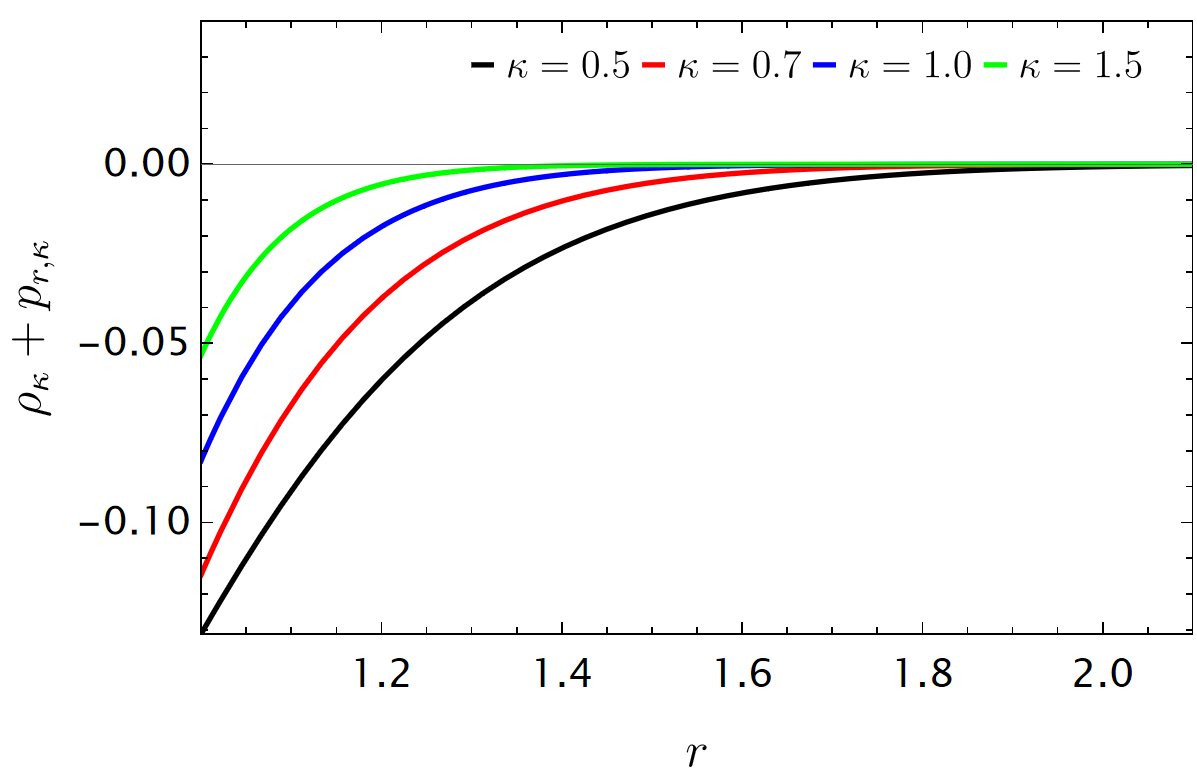}
    \includegraphics[width=0.49\linewidth]{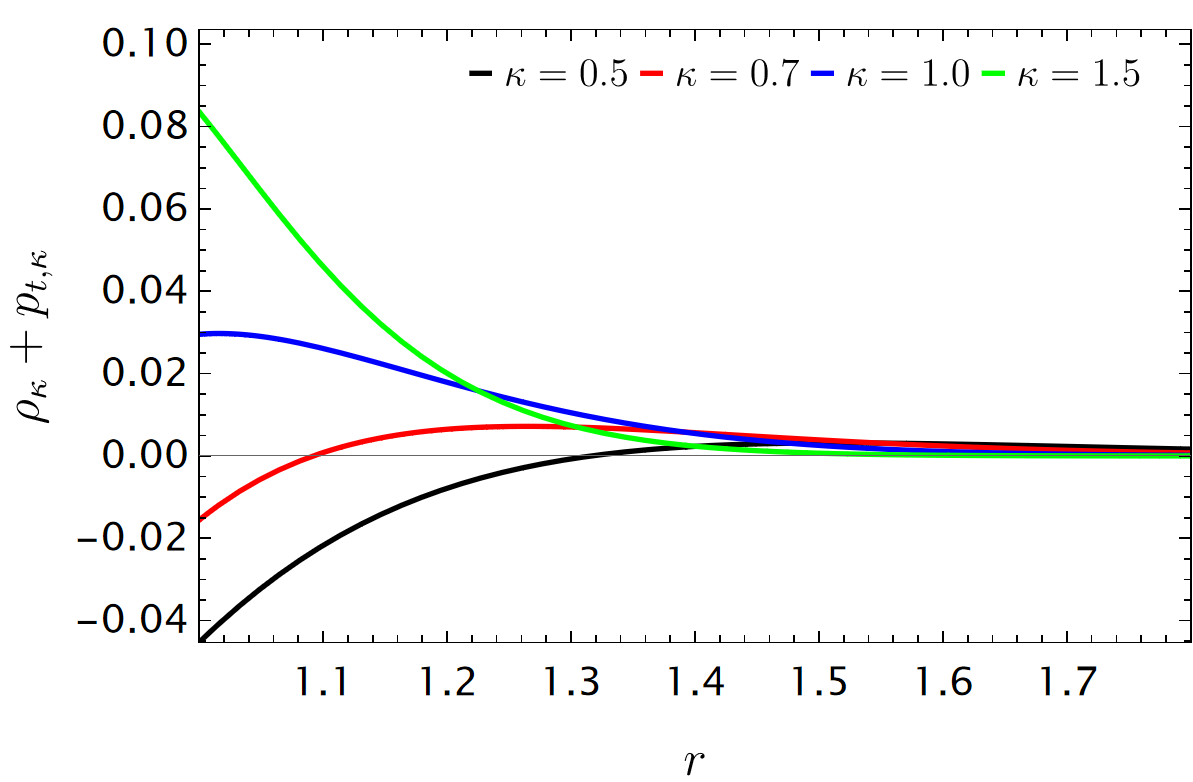}
    \includegraphics[width=0.55\linewidth]{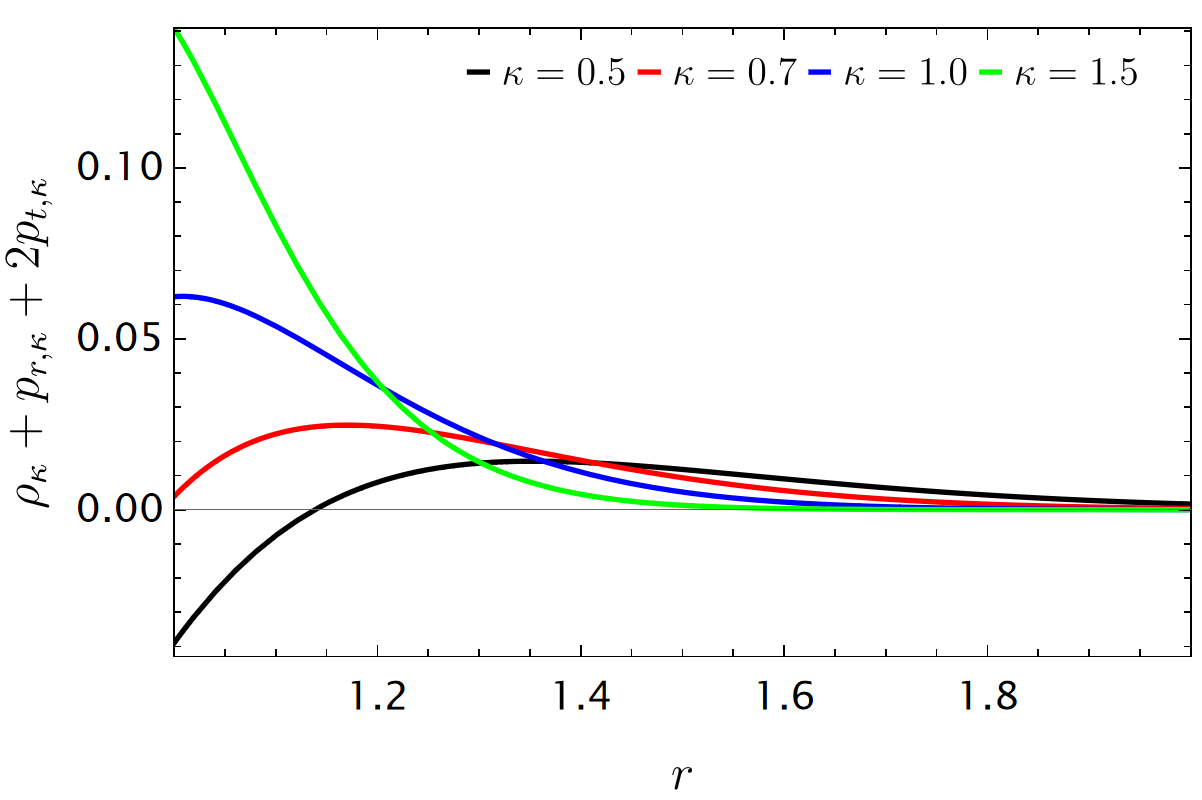}
    \caption{Behavior of the energy conditions for the Kaniadakis-inspired wormhole, with $M=1$ and $r_0=1$ fixed. The top-left panel shows the behavior of the radial NEC, the top-right panel shows the behavior of the tangential NEC, and the bottom panel shows the behavior of the SEC. We set $\kappa = 0.5$, $0.7$, $1.0$, and $1.5$. In all panels, the parameter $w_\kappa$ is fixed for each value of $\kappa$ by the throat-regularity condition \eqref{eq:wth_K}.}
    \label{fig:neckaniadakis}
\end{figure}

The energy-condition behavior is shown in Fig.~\ref{fig:neckaniadakis}. In all panels, $w_\kappa$ is fixed by Eq.~\eqref{eq:wth_K} for each value of $\kappa$. The top-left panel confirms that the radial NEC is violated throughout the plotted region, as required by the throat geometry. The top-right panel reveals a feature that is not present in the same way in the Tsallis branch: the tangential NEC may change sign depending on $\kappa$. For smaller values, such as $\kappa=0.5$, the tangential sector can also contribute to NEC violation near the throat, whereas for larger values the tangential NEC becomes positive and more strongly supports the total pressure sector. The bottom panel shows the SEC combination. Unlike Tsallis, where the SEC combination remains positive in the plotted range, the Kaniadakis sector displays a transition: for smaller $\kappa$ it may begin negative near the throat, while for larger $\kappa$ it becomes positive and can be substantially enhanced. This reflects the non-monotonic way in which the Kaniadakis parameter redistributes the effective source and the tangential stresses.

For the Kaniadakis branch, Eqs.~\eqref{eq:wth_K} and the asymptotic limit of Eq.~\eqref{eq:b_K2} give
{
\begin{equation}
 \mathcal{I}_{\mathrm{NEC},\kappa}^{\pm}
 =-M\sech(\pi\kappa r_0^2)
  -\frac{1}{4\pi\kappa r_0\tanh(\pi\kappa r_0^2)}.
 \label{eq:integrated_nec_K}
\end{equation}
}
The hyperbolic decay makes the integral finite and negative for every $\kappa>0$, consistently with the localized pointwise profiles in Fig.~\ref{fig:neckaniadakis}. Its undeformed behavior is nevertheless singular. The first term approaches $-M$, while the second grows as $-\kappa^{-2}$ when $\kappa\to0$. In this limit, the density is suppressed at fixed radius but its characteristic support moves outward, and the throat-selected pressure coefficient increases at the same time. Their combined effect makes the accumulated radial violation unbounded even though each regular Kaniadakis configuration has a convergent integral.

Analyzing the equilibrium condition, we have
\begin{equation}
 F_{h,\kappa}+F_{g,\kappa}+F_{a,\kappa}=0,
\end{equation}
where the three force contributions are
\begin{align}
 F_{h,\kappa}(r)=\;&
 -\frac{w_\kappa\kappa M}{2r^2}
 \tanh(\pi\kappa r^2)\operatorname{sech}(\pi\kappa r^2)
 \nonumber\\[0.3em]
 &+\pi w_\kappa\kappa^2M
 \left[
 \operatorname{sech}^3(\pi\kappa r^2)
 -\operatorname{sech}(\pi\kappa r^2)\tanh^2(\pi\kappa r^2)
 \right],
\end{align}
\begin{equation}
 F_{g,\kappa}(r)=
 \frac{(1+w_\kappa)\kappa M}{2r}
 \tanh(\pi\kappa r^2)\operatorname{sech}(\pi\kappa r^2)\,
 \Phi_\kappa'(r),
\end{equation}
and
\begin{align}
 F_{a,\kappa}(r)=\;&
 \frac{w_\kappa\kappa M}{2r^2}
 \tanh(\pi\kappa r^2)\operatorname{sech}(\pi\kappa r^2)
 \nonumber\\[0.3em]
 &-\pi w_\kappa\kappa^2M
 \left[
 \operatorname{sech}^3(\pi\kappa r^2)
 -\operatorname{sech}(\pi\kappa r^2)\tanh^2(\pi\kappa r^2)
 \right]
 \nonumber\\[0.3em]
 &-\frac{(1+w_\kappa)\kappa M}{2r}
 \tanh(\pi\kappa r^2)\operatorname{sech}(\pi\kappa r^2)\,
 \Phi_\kappa'(r).
\end{align}
These expressions make the qualitative difference from the algebraic sectors transparent. In the Barrow and Tsallis cases, the force terms inherit power-law scalings from the corresponding densities. Here, by contrast, the hydrostatic, gravitational, and anisotropic contributions are controlled by products of $\tanh(\pi\kappa r^2)$ and $\operatorname{sech}(\pi\kappa r^2)$. As a consequence, the TOV balance is effectively concentrated in the region where these hyperbolic factors are relevant. Changing $\kappa$ therefore modifies not only the magnitude of the forces, but also the width of the radial region where the equilibrium is dynamically significant.

\begin{figure}[!htp]
    \centering
    \includegraphics[width=0.6\linewidth]{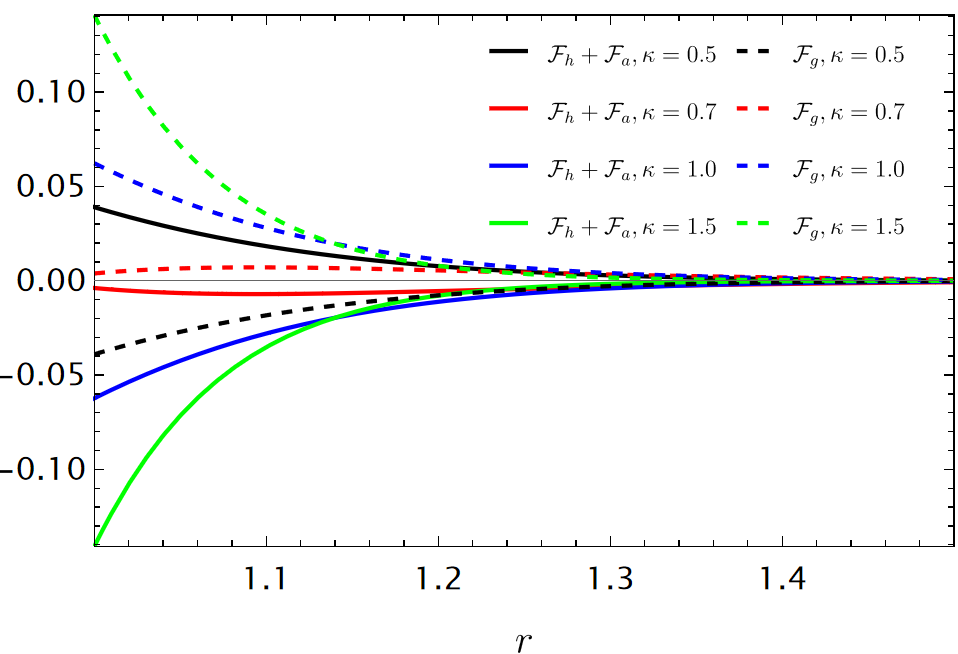}
    \caption{Equilibrium forces for the Kaniadakis-inspired wormhole as functions of the radial coordinate $r$, with $M=1$ and $r_0=1$ fixed. Solid curves represent the combined anisotropic and hydrostatic contribution $F_{a,\kappa}+F_{h,\kappa}$, while dashed curves represent the gravitational contribution $F_{g,\kappa}$. We set $\kappa=0.5$, $0.7$, $1.0$, and $1.5$, with $w_\kappa$ fixed for each curve by the throat-regularity condition \eqref{eq:wth_K}.}
    \label{fig:tovkaniadakis}
\end{figure}

The TOV balance is displayed in Fig.~\ref{fig:tovkaniadakis}. The solid curves represent the combined contribution $F_{h,\kappa}+F_{a,\kappa}$, while the dashed curves represent the gravitational contribution $F_{g,\kappa}$. For each value of $\kappa$, these two contributions have opposite signs and comparable magnitudes, showing the equilibrium relation. The vertical position of each force contribution is not uniform for all plotted values of $\kappa$: in particular, the $\kappa=0.5$ curve shows that $F_{g,\kappa}$ may lie on the negative side close to the throat. The line style, rather than the vertical position of the curve, therefore gives the unambiguous identification of the force sector. The dependence on $\kappa$ is also non-monotonic in the near-throat region, reflecting the localized hyperbolic structure of the density profile. For larger values such as $\kappa=1.0$ and $1.5$, the force separation becomes stronger and more sharply concentrated near the throat, while all contributions still decay rapidly away from it. This behavior is consistent with the energy-condition analysis, where the tangential pressure sector becomes increasingly important for larger $\kappa$ and is reflected both in the SEC combination and in the anisotropic part of the TOV balance.

From the phenomenological point of view, the Kaniadakis parameter plays a more localized role than the Barrow parameter $\Delta$ and a less purely algebraic role than the Tsallis parameter $\delta$. While increasing $\Delta$ or $\delta$ mainly changes the power-law hierarchy of the source, varying $\kappa$ reshapes a compact effective matter distribution. This makes the Kaniadakis-inspired branch especially suitable for studying wormholes whose exoticity and equilibrium are concentrated near the throat, with a rapid suppression of the effective sector in the asymptotic region.

\subsection{Logarithmic-inspired sector}

\subsubsection{Geometry induced by the logarithmic-inspired profile}

Logarithmic corrections arise naturally in several approaches to quantum gravity and horizon-state counting \cite{LogCorr}. In the entropy--geometry correspondence of Ref.~\cite{Anand}, the associated effective density takes the form
\begin{equation}
 \rho_{\log}(r)=\frac{\lambda M}{2r\left(\lambda+\pi r^2\right)^2}.
 \label{eq:rho_log2}
\end{equation}
The sign of the density is controlled by the parameter $\lambda$. For $M>0$, the choice $\lambda<0$ selects a negative-density branch, which is especially relevant for wormhole configurations supported by exotic effective matter. However, this does not exclude the possibility of obtaining geometrically viable wormhole configurations for positive values of $\lambda$, in which case the density is positive and the violation of the energy conditions becomes more dependent on the pressure sector and on the equation-of-state parameter.

In contrast to the Barrow- and Tsallis-inspired sectors, whose densities are governed by pure power laws, the logarithmic branch is controlled by a rational profile with an intrinsic length scale set by $\lambda$. It also differs from the Kaniadakis case, where the source is localized by hyperbolic functions. Here, the interplay between the constant scale $\lambda$ and the area contribution $\pi r^2$ produces a profile that is neither purely algebraic nor exponentially localized. In addition, the presence of the denominator requires the explicit condition
\begin{equation}
 \lambda+\pi r_0^2>0,
 \label{eq:lambda_cond2}
\end{equation}
in order to exclude poles from the physical region.

\begin{figure}[!htp]
    \centering
    \includegraphics[width=0.6\linewidth]{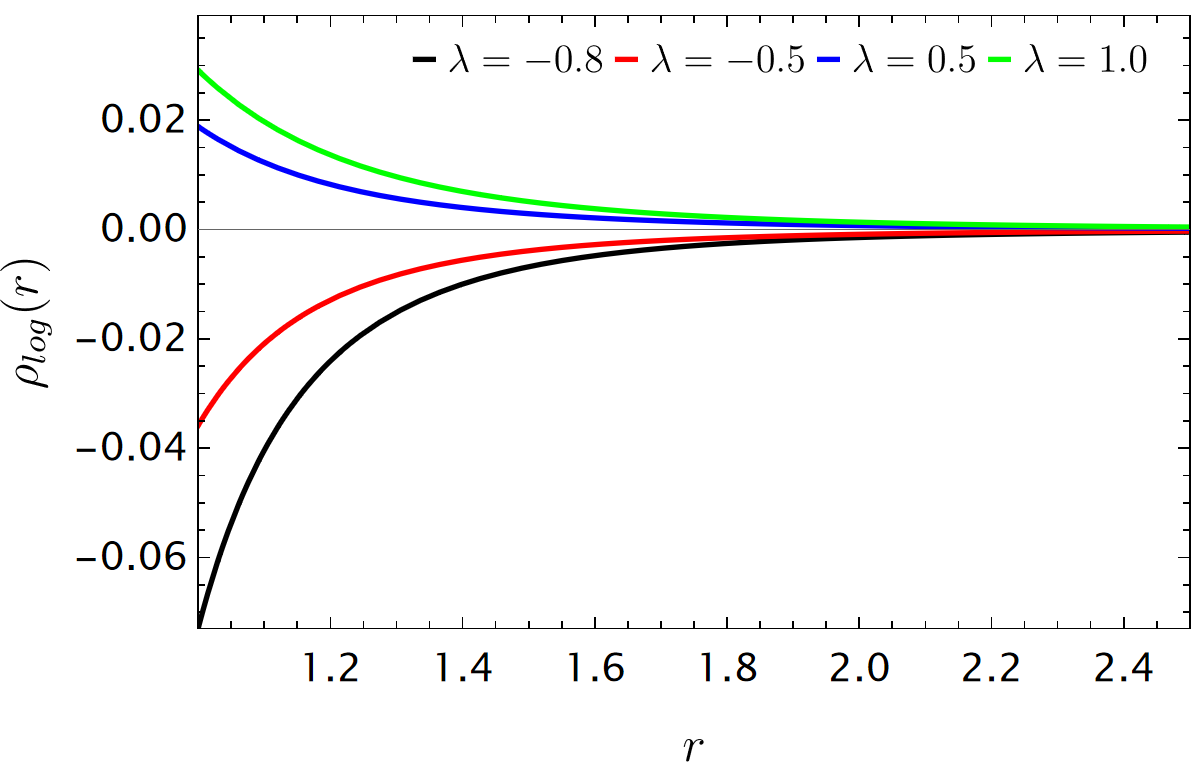}
    \caption{The Logarithmic-inspired density profile, $\rho_{\log}(r)$, for different values of the parameter $\lambda$ as a function of the radial coordinate $r$, with $M=1$ and $r_0=1$ fixed. The curves correspond to $\lambda=-0.8$, $-0.5$, $0.5$, and $1.0$.}
    \label{fig:rholog}
\end{figure}

The behavior of $\rho_{\log}(r)$ is shown in Fig.~\ref{fig:rholog}. The figure illustrates one of the distinctive features of the logarithmic branch: unlike the previous sectors, the parameter controlling the deformation may change not only the magnitude of the source, but also its sign. Negative values of $\lambda$ produce the more standard exotic branch with $\rho_{\log}<0$, while positive values generate a positive-density branch that can still sustain a wormhole geometry once the pressure sector is properly adjusted. This makes the logarithmic correction more flexible than the Barrow, Tsallis, and Kaniadakis sectors, where the sign of the density is fixed throughout the branch considered.

The corresponding shape function follows from Eq.~\eqref{eq:b_general2} and can be integrated analytically as
\begin{equation}
 b_{\log}(r)=r_0-\frac{2\lambda M}{\lambda+\pi r^2}+\frac{2\lambda M}{\lambda+\pi r_0^2}.
 \label{eq:b_log2}
\end{equation}
Thus, the logarithmic correction generates a geometry in which the throat is influenced not by a simple algebraic tail, but by the competition between the constant scale $\lambda$ and the radial growth of the horizon-area term. This is a qualitative difference from the Barrow and Tsallis sectors, where the large-$r$ behavior is dictated by fixed powers of $r$, and also from the Kaniadakis case, where the source is more sharply localized around the throat.

The logarithmic sector behaves differently in the undeformed density limit $\lambda\to0$. Equation~\eqref{eq:rho_log2} vanishes at fixed $r$, and the large-radius limit of Eq.~\eqref{eq:b_log2} gives a density contribution that also tends to zero. Reversing these operations leads to the same result. The limits therefore commute throughout the pole-free domain, and the ADM mass approaches the boundary term $r_0/2$.

\begin{figure}[!htp]
    \centering
    \includegraphics[width=0.49\linewidth]{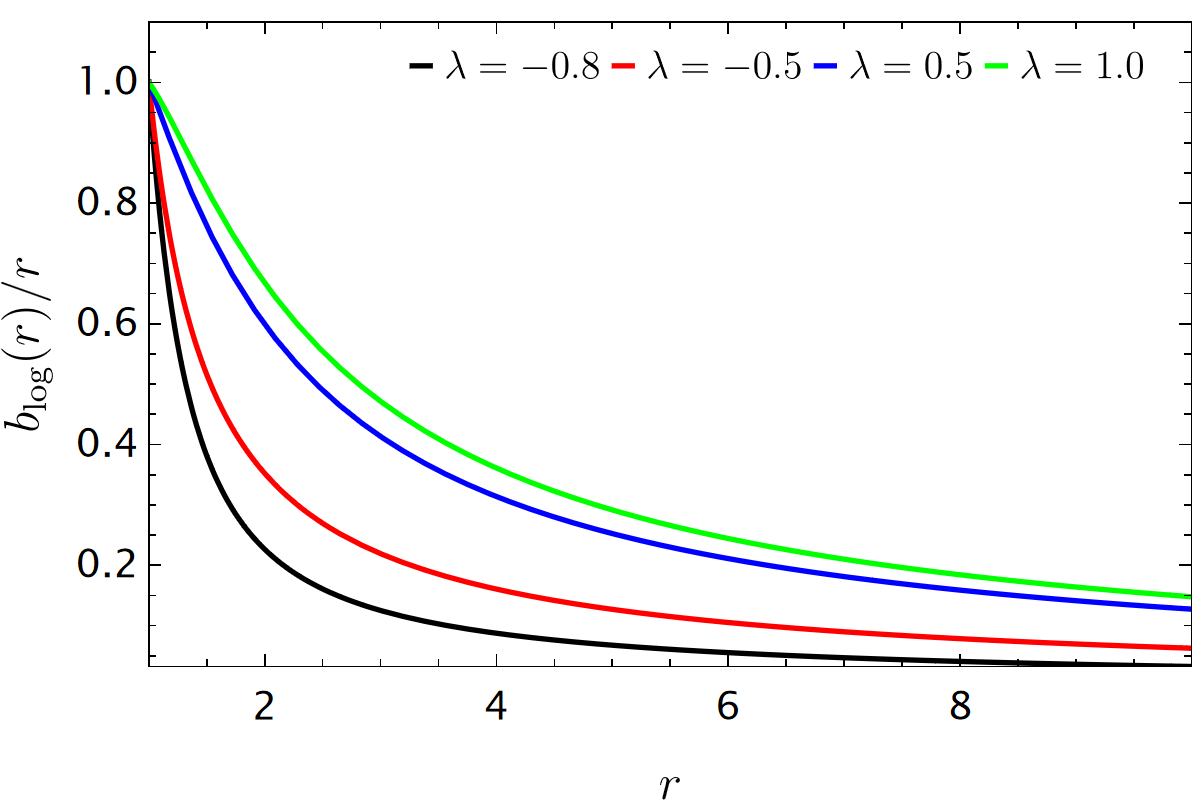}
    \includegraphics[width=0.49\linewidth]{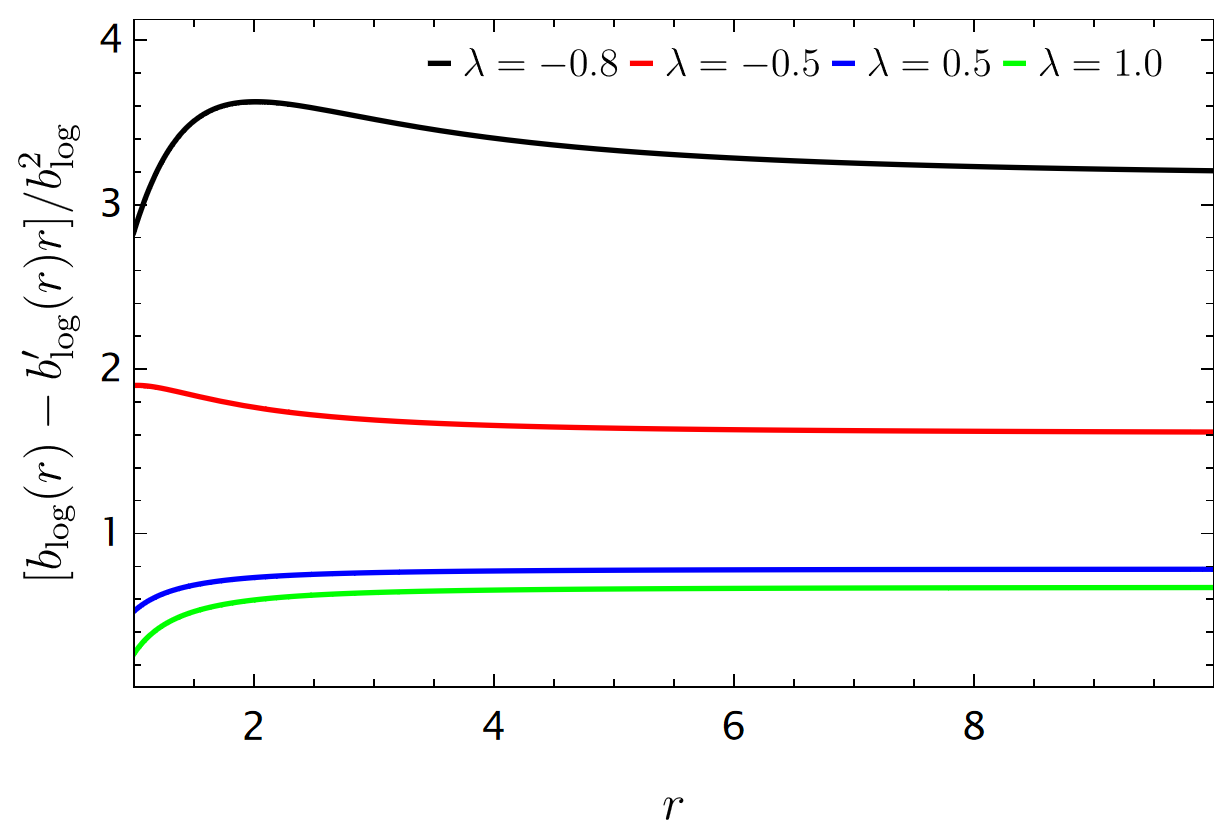}
    \caption{The geometric behavior of the Logarithmic-inspired wormhole for different values of the parameter $\lambda$, with $M=1$ and $r_0=1$ fixed. In the left panel, we show the ratio $b_{\log}(r)/r$ as a function of the radial coordinate $r$. In the right panel, we present the flare-out function $[b_{\log}(r)-b_{\log}'(r)r]/b_{\log}^2$ as a function of $r$. The curves correspond to $\lambda=-0.8$, $-0.5$, $0.5$, and $1.0$.}
    \label{fig:geolog}
\end{figure}

The geometric behavior associated with Eq.~\eqref{eq:b_log2} is displayed in Fig.~\ref{fig:geolog}. The left panel shows that the ratio $b_{\log}(r)/r$ decreases and tends to zero for all admissible values of $\lambda$ considered, indicating that the shape-function part of the asymptotic flatness condition is satisfied throughout this branch. The right panel verifies the flare-out behavior. For the values shown, the quantity $b_{\log}(r)-r b'_{\log}(r)$ remains positive in the physical domain, confirming that the spatial geometry opens outward from the throat. Compared with the Kaniadakis case, where the geometric behavior is mainly governed by the width of a localized core, the logarithmic sector is more directly controlled by the finite scale $\lambda$, which determines how strongly the denominator modifies the near-throat region.

Substituting Eq.~\eqref{eq:b_log2} into the general redshift equation \eqref{eq:redshift_general2}, one obtains
\begin{equation}
 \Phi_{\log}'(r)=
 \frac{
 r_0-\dfrac{2\lambda M}{\lambda+\pi r^2}
 +\dfrac{2\lambda M}{\lambda+\pi r_0^2}
 +\dfrac{4\pi w\lambda M r^2}{\left(\lambda+\pi r^2\right)^2}
 }
 {
 2r\left[
 r-r_0+\dfrac{2\lambda M}{\lambda+\pi r^2}
 -\dfrac{2\lambda M}{\lambda+\pi r_0^2}
 \right]
 }.
 \label{eq:phi_log2}
\end{equation}
Compared with the previous entropy sectors, the logarithmic branch stands out by combining a built-in geometric scale with a denominator structure that must remain regular in the physical region. In this sense, it is less rigid than the power-law Barrow and Tsallis branches, and less localized than the Kaniadakis one. The parameter $\lambda$ controls not only the density sector, but also how strongly the redshift derivative is affected by the finite entropy-correction scale. As in the previous cases, the regularity of Eq.~\eqref{eq:phi_log2} at the throat will be addressed in the next subsection, where the equation-of-state parameter is fixed by the throat-regularity condition.

An important difference between the logarithmic sector and the previously analyzed profiles is that the redshift function can be integrated analytically for an arbitrary admissible value of $\lambda$. After imposing the throat-regularity condition $w=w_{\log}$, it is convenient to introduce $u=r/r_0$, $\ell=\lambda/(\pi r_0^2)$, and $\mu=2M/r_0$. In terms of these quantities, and fixing the integration constant by the asymptotic normalization $\Phi_{\log}(\infty)=0$, one obtains
\begin{equation}
 \Phi_{\log}(u)=\mathcal{P}_{\log}(u)
 -\lim_{\bar u\to\infty}\mathcal{P}_{\log}(\bar u),
 \label{eq:Phi_log_integrated}
\end{equation}
where
\begin{align}
 \mathcal{P}_{\log}(u)=\;&
 -\frac{1}{2}\ln u
 -\frac{(\ell+1)^2}{4\ell\mu}\ln\left(u^2+\ell\right)
 +\frac{(\ell+1)^2+\ell\mu}{4\ell\mu}
 \ln\left|Q_{\log}(u)\right|
 \nonumber\\[0.4em]
 &+\frac{1+\ell\mu-\ell^2}{4}\,\mathcal{J}_{\log}(u),
\end{align}
with
\begin{equation}
 Q_{\log}(u)=(\ell+1)u^2-\ell\mu u+\ell(\ell+1-\mu),
\end{equation}
and
\begin{equation}
\mathcal{J}_{\log}(u)=
\begin{cases}
\dfrac{2}{\sqrt{\mathcal{D}_{\log}}}
\tan^{-1}\left[
\dfrac{2(\ell+1)u-\ell\mu}
{\sqrt{\mathcal{D}_{\log}}}
\right],
& \mathcal{D}_{\log}>0,\\[1.2em]
-\dfrac{2}{2(\ell+1)u-\ell\mu},
& \mathcal{D}_{\log}=0,\\[1.2em]
\dfrac{1}{\sqrt{-\mathcal{D}_{\log}}}
\ln\left|
\dfrac{2(\ell+1)u-\ell\mu-\sqrt{-\mathcal{D}_{\log}}}
{2(\ell+1)u-\ell\mu+\sqrt{-\mathcal{D}_{\log}}}
\right|,
& \mathcal{D}_{\log}<0,
\end{cases}
\end{equation}
where
\begin{equation}
 \mathcal{D}_{\log}
 =
 4\ell(\ell+1)(\ell+1-\mu)-\ell^2\mu^2.
\end{equation}
Therefore, unlike the Kaniadakis branch, where the hyperbolic structure prevents a compact elementary primitive, the logarithmic profile leads to a rational redshift equation and can be integrated in closed form for arbitrary $\lambda$, provided the physical-domain condition $\lambda+\pi r_0^2>0$ is respected. The different forms of $\mathcal{J}_{\log}$ simply reflect the possible factorization structures of the quadratic term $Q_{\log}(u)$. This analytic integrability is a distinctive feature of the logarithmic sector and shows that its additional scale $\lambda$ modifies the redshift geometry in a controlled, explicitly tractable way.

\begin{figure}[!htp]
    \centering
    \includegraphics[width=0.58\linewidth]{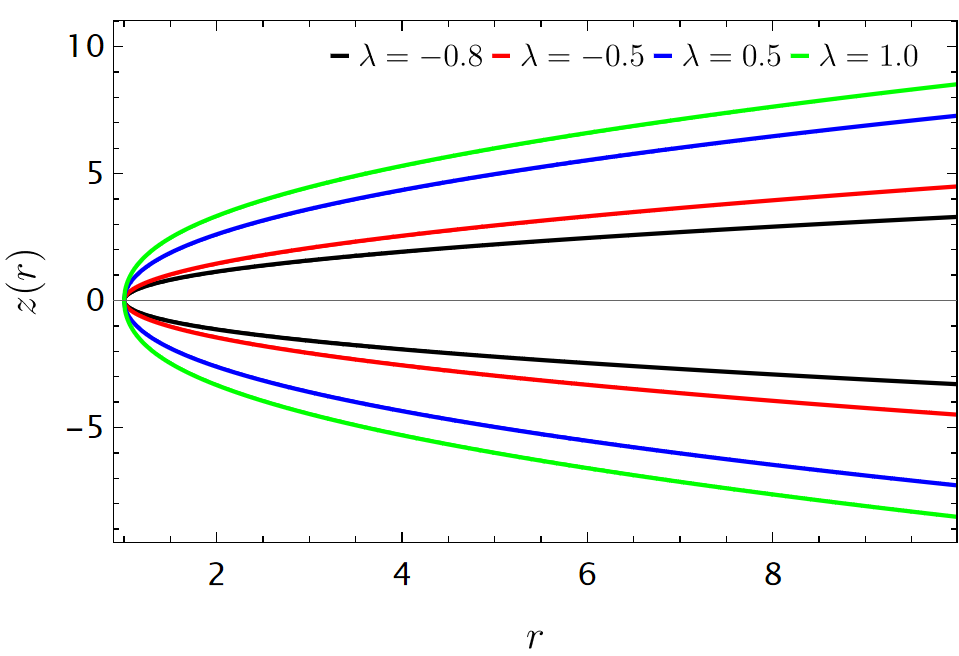}
    \includegraphics[width=0.4\linewidth]{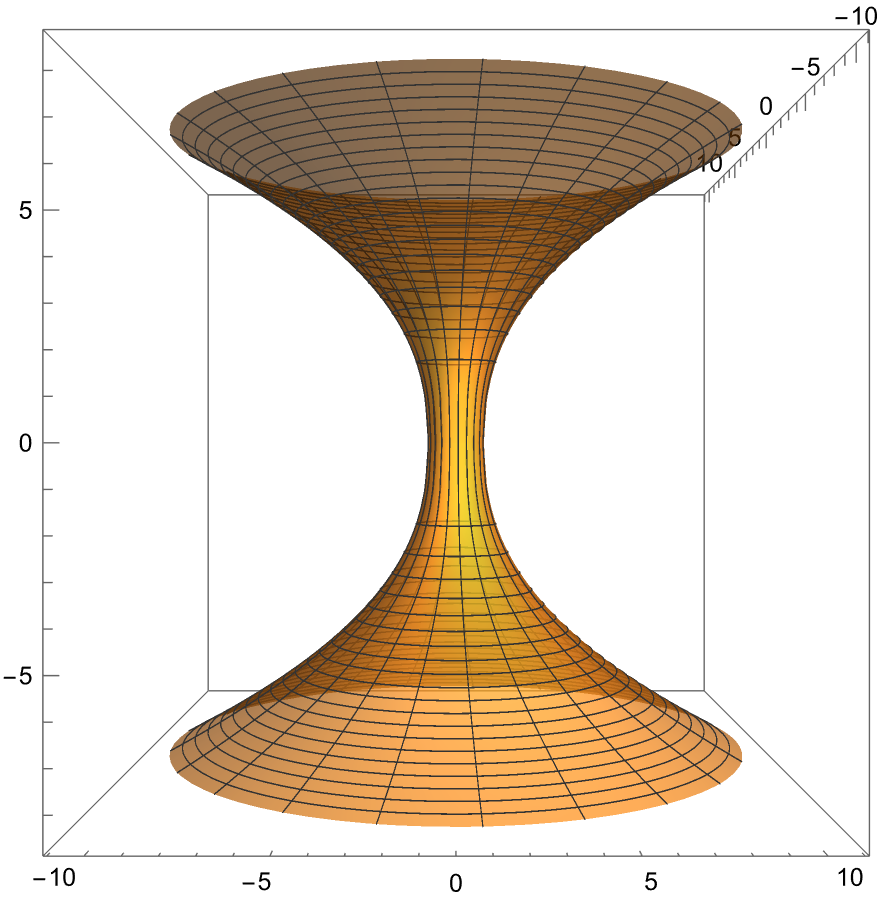}
    \caption{Embedding structure of the Logarithmic-inspired wormhole geometry for different values of the parameter $\lambda$, with $M=1$ and $r_0=1$ fixed. In the left panel, we show the embedding function $z(r)$ as a function of the radial coordinate $r$ for $\lambda=-0.8$, $-0.5$, $0.5$, and $1.0$. In the right panel, we present the corresponding three-dimensional embedded surface for the representative case $\lambda=1.0$, with $M=1$ and $r_0=1$ fixed.}
    \label{fig:embededlog}
\end{figure}

A complementary view of the geometry is provided by the embedding diagram in Fig.~\ref{fig:embededlog}. The left panel shows the embedding function $z(r)$ for different values of $\lambda$, while the right panel presents a representative three-dimensional embedded surface. As in the other sectors, the vertical tangent at the throat is the expected geometric signature of the minimum-radius surface and does not indicate a physical singularity. What distinguishes the logarithmic case is that the deformation is controlled by a finite scale rather than by a pure radial exponent or by hyperbolic localization. As a result, the embedded surfaces reveal how the parameter $\lambda$ modifies the throat neighborhood in a smoother and more scale-dependent way.

\subsubsection{Energy conditions and equilibrium in the logarithmic-inspired sector}

The radial pressure is fixed by the chosen equation of state,
\begin{equation}
 p_{r,\log}(r)=w_{\log}\rho_{\log}(r).
\end{equation}
As in the previous sectors, the equation-of-state parameter is fixed by the requirement that the redshift derivative be regular at the throat. For the logarithmic-inspired shape function, one obtains
\begin{equation}
 b_{\log}'(r)=
 \frac{4\pi\lambda M r}
 {\left(\lambda+\pi r^2\right)^2}.
\end{equation}
Therefore, the throat-regularity condition $w=-1/b'(r_0)$ gives
\begin{equation}
 w_{\log}\equiv w_{\rm th}^{(\log)}
 =
 -\frac{\left(\lambda+\pi r_0^2\right)^2}
 {4\pi\lambda M r_0}.
 \label{eq:wth_log}
\end{equation}
This expression reveals an important difference with respect to the Barrow, Tsallis, and Kaniadakis branches. For $\lambda<0$, with the pole-exclusion condition $\lambda+\pi r_0^2>0$, one has $\rho_{\log}<0$ and $w_{\log}>0$, leading again to a negative radial pressure. For $\lambda>0$, however, the density is positive while $w_{\log}<0$. In the geometrically admissible cases with $0<b_{\log}'(r_0)<1$, this implies $w_{\log}<-1$, so that the radial sector behaves in a phantom-like way: the NEC violation is driven not by negative density, but by a sufficiently negative radial pressure. Thus, the logarithmic branch naturally accommodates two physically distinct regimes within the same density profile. When $\lambda\to0$ along regular configurations, $w_{\log}$ diverges while the throat pressure remains finite. The commuting mass limits therefore do not imply a vacuum limit of the complete reconstructed source.

Using Eq.~\eqref{eq:pt_from_TOV2} together with Eqs.~\eqref{eq:rho_log2} and \eqref{eq:phi_log2}, with $w=w_{\log}$, the tangential pressure can be written explicitly as
\begin{equation}
 p_{t,\log}(r)=
 \frac{\lambda M}{4}
 \left[
 \frac{w_{\log}\left(\lambda-3\pi r^2\right)}
 {r\left(\lambda+\pi r^2\right)^3}
 +\frac{(1+w_{\log})\Phi_{\log}'(r)}
 {\left(\lambda+\pi r^2\right)^2}
 \right].
\end{equation}
Hence, the anisotropy of the logarithmic branch is controlled not only by the throat-selected equation-of-state parameter, but also by the competition between the intrinsic scale $\lambda$ and the radial area term $\pi r^2$. This is qualitatively different from the Kaniadakis case, where the anisotropy is governed by localized hyperbolic factors. Here, the denominator structure makes the pressure sector particularly sensitive to the distance from the excluded pole.

The radial NEC is
\begin{equation}
 \rho_{\log}+p_{r,\log}
 =
 \frac{4\pi\lambda M r_0-(\lambda+\pi r_0^2)^2}
 {8\pi r r_0(\lambda+\pi r^2)^2}.
\end{equation}
For $\lambda<0$, the density is negative and $w_{\log}>0$, so the radial NEC is violated throughout the physical domain. For $\lambda>0$, the density is positive, but the throat-regularity condition typically selects $w_{\log}<-1$, again producing radial NEC violation. At the throat, this result is fixed by geometry:
\begin{equation}
 \left.\left(\rho_{\log}+p_{r,\log}\right)\right|_{r_0}
 =
 \frac{4\pi\lambda M r_0-(\lambda+\pi r_0^2)^2}
 {8\pi r_0^2(\lambda+\pi r_0^2)^2}<0.
\end{equation}
Thus, the logarithmic sector is especially interesting because the same throat identity can realize the required radial exoticity either through a negative-density branch or through a positive-density branch with phantom-like radial pressure.

\begin{figure}[htp!]
    \centering
    \includegraphics[width=0.49\linewidth]{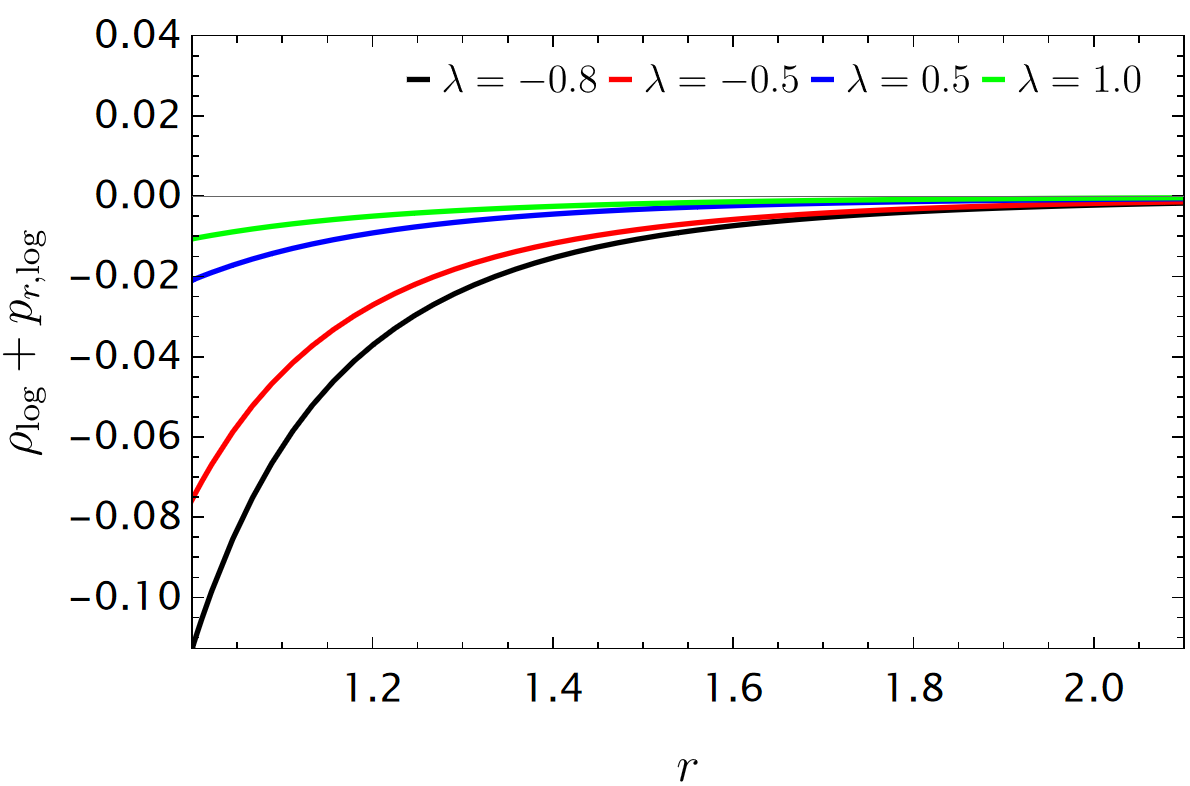}
    \includegraphics[width=0.49\linewidth]{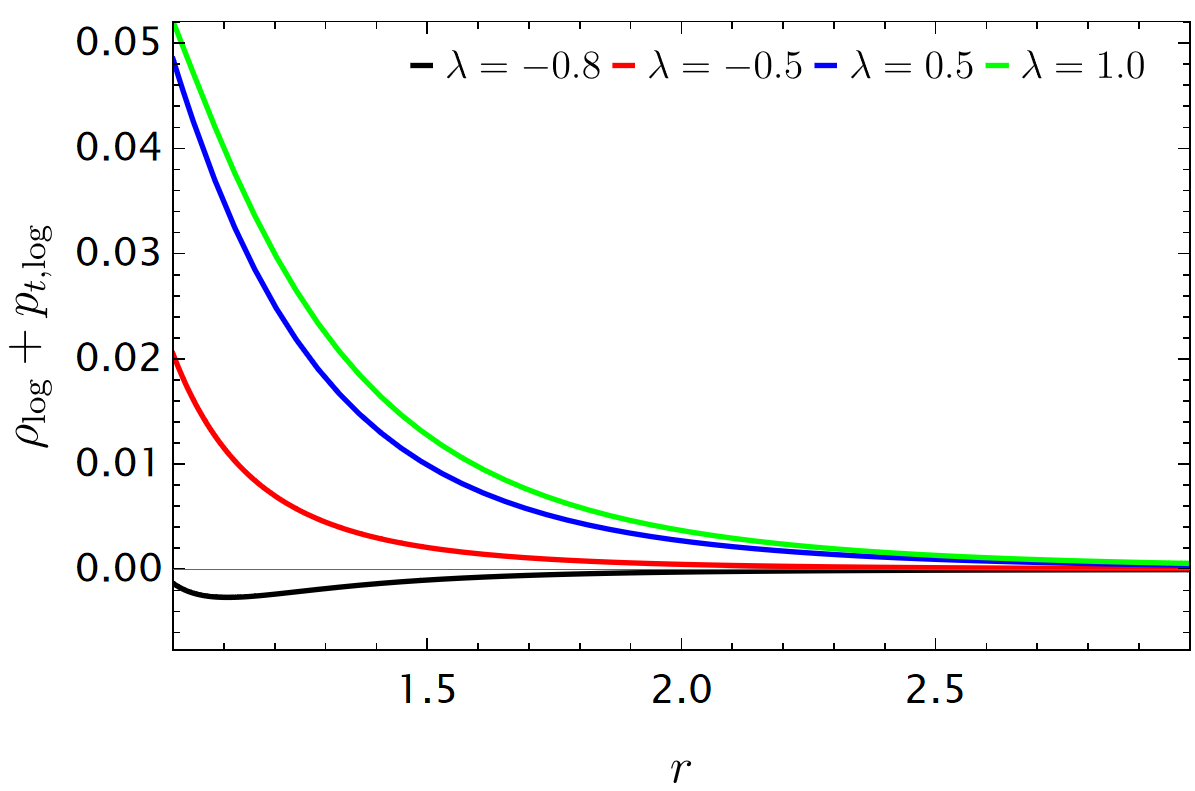}
    \includegraphics[width=0.55\linewidth]{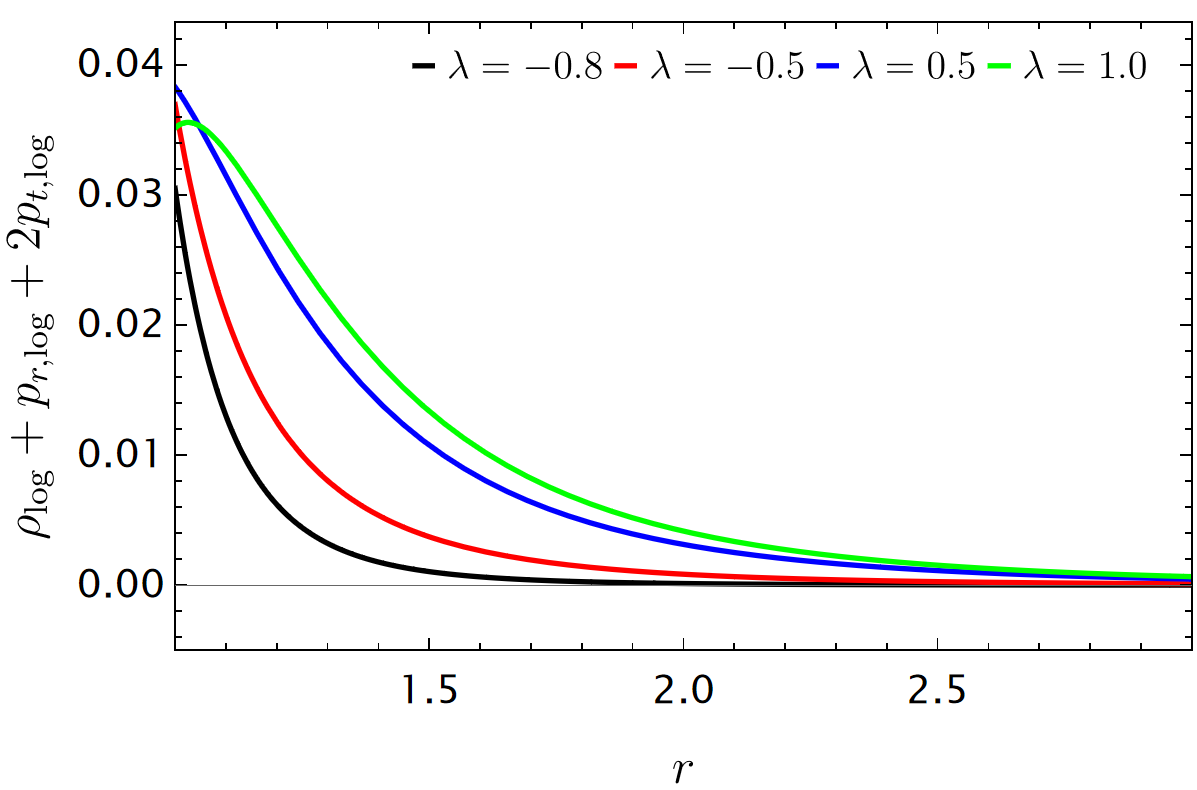}
    \caption{Behavior of the energy conditions for the logarithmic-inspired wormhole, with $M=1$ and $r_0=1$ fixed. The top-left panel shows the behavior of the radial NEC, the top-right panel shows the behavior of the tangential NEC, and the bottom panel shows the behavior of the SEC combination. We set $\lambda=-0.8$, $-0.5$, $0.5$, and $1.0$. In all panels, the parameter $w_{\log}$ is fixed for each value of $\lambda$ by the throat-regularity condition \eqref{eq:wth_log}.}
    \label{fig:neclog}
\end{figure}

Figure~\ref{fig:neclog} displays the behavior of the radial NEC, tangential NEC, and SEC combination for the logarithmic-inspired sector. In all panels, $w_{\log}$ is fixed by Eq.~\eqref{eq:wth_log} for each value of $\lambda$. The top-left panel shows that the radial NEC remains negative for all plotted values, although the physical origin of this violation differs between $\lambda<0$ and $\lambda>0$. The top-right panel shows that the tangential NEC is mostly positive, except for the more negative branch close to the throat, where the denominator structure enhances the sensitivity of the transverse pressure. The bottom panel shows that the SEC combination $\rho_{\log}+p_{r,\log}+2p_{t,\log}$ remains positive in the plotted range. This does not mean that the full SEC is satisfied, since the radial NEC is already violated; rather, it shows that the tangential sector compensates part of the radial exoticity in the total pressure combination. Compared with Kaniadakis, where the SEC combination can change sign with $\kappa$, the logarithmic sector displays a more uniformly positive SEC combination for the chosen values of $\lambda$.

The logarithmic integral follows from Eqs.~\eqref{eq:wth_log}, \eqref{eq:b_log2}, and \eqref{eq:integrated_nec_identity}:
{
\begin{equation}
 \mathcal{I}_{\mathrm{NEC},\log}^{\pm}
 =\frac{\lambda M}{\lambda+\pi r_0^2}
  -\frac{\lambda+\pi r_0^2}{4\pi r_0}.
 \label{eq:integrated_nec_log}
\end{equation}
}
The rational profile makes the weighted integrand decay as $r^{-3}$, and the integral is finite throughout the admissible pole-free branches. For $\lambda<0$, its negative sign follows from the negative density together with $w_{\log}>0$. For $\lambda>0$, the density contribution is positive, but the throat-selected phantom-like pressure is sufficiently negative to keep the total integral below zero. When $\lambda\to0$, the density contribution vanishes while the product of the diverging coefficient $w_{\log}$ with the suppressed density remains integrable. The two branches therefore approach the same finite one-ended value $-r_0/4$.

Analyzing the equilibrium condition, we have
\begin{equation}
 F_{h,\log}+F_{g,\log}+F_{a,\log}=0,
\end{equation}
where the three force contributions are
\begin{equation}
 F_{h,\log}(r)=
 \frac{w_{\log}\lambda M\left(\lambda+5\pi r^2\right)}
 {2r^2\left(\lambda+\pi r^2\right)^3},
\end{equation}
\begin{equation}
 F_{g,\log}(r)=
 -\frac{(1+w_{\log})\lambda M}
 {2r\left(\lambda+\pi r^2\right)^2}\,
 \Phi_{\log}'(r),
\end{equation}
and
\begin{equation}
 F_{a,\log}(r)=
 -\frac{w_{\log}\lambda M\left(\lambda+5\pi r^2\right)}
 {2r^2\left(\lambda+\pi r^2\right)^3}
 +\frac{(1+w_{\log})\lambda M}
 {2r\left(\lambda+\pi r^2\right)^2}\,
 \Phi_{\log}'(r).
\end{equation}
These expressions show that the logarithmic branch has a more delicate equilibrium structure than the purely algebraic sectors. In the Barrow and Tsallis cases, the force terms inherit a direct power-law hierarchy from the density profile. In the Kaniadakis case, the force balance is localized by hyperbolic functions. Here, by contrast, the factors $\left(\lambda+\pi r^2\right)^{-2}$ and $\left(\lambda+\pi r^2\right)^{-3}$ make the equilibrium sensitive to the finite logarithmic scale and to the pole-avoidance condition. Consequently, changing $\lambda$ modifies not only the magnitude of the forces, but also whether the supporting radial sector is negative-density-like or phantom-like.

\begin{figure}[!htp]
    \centering
    \includegraphics[width=0.6\linewidth]{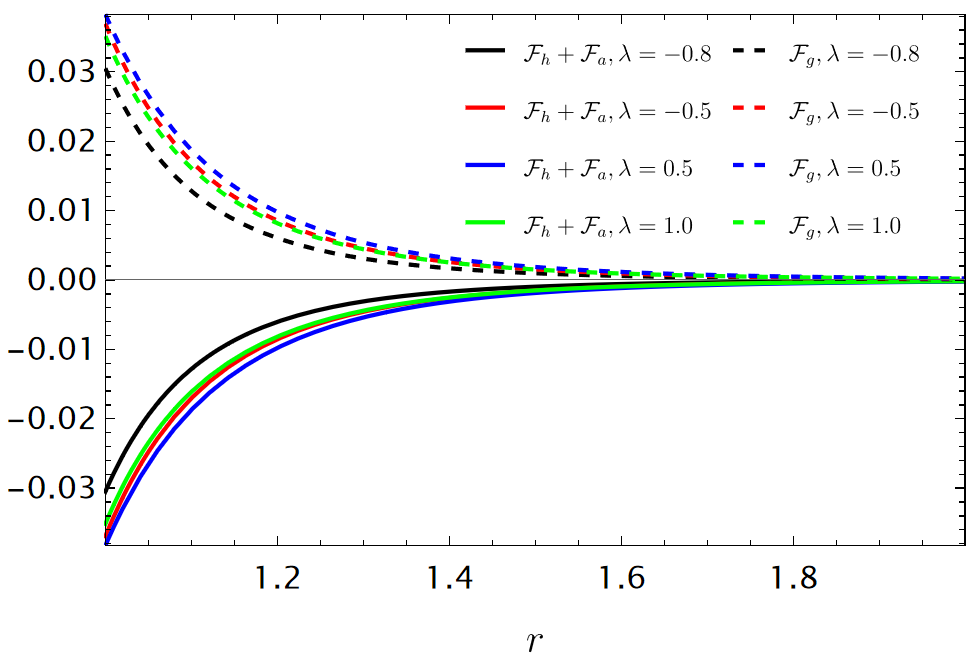}
    \caption{Equilibrium forces for the logarithmic-inspired wormhole as functions of the radial coordinate $r$, with $M=1$ and $r_0=1$ fixed. Solid curves represent the combined anisotropic and hydrostatic contribution $F_{a,\log}+F_{h,\log}$, while dashed curves represent the gravitational contribution $F_{g,\log}$. We set $\lambda=-0.8$, $-0.5$, $0.5$, and $1.0$, with $w_{\log}$ fixed for each curve by the throat-regularity condition \eqref{eq:wth_log}.}
    \label{fig:tovlog}
\end{figure}

The TOV balance is displayed in Fig.~\ref{fig:tovlog}. The solid curves represent the combined contribution $F_{h,\log}+F_{a,\log}$, while the dashed curves represent the gravitational contribution $F_{g,\log}$. For each value of $\lambda$, the two contributions have opposite signs and comparable magnitudes, confirming the equilibrium relation. The positive-$\lambda$ curves should be interpreted with special care: although the density is positive in this regime, the selected value $w_{\log}<-1$ produces a phantom-like radial pressure, allowing the radial NEC violation required by the throat. For negative $\lambda$, the interpretation is closer to the previous negative-density sectors. Thus, the logarithmic branch interpolates between two different mechanisms of support: one based on negative effective density and another based on a strongly negative radial pressure. As $r$ increases, all contributions tend to zero, consistently with the decay of the logarithmic-inspired source and the asymptotic weakening of the effective matter sector.

From the phenomenological point of view, the logarithmic sector is more flexible than the previous branches. For $\lambda<0$, it resembles the negative-density effective sources of Barrow, Tsallis, and Kaniadakis. For $\lambda>0$, however, the density becomes positive and the wormhole is supported by a phantom-like radial pressure selected by the regularity of the redshift function. This makes the logarithmic-inspired branch especially useful for comparing two distinct forms of exoticity within the same entropic profile: negative effective density versus positive density with super-negative radial pressure.

\subsection{Exponential-inspired sector}

\subsubsection{Geometry induced by the exponential-inspired profile}

The exponentially corrected entropy introduces a non-perturbative deformation of the Bekenstein--Hawking law through the parameter $\eta$. In the entropy--geometry correspondence of Ref.~\cite{Anand}, the associated density reads
\begin{equation}
 \rho_{\exp}(r)=
 -\frac{\eta M e^{\pi r^2}}
 {2r\left(e^{\pi r^2}-\eta\right)^2}.
 \label{eq:rho_exp2}
\end{equation}
For $\eta>0$, the density is negative throughout the physical branch, provided the denominator does not vanish. In contrast to the Barrow and Tsallis sectors, whose radial behavior is controlled by power laws, the exponential branch is governed by a non-perturbative suppression scale. It also differs from the Kaniadakis profile, where localization is produced by hyperbolic functions, and from the logarithmic branch, where the correction is rational in $\lambda+\pi r^2$. Here, the effective source is controlled by the competition between the exponential growth of the area term and the constant parameter $\eta$.

The regularity of the source requires the pole to remain outside the physical region. Since the wormhole domain starts at $r=r_0$, we impose
\begin{equation}
 e^{\pi r_0^2}-\eta>0.
 \label{eq:eta_cond2}
\end{equation}
For the numerical choice $r_0=1$, this means $\eta<e^\pi$. The values $\eta=1.0$, $2.0$, $4.0$, and $6.0$ therefore lie safely inside the admissible domain and also avoid the very weak-source regime, where the throat-regularity condition would require an excessively large value of the equation-of-state parameter.

\begin{figure}[!htp]
    \centering
    \includegraphics[width=0.6\linewidth]{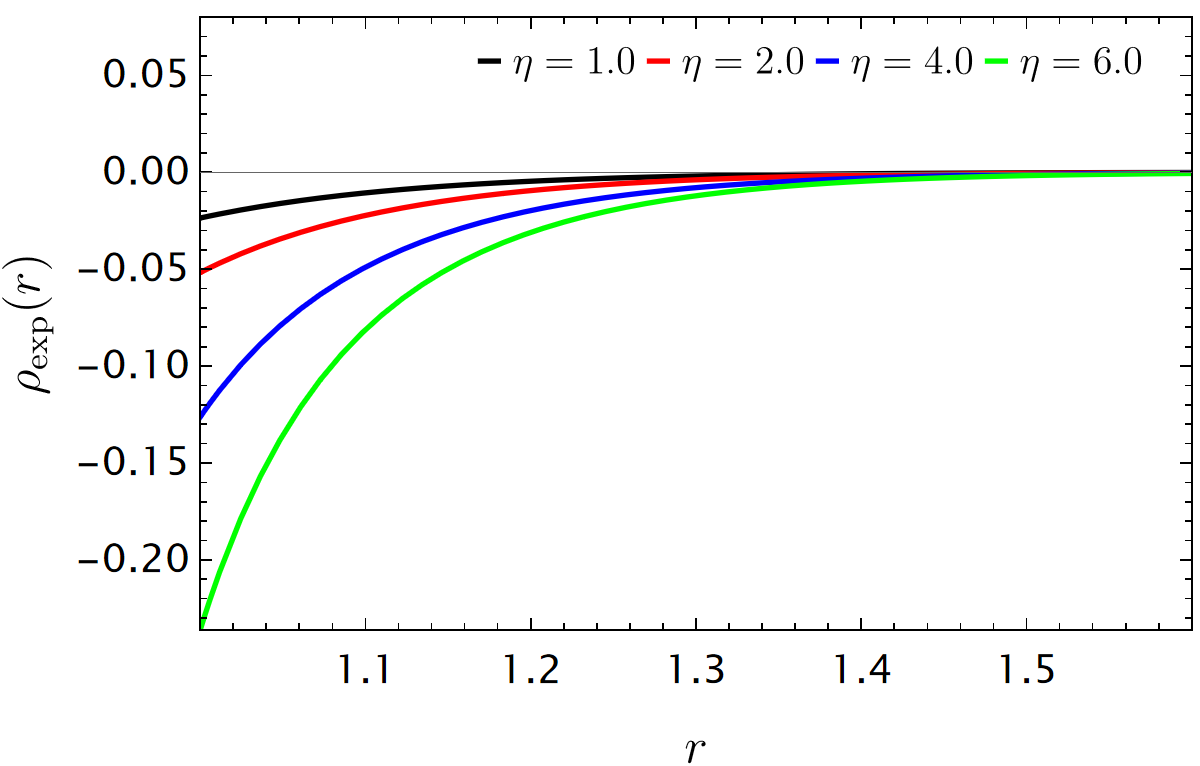}
    \caption{The exponential-inspired density profile, $\rho_{\text{exp}}(r)$, for different values of the parameter $\eta$ as a function of the radial coordinate $r$, with $M=1$ and $r_0=1$ fixed. The curves correspond to $\eta = 1.0$, $2.0$, $4.0$, and $6.0$.}
    \label{fig:rhoexp}
\end{figure}

The density profile is shown in Fig.~\ref{fig:rhoexp}. Increasing $\eta$ strengthens the magnitude of the negative density near the throat, while the exponential factor rapidly suppresses the source as $r$ grows. This makes the exponential branch more sharply concentrated than the algebraic Barrow and Tsallis sectors. Compared with the logarithmic case, the sign of the density does not change in the branch considered here; instead, the main effect of $\eta$ is to control how strongly the negative source is amplified before the exponential decay dominates.

The corresponding shape function follows from Eq.~\eqref{eq:b_general2} and can be integrated analytically as
\begin{equation}
 b_{\exp}(r)=r_0+2\eta M
 \left[
 \frac{1}{e^{\pi r^2}-\eta}
 -\frac{1}{e^{\pi r_0^2}-\eta}
 \right].
 \label{eq:b_exp2}
\end{equation}
Unlike the Barrow- and Tsallis-inspired branches, where the shape function inherits an algebraic radial dependence, the exponential sector is controlled by a rapidly suppressed denominator. Therefore, the deformation has its largest geometric effect close to the throat and becomes negligible much faster in the asymptotic region.

The exponential sector has the same limit-order classification as the logarithmic one. At fixed $r$, Eq.~\eqref{eq:rho_exp2} vanishes as $\eta\to0$, while Eq.~\eqref{eq:b_exp2} approaches $r_0$. Taking the large-radius limit first also gives $b_\infty\to r_0$, so the density contribution to the ADM mass vanishes in either order. The pole-free exponential limits therefore commute, leaving the boundary term $r_0/2$.

\begin{figure}[!htp]
    \centering
    \includegraphics[width=0.49\linewidth]{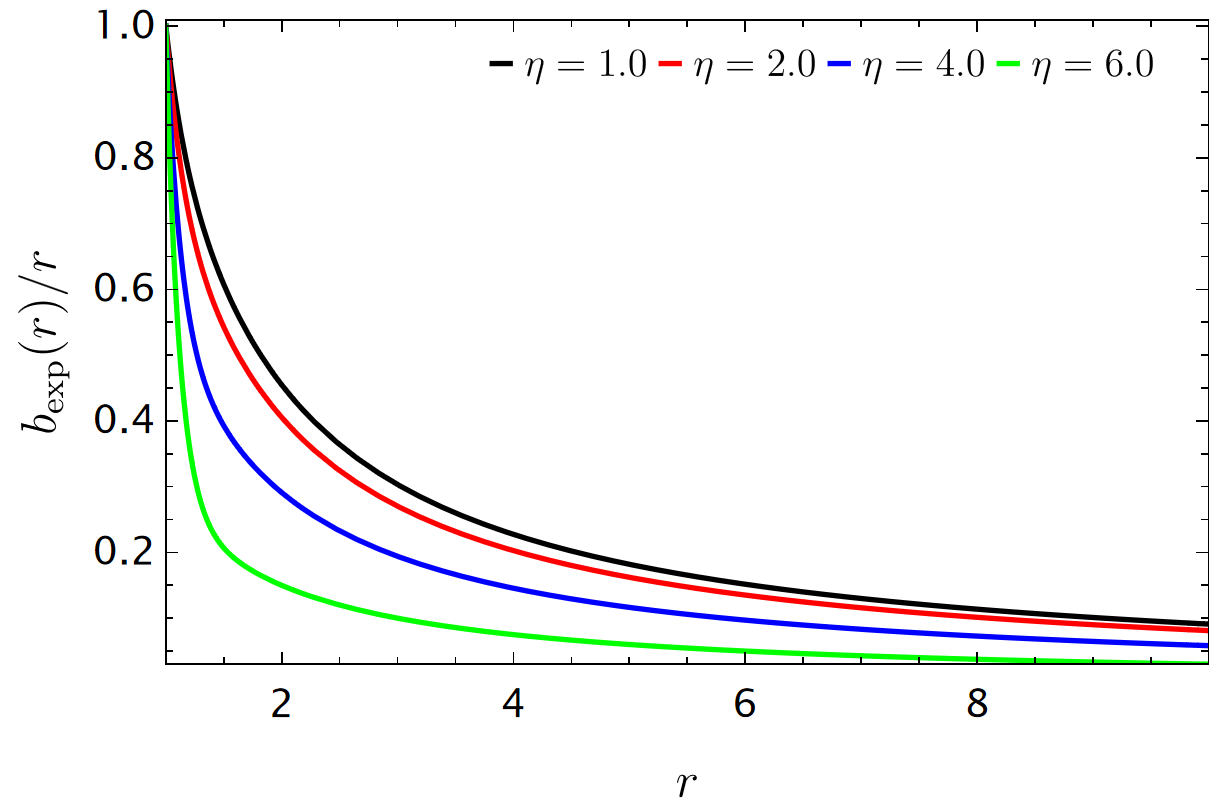}
    \includegraphics[width=0.49\linewidth]{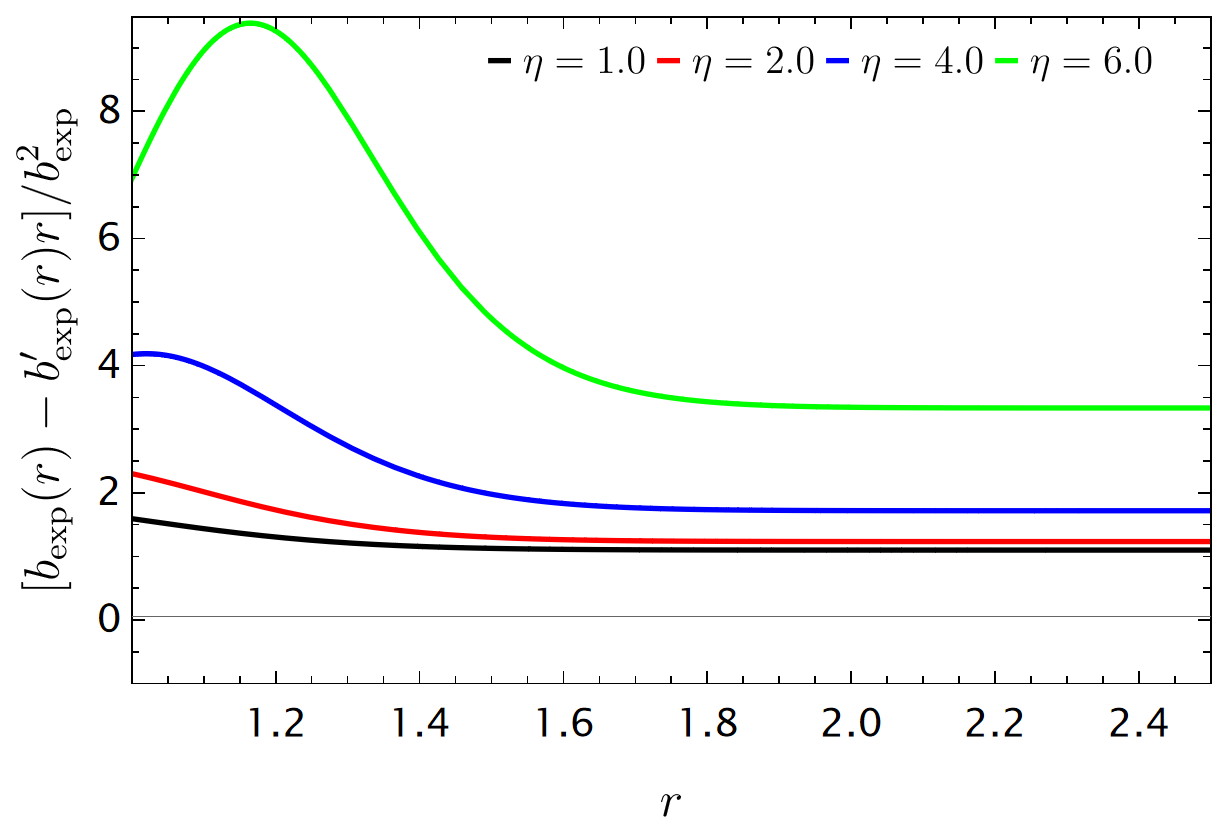}
    \caption{The geometric behavior of the exponential-inspired wormhole for different values of the parameter $\eta$, with $M=1$ and $r_0=1$ fixed. In the left panel, we show the ratio $b_{\text{exp}}(r)/r$ as a function of the radial coordinate $r$. In the right panel, we present the flare-out function $[b_{\text{exp}}(r)-b_{\text{exp}}'(r)r]/b_{\text{exp}}^2$ as a function of $r$. The curves correspond to $\eta=1.0$, $2.0$, $4.0$, and $6.0$.}
    \label{fig:geoexp}
\end{figure}

The geometric behavior associated with Eq.~\eqref{eq:b_exp2} is displayed in Fig.~\ref{fig:geoexp}. The left panel shows that $b_{\exp}(r)/r$ decreases and tends to zero for all values of $\eta$ considered, confirming the shape-function part of the asymptotic flatness condition. The right panel verifies the flare-out behavior. For the plotted values of $\eta$, the quantity $b_{\exp}(r)-r b'_{\exp}(r)$ remains positive in the physical region, indicating that the spatial geometry opens outward from the throat. The larger values of $\eta$ produce a stronger near-throat modification, but this effect is rapidly suppressed away from the throat because of the exponential dependence.

Substituting Eq.~\eqref{eq:b_exp2} into the general redshift equation \eqref{eq:redshift_general2}, one obtains
\begin{equation}
 \Phi_{\exp}'(r)=
 \frac{
 r_0+2\eta M
 \left[
 \dfrac{1}{e^{\pi r^2}-\eta}
 -\dfrac{1}{e^{\pi r_0^2}-\eta}
 \right]
 -\dfrac{4\pi \eta w M r^2 e^{\pi r^2}}{\left(e^{\pi r^2}-\eta\right)^2}
 }
 {
 2r\left[
 r-r_0-2\eta M
 \left(
 \dfrac{1}{e^{\pi r^2}-\eta}
 -\dfrac{1}{e^{\pi r_0^2}-\eta}
 \right)
 \right]
 }.
 \label{eq:phi_exp2}
\end{equation}
Compared with the other entropy sectors, the exponential branch stands out as the most strongly suppressed profile in the asymptotic region. While the logarithmic case introduces a finite correction scale through a rational denominator, the present branch depends on the exponential area factor $e^{\pi r^2}$. As a consequence, the redshift response is expected to be highly localized near the throat. The regularity of Eq.~\eqref{eq:phi_exp2} at $r=r_0$ will be addressed in the next subsection, where the equation-of-state parameter is fixed by the throat-regularity condition.

Although Eq.~\eqref{eq:phi_exp2} is sufficient for the geometric and matter-sector analysis, it is useful to comment on the corresponding redshift function. Unlike the logarithmic sector, where the redshift equation becomes rational after a suitable reparametrization, the exponential branch does not generally admit a compact elementary primitive for arbitrary $\eta$. Nevertheless, $\Phi_{\exp}(r)$ can be consistently obtained by numerical integration for all admissible values satisfying Eq.~\eqref{eq:eta_cond2}, with the integration constant fixed by the asymptotic normalization $\Phi_{\exp}(\infty)=0$. This does not limit the subsequent analysis, since the relevant physical quantities depend on $\Phi_{\exp}'(r)$ and on its finite limiting behavior at the throat.

\begin{figure}[!htp]
    \centering
    \includegraphics[width=0.58\linewidth]{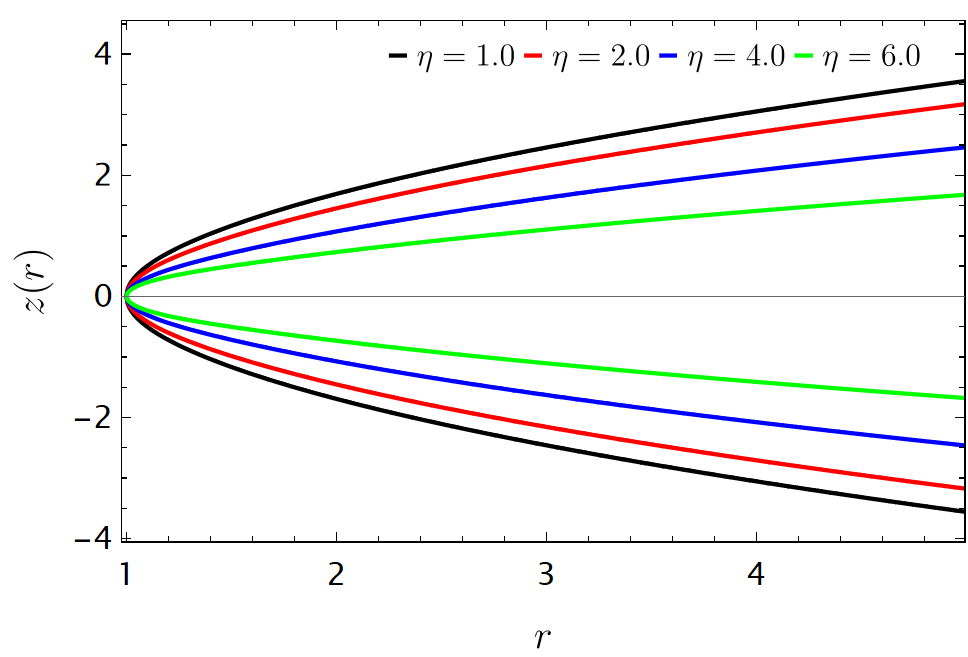}
    \includegraphics[width=0.4\linewidth]{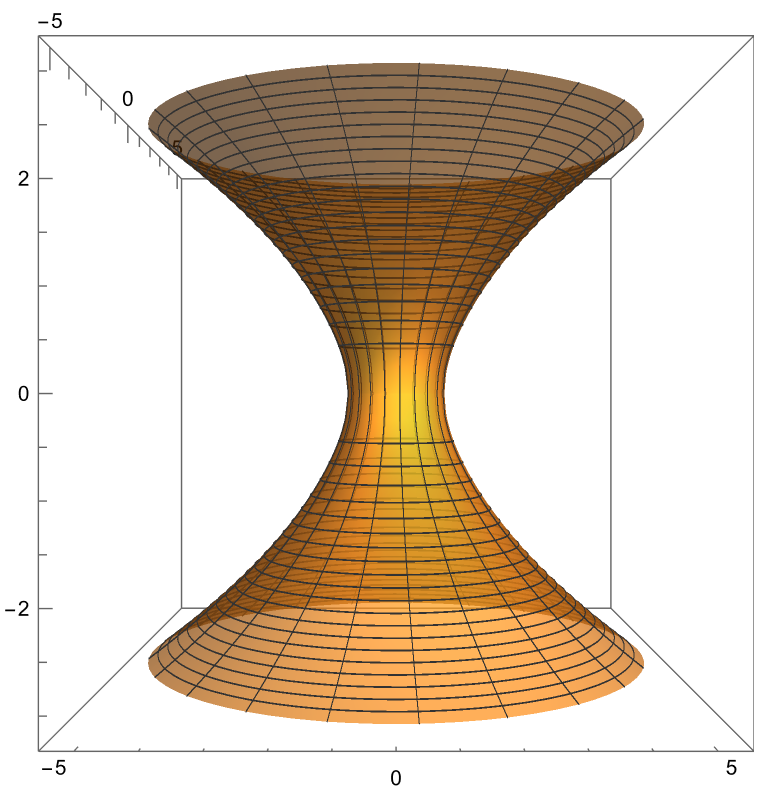}
    \caption{Embedding structure of the exponential-inspired wormhole geometry for different values of the parameter $\eta$, with $M=1$ and $r_0=1$ fixed. In the left panel, we show the embedding function $z(r)$ as a function of the radial coordinate $r$ for $\eta=1.0$, $2.0$, $4.0$, and $6.0$. In the right panel, we present the corresponding three-dimensional embedded surface for the representative case $\eta=2.0$, with $M=1$ and $r_0=1$ fixed.}
    \label{fig:embeddedexp}
\end{figure}

The embedding diagram in Fig.~\ref{fig:embeddedexp} provides a complementary visualization of the exponential-inspired spatial geometry. The left panel shows the embedding function $z(r)$ for different values of $\eta$, while the right panel presents a representative three-dimensional embedded surface. As in the previous sectors, the vertical tangent at the throat is the expected signature of a minimum-radius surface and does not indicate a physical singularity. The exponential branch modifies the opening of the embedded surface mainly through a sharply localized near-throat contribution. This distinguishes it from the power-law sectors, where the deformation is distributed over a longer radial range, and from the logarithmic sector, where the finite scale produces a smoother rational modification.

\subsubsection{Energy conditions and equilibrium in the exponential-inspired sector}

The radial pressure is fixed by the chosen equation of state,
\begin{equation}
 p_{r,\exp}(r)=w_{\exp}\rho_{\exp}(r).
\end{equation}
As in the previous sectors, the equation-of-state parameter is selected by the regularity of the redshift derivative at the throat. For the exponential-inspired shape function, one obtains
\begin{equation}
 b_{\exp}'(r)=
 -\frac{4\pi\eta M r e^{\pi r^2}}
 {\left(e^{\pi r^2}-\eta\right)^2}.
\end{equation}
Therefore, the throat-regularity condition $w=-1/b'(r_0)$ gives
\begin{equation}
 w_{\exp}\equiv w_{\rm th}^{(\exp)}
 =
 \frac{\left(e^{\pi r_0^2}-\eta\right)^2}
 {4\pi\eta M r_0 e^{\pi r_0^2}}
 =
 \frac{e^{-\pi r_0^2}\left(e^{\pi r_0^2}-\eta\right)^2}
 {4\pi\eta M r_0}.
 \label{eq:wth_exp}
\end{equation}
For $\eta>0$, $M>0$, and within the pole-free domain $e^{\pi r_0^2}-\eta>0$, one has $w_{\exp}>0$. Since $\rho_{\exp}<0$, this again corresponds to a negative radial pressure, namely a radial tension. However, differently from the logarithmic sector, the exponential branch does not allow a positive-density phantom-like regime within the parameter range considered here. The required radial exoticity is instead generated by a negative effective density whose magnitude is controlled by the competition between $\eta$ and the exponential denominator. As $\eta\to0$ along the regular branch, $w_{\exp}$ diverges but the throat pressure approaches a finite value. Thus, as in the logarithmic sector, commuting mass limits do not make the full source vacuum.

Using Eq.~\eqref{eq:pt_from_TOV2} together with Eqs.~\eqref{eq:rho_exp2} and \eqref{eq:phi_exp2}, with $w=w_{\exp}$, the tangential pressure can be written explicitly as
\begin{align}
 p_{t,\exp}(r)=\;&
 -\frac{w_{\exp}\eta M e^{\pi r^2}}
 {4r\left(e^{\pi r^2}-\eta\right)^2}
 +\frac{\pi w_{\exp}\eta M r e^{\pi r^2}
 \left(e^{\pi r^2}+\eta\right)}
 {2\left(e^{\pi r^2}-\eta\right)^3}
 \nonumber\\[0.3em]
 &-\frac{(1+w_{\exp})\eta M e^{\pi r^2}}
 {4\left(e^{\pi r^2}-\eta\right)^2}
 \,\Phi_{\exp}'(r).
\end{align}
This shows that the anisotropy of the exponential branch is driven by two competing effects: the rapid suppression produced by the exponential factor and the enhancement associated with the denominator $e^{\pi r^2}-\eta$. Thus, although this sector shares with the logarithmic branch the presence of a pole-avoidance condition, its physical behavior is more strongly localized. In the logarithmic case the correction scale modifies the source through a rational profile; here, the exponential dependence confines the relevant pressure response much more sharply around the throat.

The radial NEC is
\begin{equation}
 \rho_{\exp}+p_{r,\exp}
 =-
 \frac{e^{\pi r^2}\left[4\pi\eta M r_0e^{\pi r_0^2}+
 \left(e^{\pi r_0^2}-\eta\right)^2\right]}
 {8\pi r r_0 e^{\pi r_0^2}\left(e^{\pi r^2}-\eta\right)^2}.
\end{equation}
Since $\rho_{\exp}<0$ and $w_{\exp}>0$ in the branch considered here, the radial NEC is violated throughout the physical domain. At the throat, this violation is again fixed by the flare-out condition,
\begin{equation}
 \left.\left(\rho_{\exp}+p_{r,\exp}\right)\right|_{r_0}
 =-
 \frac{4\pi\eta M r_0e^{\pi r_0^2}+\left(e^{\pi r_0^2}-\eta\right)^2}
 {8\pi r_0^2\left(e^{\pi r_0^2}-\eta\right)^2}<0.
\end{equation}
Thus, the exponential sector provides another negative-density realization of Eq.~\eqref{eq:NEC_throat_geometry}: the radial NEC violation is fixed by the throat geometry, while the exponential profile controls how rapidly the exoticity is suppressed away from the throat.

\begin{figure}[!htp]
    \centering
    \includegraphics[width=0.49\linewidth]{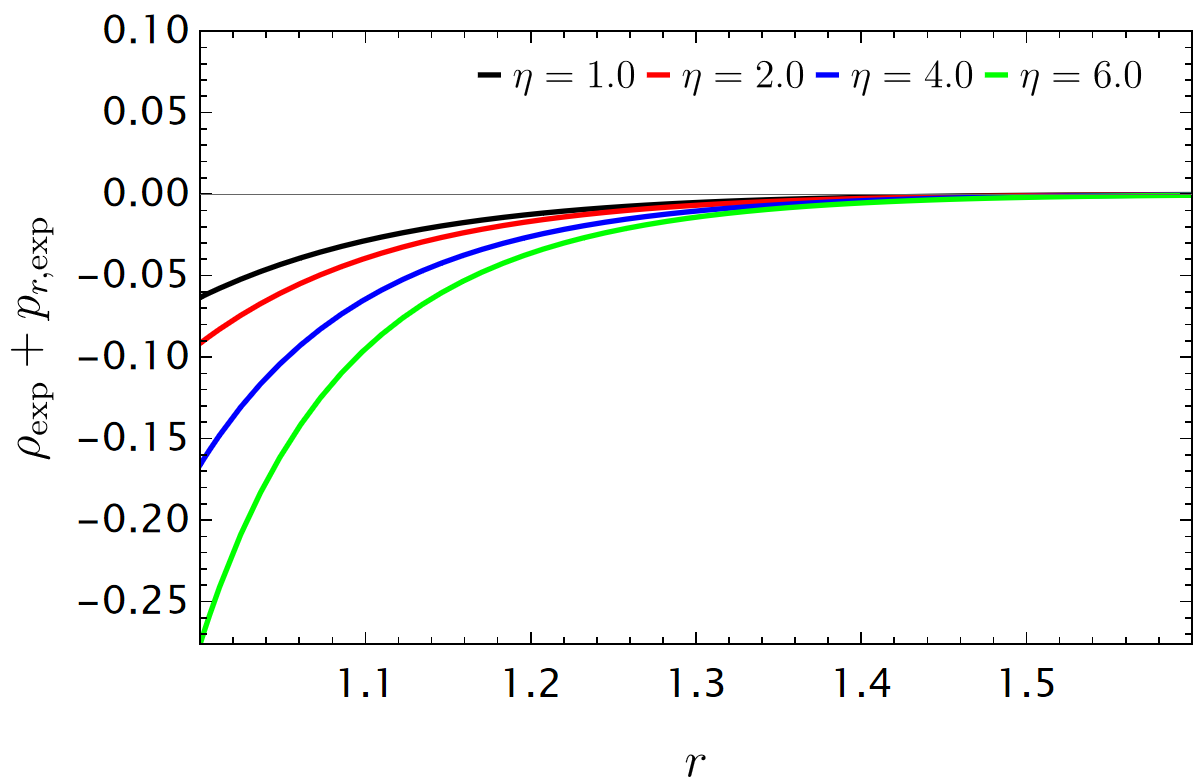}
    \includegraphics[width=0.49\linewidth]{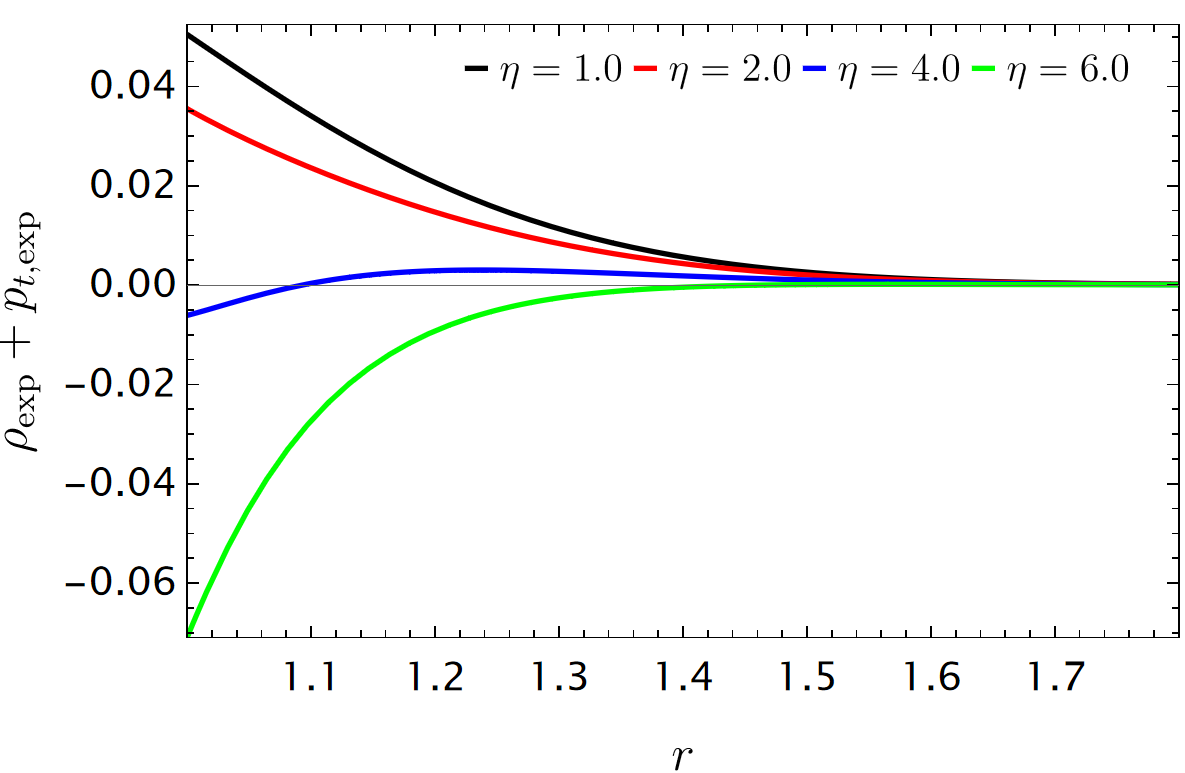}
    \includegraphics[width=0.55\linewidth]{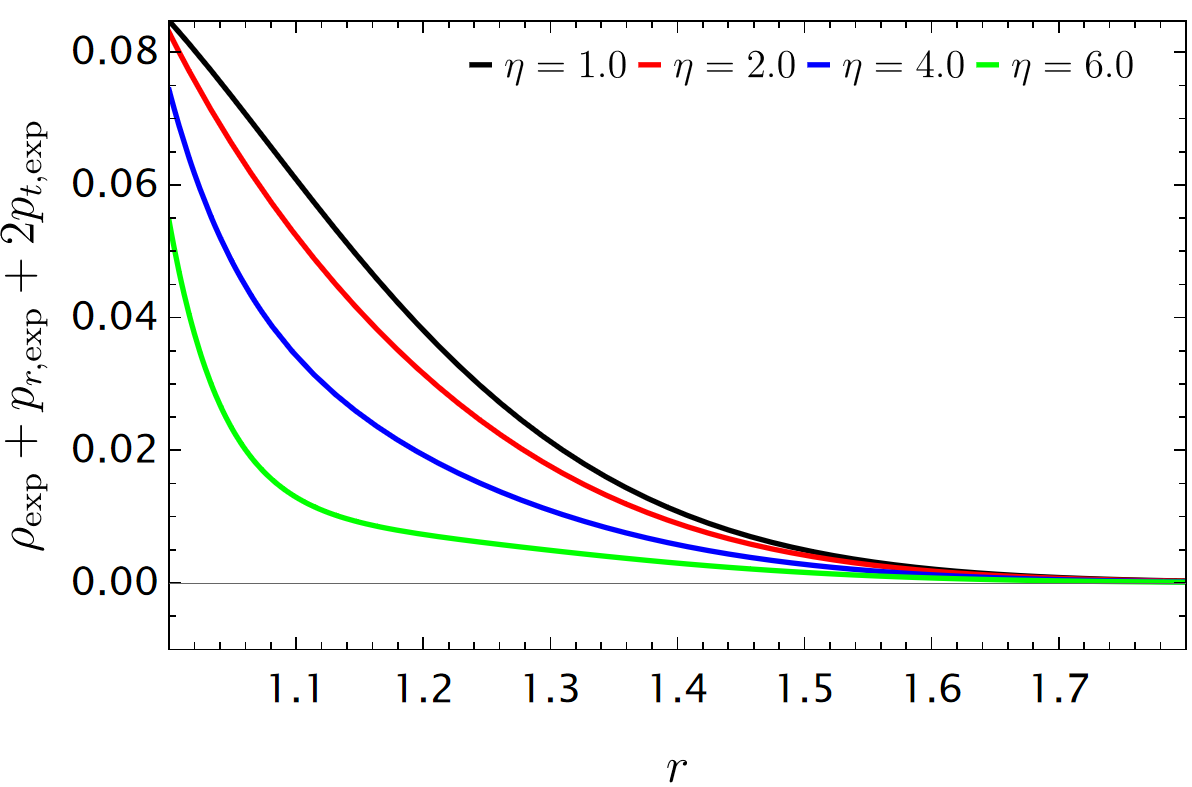}
    \caption{Behavior of the energy conditions for the exponential-inspired wormhole, with $M=1$ and $r_0=1$ fixed. The top-left panel shows the behavior of the radial NEC, the top-right panel shows the behavior of the tangential NEC, and the bottom panel shows the behavior of the SEC combination. We set $\eta=1.0$, $2.0$, $4.0$ and $6.0$. In all panels, the parameter $w_{\exp}$ is fixed for each value of $\eta$ by the throat-regularity condition \eqref{eq:wth_exp}.}
    \label{fig:necexp}
\end{figure}

Figure~\ref{fig:necexp} displays the radial NEC, tangential NEC, and SEC combination for the exponential-inspired sector. In all panels, $w_{\exp}$ is fixed by Eq.~\eqref{eq:wth_exp} for each value of $\eta$. The top-left panel shows that the radial NEC is negative for all plotted configurations, with a stronger near-throat violation as $\eta$ increases. This reflects the amplification of the negative density when the denominator becomes smaller. The top-right panel reveals a more delicate behavior: for smaller values of $\eta$, the tangential NEC remains positive near the throat, while larger values can drive the tangential sector toward negative values. This indicates that, in the exponential branch, the exoticity may become less purely radial as the source becomes more concentrated. The bottom panel shows that the SEC combination remains positive in the plotted range. As in the logarithmic sector, this should not be interpreted as satisfaction of the full SEC, since the radial NEC is already violated. Rather, it indicates that the tangential pressure can compensate part of the radial exoticity in the total pressure combination.

Combining Eqs.~\eqref{eq:wth_exp}, \eqref{eq:b_exp2}, and \eqref{eq:integrated_nec_identity} yields
{
\begin{equation}
 \mathcal{I}_{\mathrm{NEC},\exp}^{\pm}
 =-\frac{\eta M}{e^{\pi r_0^2}-\eta}
  -\frac{e^{\pi r_0^2}-\eta}{4\pi r_0e^{\pi r_0^2}}.
 \label{eq:integrated_nec_exp}
\end{equation}
}
Both terms are negative for the pole-free branch with $\eta>0$, and the exponential suppression guarantees convergence for every admissible configuration. In the limit $\eta\to0$, the density contribution vanishes, whereas the factor $w_{\exp}\propto\eta^{-1}$ compensates the decreasing source amplitude in the radial pressure. The accumulated violation consequently approaches the finite one-ended value $-1/(4\pi r_0)$. Thus, rapid localization prevents the undeformed limit from developing the long-range divergence found in the algebraic and Kaniadakis branches.

Analyzing the equilibrium condition, we have
\begin{equation}
 F_{h,\exp}+F_{g,\exp}+F_{a,\exp}=0,
\end{equation}
where the three force contributions are
\begin{equation}
 F_{h,\exp}(r)=
 -\frac{w_{\exp}\eta M e^{\pi r^2}}
 {2r^2\left(e^{\pi r^2}-\eta\right)^2}
 -\frac{\pi w_{\exp}\eta M e^{\pi r^2}
 \left(e^{\pi r^2}+\eta\right)}
 {\left(e^{\pi r^2}-\eta\right)^3},
\end{equation}
\begin{equation}
 F_{g,\exp}(r)=
 \frac{(1+w_{\exp})\eta M e^{\pi r^2}}
 {2r\left(e^{\pi r^2}-\eta\right)^2}
 \,\Phi_{\exp}'(r),
\end{equation}
and
\begin{equation}
 F_{a,\exp}(r)=
 \frac{w_{\exp}\eta M e^{\pi r^2}}
 {2r^2\left(e^{\pi r^2}-\eta\right)^2}
 +\frac{\pi w_{\exp}\eta M e^{\pi r^2}
 \left(e^{\pi r^2}+\eta\right)}
 {\left(e^{\pi r^2}-\eta\right)^3}
 -\frac{(1+w_{\exp})\eta M e^{\pi r^2}}
 {2r\left(e^{\pi r^2}-\eta\right)^2}
 \,\Phi_{\exp}'(r).
\end{equation}
These expressions make clear that the exponential branch has one of the most sharply localized equilibrium structures among the sectors considered here. In the Barrow and Tsallis cases, the force terms inherit algebraic tails from the density profile. In the Kaniadakis case, the support is localized by hyperbolic functions. Here, the exponential denominator produces a rapid suppression away from the throat, while the same denominator can strongly enhance the near-throat force amplitudes as $\eta$ increases within the admissible domain.

\begin{figure}[!htp]
    \centering
    \includegraphics[width=0.6\linewidth]{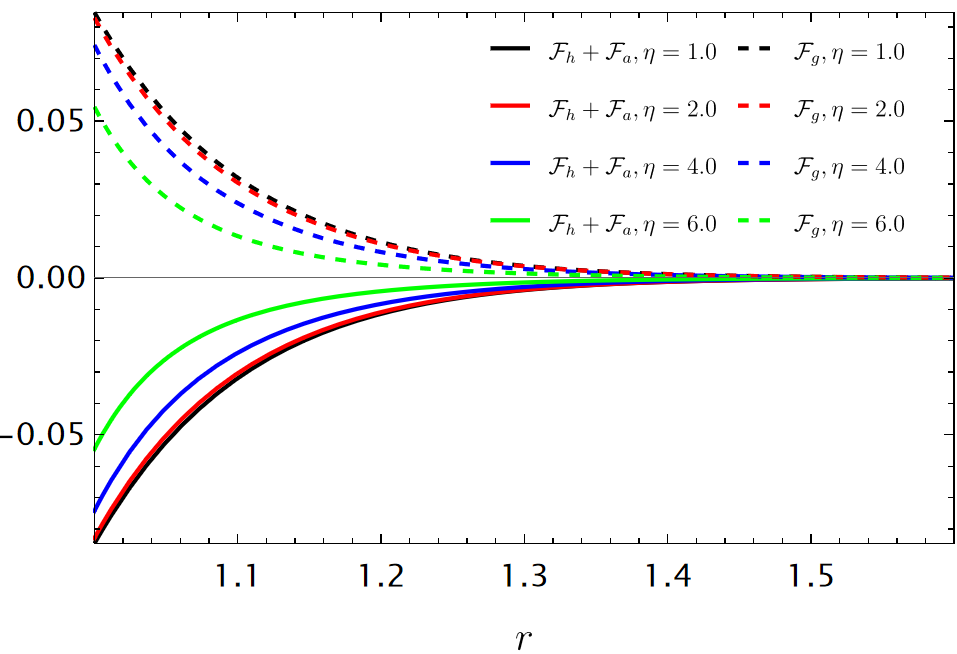}
    \caption{Equilibrium forces for the exponential-inspired wormhole as functions of the radial coordinate $r$, with $M=1$ and $r_0=1$ fixed. Solid curves represent the combined anisotropic and hydrostatic contribution $F_{a,\exp}+F_{h,\exp}$, while dashed curves represent the gravitational contribution $F_{g,\exp}$. We set $\eta=1.0$, $2.0$, $4.0$, and $6.0$, with $w_{\exp}$ fixed for each curve by the throat-regularity condition \eqref{eq:wth_exp}.}
    \label{fig:tovexp}
\end{figure}

The TOV balance is displayed in Fig.~\ref{fig:tovexp}. The solid curves represent the combined contribution $F_{h,\exp}+F_{a,\exp}$, while the dashed curves represent the gravitational contribution $F_{g,\exp}$. For each value of $\eta$, the two contributions have opposite signs and comparable magnitudes, confirming the equilibrium relation. Increasing $\eta$ strengthens the near-throat response of the force terms, but the exponential profile also makes this response decay rapidly with $r$. This behavior distinguishes the exponential sector from the logarithmic one: both contain a denominator-controlled enhancement, but the exponential branch confines the equilibrium to a narrower radial region. Therefore, the TOV analysis indicates that the exotic support is not only negative-density driven, but also strongly localized around the throat.

From the phenomenological point of view, the exponential-inspired sector provides a particularly clean example of a localized negative-density source. It does not interpolate between negative-density and phantom-like positive-density regimes, as the logarithmic branch does. Instead, the parameter $\eta$ controls how intense and narrow the negative effective matter distribution becomes. For the values considered here, the configurations remain safely below the pole threshold while showing that the exponential correction can sustain the throat with rapidly decaying matter and force contributions in the exterior region.

\section{Comparative global properties}
\label{sec:mass_comparison}

The local and asymptotic analyses of the five sectors provide two complementary kinds of global information. The asymptotic shape coefficient determines the mass carried by each end and tests whether removal of the entropy deformation commutes with the infinite-radius limit. The radial NEC integral instead measures the accumulated pressure-weighted exoticity. These quantities are connected through the shape-function reconstruction, but they probe different properties of the effective source and are therefore compared separately below.

\subsection{Asymptotic masses and undeformed limits}

Table~\ref{tab:asymptotic_limits} collects the finite asymptotic coefficients obtained directly from the five shape functions and shows the two successive limit orders. The first limit column takes $r\to\infty$ before removing the deformation, whereas the second removes the deformation at fixed $r$ before taking the asymptotic limit.

For each fixed, nonzero deformation parameter, the five solutions also satisfy $r b'(r)\to0$. The redshift equation~\eqref{eq:redshift_general2}, with $\Phi(\infty)=0$, then gives
{
\begin{equation}
 \Phi(r)=-\frac{b_\infty}{2r}+o\!\left(r^{-1}\right),
 \qquad r\to\infty,
 \label{eq:phi_asymptotic_mass}
\end{equation}
}
so the temporal and spatial asymptotic coefficients provide the same mass for every regular member of the families considered.

\begin{table*}[!htb]
\centering
\renewcommand{\arraystretch}{1.35}
\begin{tabular}{lccc}
\hline\hline
Sector & $b_\infty$ & \shortstack{asymptotic\\first} & \shortstack{undeformed\\first}\\
\hline
Barrow & $r_0-\dfrac{4M\pi^{-\Delta/2}}{\Delta+2}r_0^{-\Delta}$ & $r_0-2M$ & $r_0$\\
Tsallis & $r_0-\dfrac{2M\pi^{1-\delta}}{\delta}r_0^{2-2\delta}$ & $r_0-2M$ & $r_0$\\
Kaniadakis & $r_0-2M\operatorname{sech}(\pi\kappa r_0^2)$ & $r_0-2M$ & $r_0$\\
Logarithmic & $r_0+\dfrac{2\lambda M}{\lambda+\pi r_0^2}$ & $r_0$ & $r_0$\\
Exponential & $r_0-\dfrac{2\eta M}{e^{\pi r_0^2}-\eta}$ & $r_0$ & $r_0$\\
\hline\hline
\end{tabular}
\caption{Asymptotic shape coefficients and successive undeformed limits for the five entropy-inspired sectors. The undeformed values are $\Delta\to0$, $\delta\to1$, $\kappa\to0$, $\lambda\to0$, and $\eta\to0$, respectively. The ADM mass at either end of the symmetric construction is obtained from Eq.~\eqref{eq:madm_general}.}
\label{tab:asymptotic_limits}
\end{table*}

The comparison separates the sectors into two classes. For Barrow, Tsallis, and Kaniadakis, the two limit orders do not commute: although the density vanishes pointwise and the fixed-radius shape function approaches $r_0$, the infinity-first limit retains the same density-sector contribution $-M$ to the mass at each symmetric end. The mechanism is algebraic in the Barrow and Tsallis cases, whereas in the Kaniadakis sector the characteristic support moves outward as $\kappa\to0$. For the logarithmic and exponential sectors, the limits commute in their pole-free domains and the density contribution vanishes, leaving only the boundary term $r_0/2$, fixed by $b(r_0)=r_0$, in the ADM mass. Thus asymptotic flatness is compatible with a finite, parameter-dependent mass, and the imported amplitude $M$ should not be interpreted as the mass of either wormhole end.

These results also delimit the meaning of the undeformed limit used here. They classify the pointwise density, improper integral, and asymptotic coefficient of the explicit solutions, but they do not turn the full reconstructed source into vacuum at fixed throat radius and do not establish a complete Geroch limit of the spacetime family.

\subsection{Integrated radial exoticity}

The sector-specific results in Eqs.~\eqref{eq:integrated_nec_B}, \eqref{eq:integrated_nec_T}, \eqref{eq:integrated_nec_K}, \eqref{eq:integrated_nec_log}, and \eqref{eq:integrated_nec_exp} share one basic property: for every admissible nonzero deformation, the one-ended integral converges and is negative. Thus none of the regular configurations considered has an infinite accumulated radial NEC violation. The distinction appears only when the entropy deformation is removed along these regular branches.

Barrow and Tsallis form an algebraic class. Their weighted NEC profiles approach a marginal $r^{-1}$ tail as $\Delta\to0$ and $\delta\to1^{+}$, respectively, while the throat-selected pressure coefficient diverges. Their integrated exoticity is therefore unbounded from below. Kaniadakis belongs to the same divergent class, but through a different mechanism: the hyperbolic profile is integrable for every $\kappa>0$, yet its support expands and the pressure weighting grows as $\kappa\to0$, producing an even stronger inverse-power divergence.

The logarithmic and exponential sectors define the finite class. Their density contribution to the asymptotic mass vanishes when the deformation is removed, but the simultaneous growth of the throat-selected coefficient leaves a finite pressure contribution. The logarithmic branches approach $-r_0/4$ per end, independently of whether the density is negative or positive and phantom supported, whereas the exponential branch approaches $-1/(4\pi r_0)$ per end. Hence, commuting density-mass limits do not imply vanishing integrated exoticity; they show instead that the suppressed density and the reconstructed pressure have a finite combined radial effect.

All these statements concern limits of regular configurations with nonzero deformation. The exactly undeformed endpoint is not assigned an integrated value within the constant-$w$ family, because no finite throat-regular coefficient exists there. This closure limitation is distinct from the convergence of every deformed solution and from the separate question of defining another undeformed wormhole with a different pressure prescription.

\section{Conclusions}
\label{sec:conclusions}
\label{sec:conclusion}

In this work, we have investigated static and spherically symmetric traversable wormhole geometries using effective density profiles inspired by modified black-hole entropies. The construction follows a density-sector reconstruction: the entropy--geometry correspondence supplies the radial source profile, and the Morris--Thorne field equations determine the pressures required by the traversable wormhole geometry. Accordingly, our results do not show that generalized entropy itself, or the complete entropy-derived black-hole stress tensor, directly supports a traversable wormhole. They show instead that the selected entropy-induced density profiles can be used as phenomenological inputs when the wormhole pressures are reconstructed independently from the Morris--Thorne equations and throat regularity.

A central outcome of the analysis is that the equation-of-state parameter cannot be treated as an arbitrary free parameter once the redshift function is required to be regular at the throat. The throat condition fixes this parameter in terms of the derivative of the shape function, or equivalently in terms of the density evaluated at the throat. As a consequence, the apparent indeterminate behavior of the redshift derivative at the throat is not a pathology, but a removable feature, provided the density profile is sufficiently smooth and the flare-out condition is not saturated. This observation is essential for the consistency of the construction, since it connects the regularity of the temporal metric component directly to the local wormhole geometry.

The asymptotic analysis supplies the complementary global interpretation. The imported source amplitude is not generally the ADM mass of the reconstructed geometry, and asymptotic flatness does not imply vanishing mass. The undeformed and infinite-radius limits fail to commute for the Barrow, Tsallis, and Kaniadakis sectors, while they commute for the logarithmic and exponential sectors in their pole-free domains. Moreover, removal of the density deformation does not by itself define a vacuum limit of the complete reconstructed source.

The radial null energy condition is violated in all regular configurations considered. After imposing the throat-regularity condition, this violation is no longer the result of an arbitrary choice of the equation-of-state parameter; rather, it follows directly from the flare-out condition at the throat. The tangential NEC and the strong energy condition combination, however, distinguish the different entropy-inspired sectors. In this sense, the radial NEC diagnoses the exoticity required by the throat, whereas the tangential NEC and the SEC combination reveal how this exoticity is distributed through anisotropic stresses and redshift gradients.

The areal-volume integral of the radial NEC combination is finite and negative for every nonzero deformation considered. Along the regular undeformed limits, it becomes unbounded in the Barrow, Tsallis, and Kaniadakis sectors, whereas the logarithmic and exponential sectors approach finite negative values. The integrated exoticity is therefore controlled by the reconstructed pressure sector as well as by the density profile and cannot be inferred from the density contribution to the asymptotic mass alone.

The Barrow-inspired sector provides a power-law negative-density source controlled by the fractal deformation parameter. The Tsallis-inspired sector is also algebraic, but its non-extensive parameter controls the radial decay more strongly. In the parameter range considered, the Tsallis branch displays a more robust positive behavior of the SEC combination, indicating that the tangential pressure plays a stronger compensating role than in the Barrow case. Their equilibrium phenomenology is also distinct: in the Barrow branch, the gravitational and compensating hydrostatic--anisotropic contributions can reverse their signs together while preserving their mutual cancellation, whereas the Tsallis branch retains the same sign pattern while developing a progressively more localized near-throat balance.

The Kaniadakis-inspired sector differs qualitatively from the algebraic branches. Its density is governed by hyperbolic functions and is therefore localized in a finite radial region. The dependence on the deformation parameter is not simply monotonic: the source may be suppressed both for small and large values of the parameter, with an intermediate range where the effective matter sector is most relevant. This behavior is reflected in the energy conditions and in the TOV balance, where the tangential NEC and the SEC combination can change qualitatively depending on the value of the Kaniadakis parameter. The same localized redistribution can also drive a joint reversal of the gravitational and compensating pressure contributions without spoiling their mutual balance, showing that the sign pattern of the equilibrium is itself sensitive to the entropy deformation. The corresponding redshift function does not admit a compact elementary primitive, but it can be consistently obtained by numerical integration.

The logarithmic-inspired sector is the most flexible among the profiles analyzed. Unlike the negative-density branches of Barrow, Tsallis, Kaniadakis, and exponential corrections, the logarithmic density can change sign depending on the correction parameter. For negative values of this parameter, the wormhole is supported by a negative effective density, similarly to the previous sectors. For positive values, however, the density becomes positive and the throat is supported by a phantom-like radial pressure selected by the redshift regularity condition. Thus, the logarithmic branch provides two distinct mechanisms for sustaining the required radial NEC violation within the same entropic profile. It is also analytically distinctive, since the redshift function can be integrated in closed form for an arbitrary admissible value of the logarithmic parameter. Despite this change in the physical origin of the support, the TOV balance preserves the same sign pattern across the branches considered.

The exponential-inspired sector represents a sharply localized negative-density source. Its behavior is controlled by the competition between the exponential growth of the area term and the correction parameter. Although it shares with the logarithmic branch the presence of a denominator that restricts the admissible parameter range, its physical interpretation is different: the density remains negative in the branch considered, and the exotic support is increasingly concentrated near the throat as the correction parameter grows. The TOV analysis shows that the corresponding force balance is also strongly localized and does not reverse its sign pattern in the configurations considered, with all contributions rapidly suppressed away from the throat.

Taken together, the sector-by-sector results establish compatibility between the selected entropy-induced density profiles and regular traversable configurations within the adopted reconstruction and parameter ranges. The physical mechanism depends strongly on the entropy deformation. Algebraic sectors distribute the effective source through power-law tails, Kaniadakis produces a localized hyperbolic distribution, the logarithmic profile can interpolate between negative-density and phantom-like regimes, and the exponential correction confines the exotic matter more sharply near the throat. The embedding diagrams confirm the geometric opening of the spatial slices.

The present construction provides a controlled phenomenological bridge between modified entropy profiles and traversable wormhole matter sectors. Several extensions are therefore natural. One possibility is to relax the constant equation-of-state assumption and investigate whether a radially dependent closure yields a regular limit of the complete stress tensor when the density deformation is removed. Another direction is to compare the present density-sector reconstruction with models in which the full anisotropic effective sector generated by the entropy--geometry correspondence is mapped into a wormhole spacetime. It would also be useful to perform a systematic parameter scan including tidal constraints, stability under perturbations, and possible junction conditions with exterior vacuum geometries. Finally, the same framework can be extended to other entropy functionals, to rotating or time-dependent wormholes, and to observational probes such as lensing, shadows, and accretion signatures, allowing one to test how microscopic entropy deformations might leave macroscopic imprints on nontrivial spacetime geometries.

\section*{Acknowledgments}
CRM would like to thank Conselho Nacional de Desenvolvimento Cient\'{i}fico e Tecnol\'ogico (CNPq) for the partial financial support, through grant 301122/2025-3. FBL is funded by Fundação Cearense de Apoio ao Desenvolvimento Científico e Tecnológico (FUNCAP) and by  Conselho Nacional de Desenvolvimento Científico e Tecnológico (CNPq), grant number 305947/2024-9.

\bibliographystyle{apsrev4-1}
\bibliography{ref.bib}
\end{document}